\documentclass{article}

\usepackage{setspace}
\usepackage[final,nonatbib]{neurips_2023}
\makeatletter
\def\@noticestring{}
\makeatother

\usepackage[utf8]{inputenc}
\usepackage[T1]{fontenc}
\usepackage{xurl}
\usepackage{hyperref}
\usepackage{booktabs}
\usepackage{amsfonts}
\usepackage{nicefrac}
\usepackage{microtype}
\usepackage{graphicx}
\usepackage{tabularx}
\usepackage{longtable}
\usepackage[dvipsnames,table,xcdraw]{xcolor}

\definecolor{light-gray}{HTML}{E5E4E2}
\usepackage[most]{tcolorbox}
\usepackage{enumitem}
\usepackage{arabtex}
\usepackage{utf8}
\hypersetup{
  linkcolor  = violet!85!black,
  citecolor  = Magenta,
  urlcolor   = Aquamarine!85!black,
  colorlinks = true,
}
\hypersetup{breaklinks=true}
\usepackage{xurl}
\usepackage[hypcap=false,font={small,bf}]{caption}

\setlist[enumerate]{leftmargin=*, label=\arabic*.}
\setlist[itemize]{leftmargin=*}
\setlist[itemize,1]{label=\textbullet}
\setlist[itemize,2]{label=\textopenbullet}

\usepackage[style=trad-abbrv,urldate=long,dateabbrev=false,sorting=none,maxbibnames=9,maxcitenames=2,citestyle=numeric-comp,%
]{biblatex}
\AtEveryBibitem{%
\ifentrytype{article}{%supresses the fields for those entrytypes
\clearfield{urlyear}%
\clearfield{pubstate}%
}{}%the other entrytypes are nore affected
}
\usepackage{array,booktabs,calc}
\usepackage{titlesec}
\titleformat{name=\subsubsection}
  {\normalfont\bfseries}{\thesubsubsection}{2em}{\hspace{-1em}}{}
\titleformat{name=\subsubsection,numberless}
  {\normalfont\bfseries}{}{1cm}{\hspace{0cm}}{}

\title{Open-Sourcing Highly Capable Foundation Models: An evaluation of risks, benefits, and alternative methods for pursuing open-source objectives}

\author{%
  \parbox{0.88\linewidth}{\centering\bfseries%
  Elizabeth Seger$^{1,2,}$\rule[-1.5pt]{0pt}{12pt}\thanks{Corresponding author: \href{mailto:elizabeth.seger@governance.ai}{elizabeth.seger@governance.ai}\\
  Please cite as Seger, Dreksler, Moulange, Dardaman, Schuett, Wei, et al, ‘Open-Sourcing Highly Capable Foundation Models: An Evaluation of Risks, Benefits, and Alternative Methods for Pursuing Open-Source Objectives’, Centre for the Governance of AI, 2023.\vspace{-12mm}}\quad Noemi Dreksler$^{1}$\quad Richard Moulange$^{1,3}$ \quad Emily Dardaman$^{4}$ \quad Jonas Schuett$^{1}$\quad K. Wei$^{1,5}$\quad \rule[-1.5pt]{0pt}{12pt}Christoph Winter$^{6,7,8}$\quad Mackenzie Arnold$^{8}$ 
  \quad Seán Ó hÉigeartaigh$^{2}$ \quad Anton Korinek$^{1,9,10}$ \quad Markus Anderljung$^{1}$ \quad 
  Ben Bucknall$^{11}$ \quad Alan Chan$^{12,13}$\rule[-1.5pt]{0pt}{12pt} \quad Eoghan Stafford$^{1}$ \quad Leonie Koessler$^{1}$ \quad
  Aviv Ovadya$^{14}$ \\ Ben Garfinkel$^{1}$ \quad Emma Bluemke$^{1}$ \quad Michael Aird$^{15}$\rule[-1.5pt]{0pt}{12pt} \quad  Patrick Levermore$^{15}$ \quad
  Julian Hazell$^{1,16}$\rule[-1.5pt]{0pt}{12pt} \quad Abhishek Gupta$^{3,17}$\rule[-1.5pt]{0pt}{13pt}} \\\\
  \parbox{0.97\linewidth}{
  $^1$Centre for the Governance of AI\quad $^2$AI: Futures and Responsibility Programme, University of Cambridge \quad $^3$MRC Biostatistics Unit, University of Cambridge\quad  $^4$BCG Henderson Institute \quad $^5$Harvard Law School \quad $^6$Harvard University \quad $^7$Instituto Tecnológico Autónomo de México \quad $^8$Legal Priorities Project \quad
  $^9$University of Virginia\quad $^{10}$Brookings Institution \quad $^{11}$Independent \quad
  $^{12}$Mila \quad $^{13}$University of Montreal \quad $^{14}$Thoughtful Technology Project \quad
  $^{15}$Institute for AI Policy \& Strategy \quad $^{16}$Oxford Internet Institute, University of Oxford \quad $^{17}$Montreal AI Ethics Institute} \\\\
  \parbox{0.88\linewidth}{\textbf{Acknowledgements:} We thank the following people for feedback and comments: Andrew Trask, Ben Cottier, Herbie Bradley, Irene Solaiman, Norman Johnson, Peter Cihon, Shahar Avin, Stella Biderman, Toby Shevlane.} \\\\
  \parbox{0.94\linewidth}{\emph{
  Given the size of the group, inclusion as an author does not entail endorsement of all claims in the paper, nor does inclusion entail an endorsement on the part of any individual’s organization.\\
  }}% \vspace{-2mm}
}

\begin{document}
\maketitle

\begin{abstract}
Recent decisions by leading AI labs to either open-source their models or to restrict access to their models has sparked debate about whether, and how, increasingly capable AI models should be shared. Open-sourcing in AI typically refers to making model architecture and weights freely and publicly accessible for anyone to modify, study, build on, and use. This offers advantages such as enabling external oversight, accelerating progress, and decentralizing control over AI development and use. However, it also presents a growing potential for misuse and unintended consequences. This paper offers an examination of the risks and benefits of open-sourcing highly capable foundation models. While open-sourcing has historically provided substantial net benefits for most software and AI development processes, we argue that for some highly capable foundation models likely to be developed in the near future, open-sourcing may pose sufficiently extreme risks to outweigh the benefits. In such a case, highly capable foundation models should not be open-sourced, at least not initially. Alternative strategies, including non-open-source model sharing options, are explored. The paper concludes with recommendations for developers, standard-setting bodies, and governments for establishing safe and responsible model sharing practices and preserving open-source benefits where safe.
\end{abstract}

\clearpage
\setcounter{footnote}{0}

\addcontentsline{toc}{section}{Executive Summary}
\section*{Executive Summary}\label{executive-summary}

Recent decisions by AI developers to open-source foundation models have
sparked debate over the prudence of open-sourcing increasingly capable
AI systems. Open-sourcing in AI typically involves making model
architecture and weights freely and publicly accessible for anyone to
modify, study, build on, and use. On the one hand, this offers \textbf{clear
advantages} including enabling external oversight, accelerating progress,
and decentralizing AI control. On the other hand, it presents \textbf{notable
risks}, such as allowing malicious actors to use AI models for harmful
purposes without oversight and to disable model safeguards designed to
prevent misuse.

This paper attempts to clarify open-source terminology and to offer a
thorough analysis of risks and benefits from open-sourcing AI. While
open-sourcing has, to date, provided substantial net benefits for most
software and AI development processes, we argue that for some highly
capable models likely to emerge in the near future, the risks of open
sourcing may outweigh the benefits.

There are three main factors underpinning this concern:

\begin{enumerate}

\item \textbf{Highly capable models have the potential for extreme risks.}
      Of primary concern is diffusion of dangerous AI capabilities that
      could pose extreme risks---risk of significant physical harm or
      disruption to key societal functions. Malicious actors might apply
      highly capable systems, for instance, to help build new biological and
      chemical weapons, or to mount cyberattacks against critical
      infrastructures and institutions. We also consider other risks such as
      models helping malicious actors disseminate targeted misinformation at
      scale or to enact coercive population surveillance.

    Arguably, current AI capabilities do not yet surpass a critical
    threshold of capability for the most extreme risks. However, we are
    already seeing nascent dangerous capabilities emerge, and this trend is
    likely to continue as models become increasingly capable and it becomes
    easier and requires less expertise and compute resources for users to
    deploy and fine-tune these models. (Section~\ref{sec3:risks-of-open-sourcing-foundation-models})

\item
      \textbf{Open-sourcing is helpful in addressing some risks, but could---overall---exacerbate the extreme risks that highly capable AI models may pose.} For traditional software, open-sourcing
      facilitates defensive activities to guard against misuse more so than
      it facilitates offensive misuse by malicious actors. However, the
      offense-defense balance is likely to skew more towards offense for
      increasingly capable foundation models for a variety of reasons
      including: (i) Open-sourcing allows malicious actors to disable safeguards against misuse and to possibly introduce new dangerous capabilities via fine-tuning. (ii) Open-sourcing greatly increases attacker knowledge of
      possible exploits beyond what they would have been able to easily
      discover otherwise. (iii) Researching safety vulnerabilities is
      comparatively time consuming and resource intensive, and fixes are
      often neither straightforward nor easily implemented. (iv) It is more
      difficult to ensure improvements are implemented downstream, and flaws
      and safety issues are likely to perpetuate further due to the general
      use nature of the foundation models. (Section~\ref{sec3:risks-of-open-sourcing-foundation-models})
\item
      \textbf{There are alternative, less risky methods for pursuing
      open-source goals.} There are a variety of strategies that might be
      employed to work towards the same goals as open-sourcing for highly
      capable foundation models but with less risk, albeit with their own
      shortcomings. These alternative methods include more structured model
      access options catered to specific research, auditing, and downstream
      development needs, as well as proactive efforts to organize secure
      collaborations, and to encourage and enable wider involvement in AI
      development, evaluation, and governance processes. (Section~\ref{sec4:benefits-of-open-sourcing-foundation-models-and-alternative-methods-for-achieving-them})
\end{enumerate}

In light of these potential risks, limitations, and alternatives,
\textbf{we offer the following recommendations} for developers,
standards setting bodies, and governments. These recommendations are to
help establish safe and responsible model sharing practices and to
preserve open-source benefits where safe. They also summarize the
paper's main takeaways. (Section~\ref{sec5:recommendations})

\begin{enumerate}
\item
      \textbf{Developers and governments should recognize that some highly
      capable models will be too risky to open-source, at least initially.}
      These models may become safe to open-source in the future as societal
      resilience to AI risk increases and improved safety mechanisms are
      developed.
\item
      \textbf{Decisions about open-sourcing highly capable foundation models
      should be informed by rigorous risk assessments.} In addition to
      evaluating models for dangerous capabilities and immediate misuse
      applications, risk assessments must consider how a model might be
      fine-tuned or otherwise amended to facilitate misuse.
\item
      \textbf{Developers should consider alternatives to open-source release
      that capture some of the same distributive, democratic, and societal
      benefits, without creating as much risk.} Some promising alternatives
      include gradual or ``staged'' model release, structured model access
      for researchers and auditors, and democratic oversight of AI
      development and governance decisions.
\item
      \textbf{Developers, standards setting bodies, and open-source
      communities should engage in collaborative and multi-stakeholder
      efforts to define fine-grained standards for when model components
      should be released.} These standards should be based on an
      understanding of the risks posed by releasing different combinations
      of model components.
\item
      \textbf{Governments should exercise oversight of open-source AI models
      and enforce safety measures when stakes are sufficiently high.} AI
      developers may not voluntarily adopt risk assessment and model sharing
      standards. Governments will need to enforce such measures through
      options such as liability law and regulation, licensing requirements,
      fines, or penalties. They will also need to build the capacity to
      enforce such oversight mechanisms effectively. Immediate work is
      needed to evaluate the costs, consequences, and legal feasibility of
      various policy interventions and enforcement mechanisms we list.
\end{enumerate}

\clearpage 
\tableofcontents
\clearpage

\section{Introduction}\label{sec1:introduction}

As AI developers build increasingly capable models, they face a dilemma
about whether and how they should share their models. One foundational
decision they must make is whether to open-source their models---that
is, make their models freely and publicly accessible for anyone to use,
study, modify, and share.\footnote{We use the term open-source without
  precise requirements on license permissions, but more generally to
  mean making a model publicly and freely available. See section~\ref{sec2:what-do-we-mean-by-open-source-highly-capable-foundation-models} for
  further discussion on open-source meaning and terminology.}

Software development communities have traditionally enjoyed strong norms
for sharing and open-source publication. Accordingly, for many AI
researchers and developers open-sourcing is a deeply held professional
and personal value. However, this value can sit in tension with others,
like growing a profitable organization may contradict protecting
consumers from harm
\cite{thecollectiveintelligenceproject2023}.
Debate continues about the risks, benefits, and tradeoffs of open-source
model release.

Recently, some large AI labs have decided that open-sourcing foundation
models involves unacceptable trade-offs and have chosen to restrict
model access out of competitive concerns and worries about model misuse.
These labs are either keeping their models completely private (e.g.,
DeepMind's Chinchilla
\cite{hoffmann2022}) or employing
a structured access approach to model sharing (e.g., OpenAI's GPT-4
\cite{openai} and
Anthropic's Claude 2
\cite{anthropic2023} via their
APIs \cite{brockman2023}, which
enable the enforcement of user restrictions and implementation of
controls such as safety filters in order to manage harms.

There has been Pushback against this trend to restrict model access and
calls to reinforce traditional software development community norms for
sharing and openness is common.. The concerns are that model access
restriction stifles innovation, disallows external oversight, hinders
the distribution of AI benefits, and concentrates control over
AI's future to a small number of major AI labs \cite{goldman2023,creativecommons2023}.
Labs such as Hugging Face, Allen Institute for AI, EleutherAI,
RedPajama, LAION, Together.xyz, Mosaic, and Stability AI have recently
chosen to open-source large models. Meta has been a particularly vocal
open-source proponent with its release of I-JEPA
\cite{assran2023}, an efficient
and visual transformer in June 2023, followed closely by Llama 2 \cite{metaai2023,inskeep2023,milmo2023}, in
July 2023.

There are many considerable benefits of open-source software (OSS)
development. For thirty years, OSS has proliferated alongside, and often
inside, of commercial software, encouraging cooperation, promoting
software adoption via lowered costs, reducing monopolistic control by
major software companies, fostering rapid innovation, growing talent,
and improving software quality through community review \cite{langenkamp2022,engler2021,engler2022}.
The academic tradition in which many machine learning researchers are
trained also enjoys strong norms of open research publication. It is
only natural that many machine learning developers and researchers
follow suit, creating groups and organizations like Hugging Face,
Stability AI, RedPajama, and EleutherAI in order to build and release
increasingly capable AI models.

However, we will explain that there is a disanalogy between OSS and
open-source AI, and that we should not expect these same benefits to
seamlessly translate from OSS to cutting-edge AI development efforts.
While it is natural that an OSS lens has been used to motivate the
open-sourcing of AI systems, continuing to do so could come with
significant downsides. The rapid increase in capabilities that we have
observed, and likely will continue to see, mean that open-sourcing AI
systems come with higher risks of misuse, accidents, and dangerous
structural effects than traditional software \cite{zwetsloot2019}.

In comparative terms, open-sourcing a model will tend to present greater
risks than releasing it using a structured access approach whereby model
access is mediated, for example, through an API
\cite{shevlane2022}. First, once
a model is open-sourced, any safeguards put in place by an AI lab to
prevent its misuse can be circumvented (see Section~\ref{sec3.1:malicious-use}). No methods
currently exist to reliably prevent this. Second, once a model is
open-sourced, those with sufficient expertise and computing resources
can, without oversight, "fine-tune" it to introduce and enhance
capabilities that can be misused. These two possibilities mean that any
threshold of safe behavior observed and evaluated under closed or
restricted contexts cannot necessarily be assumed to hold once the model
is made publicly available.\footnote{Since it is difficult to verify the
  safety of any model and ensure that you have observed the true range
  of possible behaviors, this also holds true for models that are not
  open-sourced. However, the fact models can be further fine-tuned,
  adapted, and integrated with other systems upon release means that the
  true range of possible behaviors can shift in unpredictable ways
  untestable at the pre-release stage.}

Furthermore, open-source AI model release is irreversible; there is no
``undo'' function if significant harms materialize. If a model has a
flaw---some exploit that elicits undesirable capabilities---or grave
misuse potential, there is nothing to stop users from continuing to use
the model once released. Similarly, if developers release patches or
updated model versions to remedy flaws, there is no way to ensure users
will implement the patches or operate the most up-to-date version. For
malicious users who seek to exploit model vulnerabilities that allow for
harmful applications, they are incentivized not to adopt any safety
improvements.

Ultimately, as AI labs push the boundaries of foundation model
development, the risks of open-sourcing will grow as models become
increasingly capable. The risks from such capability improvements could
become sufficiently severe that the benefits of open-sourcing outweigh
the costs. We therefore recommend that decisions to open-source highly
capable foundation models should be made only after careful deliberation
that considers (i) the range of misuse risks the open-source model may
present and (ii) the potential for open-source benefits to be provided
through alternative means. We expect that in the future some highly
capable foundation models should not be open-sourced.

We begin by defining highly capable foundation models (section~\ref{sec2:what-do-we-mean-by-open-source-highly-capable-foundation-models}) and
the risks presented by open-sourcing them (Section~\ref{sec3:risks-of-open-sourcing-foundation-models}). The harms are
significant and plausibly, in certain cases, justify foundation model
access restrictions. We then turn to three key arguments for open-source
model sharing and explore alternative mechanisms for achieving the
desired end with significantly less risk (Section~\ref{sec4:benefits-of-open-sourcing-foundation-models-and-alternative-methods-for-achieving-them}). Finally, we
present recommendations for AI developers and policymakers in light of
our discussion (Section~\ref{sec5:recommendations}).

\section{What Do We Mean by ``Open-Source Highly Capable Foundation Models''?}\label{sec2:what-do-we-mean-by-open-source-highly-capable-foundation-models}

\subsection{What are Highly Capable Foundation Models?}\label{what-are-highly-capable-foundation-models}

\paragraph{Foundation models.} Foundation models, sometimes referred to as
\emph{general-purpose} AI models, are machine learning models like GPT-4
that demonstrate a base of general capabilities that allow them to be
adapted to perform a wide range of downstream tasks
\cite{bommasani2022,jones2023}.
These capabilities can include natural language conversation, behavior
prediction, image analysis, and media generation\footnote{Today, many of
  the most discussed foundation models are generative AI systems that
  are variants of large language models (LLMs) like GPT-4 (the model
  which forms the base of the conversational ChatGPT interface). LLMs
  are machine learning models with complex architectures that generate
  plausible text or visual content in response to user prompts (that are
  often text-based). To do so, they are first trained on vast amounts of
  text, where they learn to predict the next token (or word). Additional
  training then steers the LLM towards providing outputs that humans
  rate highly---this makes it more likely that the LLM will provide
  helpful, non-toxic responses.}, which can be used to develop or be
directly integrated into other AI systems, products, and
models.\footnote{We are already seeing current-generation foundation
  models, like GPT-4, being integrated into clinical diagnoses in
  healthcare
  \cite{shea2023}, visual
  web accessibility tooling \cite{openai2023},
  qualitative data analysis
  \cite{openai2023a}, video
  game character development
  \cite{openai2023b}, customer
  assistance and support
  \cite{altmann2023}, foreign
  language education
  \cite{marr2023}, financial
  fraud detection
  \cite{openai2023c}, legal
  tools \cite{harvey.ai}, and
  many other industries. As their capabilities increase, future
  generations of foundation models will continue to be deployed across
  industry and government, integrating them into many downstream
  applications across a wide-range of sectors, including safety-critical
  applications.}

When modalities are combined, \emph{multimodal foundation models} can
integrate and respond to numerous data types (e.g., text, audio, images,
etc.). For instance, Stable Diffusion
\cite{rombach2022} and DALL·E 2
\cite{ramesh2022} combine
natural language processing capabilities with image generation
capabilities to translate natural language prompts into image outputs.
GPT-4 is also multimodal, though that functionality is not made widely
available
\cite{openai2023d},\footnote{Multimodal
  functionality is now available to some Microsoft Enterprise customers
  via BingChat
  \cite{mehdi2023}.} and
Meta's open-source ImageBind project aims to link up numerous streams of
data including audio, text, visual data, movement and temperature
readings to produce immersive, multi-sensory experiences
\cite{vincent2023}.

Foundation models can be used positively in healthcare
\cite{fries2022}, for data
analysis \cite{openai2023a},
customer support
\cite{openai2023b}, immersive
gaming \cite{marr2023a}, or
personalized tutoring
\cite{marr2023}. But they
can also be misused and deployed by bad actors, for example, to generate
child sexual abuse material
\cite{milmo2023a}, create fake
real-time interviews or recorded histories for influential politicians
\cite{horvitz2022}, or to
conduct highly-effective targeted scams convincing victims that they are
calling with trusted friends and family \cite{verma2023,brewster2021}.
Other current and ongoing harms posed by foundation models include, but
are not limited to, bias, discrimination, representational harms, hate
speech and online abuse, and privacy-invading information hazards \cite{bommasani2022,weidinger2022,solaiman2023,shelby2023}.

Foundation models have also been associated with upstream harms
including poor labor conditions in the supply chain and for those hired
to label data \cite{crawford2021,gray2019}
as well as putting strain on the environment through high energy and
resource usage during training, deployment, and the production of the
required hardware \cite{li2023,strubell2019,patterson2021}.

\paragraph{``Highly capable'' foundation models.} We define \emph{highly
capable foundation models} as foundation models that exhibit high
performance across a broad domain of cognitive tasks, often performing
the tasks as well as, or better than, a human.\footnote{We intentionally
  speak about ``highly-capable models'' instead of ``frontier models''.
  The ``frontier'' refers to the cutting-edge of AI development
  \cite{jones2023}, however
  the frontier of cutting-edge AI moves forward as AI research
  progresses. This means that some highly capable systems of
  concern---those capable of exhibiting dangerous capabilities with the
  potential to cause significant physical and societal-scale harm---will
  sit behind the frontier of AI capability. Even if these models are
  behind the frontier, we should still exercise caution in deciding to
  release such models, all else being equal.}

Researchers are working to develop suitable benchmarks to track the
increase in such general-purpose capabilities by measuring performance
of such models holistically (e.g., in regards to language, reasoning,
and robustness
\cite{liang2022} and across
a spectrum of specific areas of knowledge, from professional medicine
and jurisprudence to electrical engineering and formal logic
\cite{hendrycks2021}.

\paragraph{Extreme risks and harms.} In this paper we are particularly
concerned with the possibility probability that highly capable models
may come to exhibit dangerous capabilities causing extreme risks and
harms such as significant physical harm or disruption to key societal
functions.\footnote{Shevlane et al. \cite{shevlane2023} operationalise such extreme risks and harms in terms of the scale of
  the impact they could have---e.g., killing tens of thousands of people
  or causing hundreds of billions of dollars of economic or
  environmental damage---or the level of disruption this would cause to
  society and the political order.\\
  In their recently released \emph{Responsible Scaling Policy}
  \cite{anthropic2023a}, Anthropic
  distinguishes between four AI Safety Levels (ASL's). Like the
  Anthropic document, this paper is primarily focused on the likely near
  future development of ASL-3 models which are those that show
  ``\emph{low level autonomous capabilities}'' or for which
  ``\emph{access to the model would substantially increase the risk of
  catastrophic misuse, either by proliferating capabilities, lowering
  costs, or enabling new methods of attack as compared to non-LLM
  baseline of risk.}''}

Dangerous capabilities that highly capable foundation models could
possess include making it easier for non-experts to access known
biological weapons or aid in the creation of new ones
\cite{sandbrink2023}, or giving
unprecedented offensive cyberattack capabilities to malicious actors
\cite{mirsky2021,centerforsecurityandemergingtechnology2020}.
Being able to produce highly persuasive personalized disinformation at
scale, effectively produce propaganda and influence campaigns, or act
deceptively towards humans, could also present extreme risks
\cite{anderljung2023}.
Self-proliferation abilities, such as evading post-deployment monitoring
systems, gaining financial and computing resources without user or
developer consent, or a model exfiltrating its own trained weights, are
more speculative but might also facilitate extreme risks \cite{anthropic2023a,kinniment2023}.
This is particularly the case if models are embedded within critical
infrastructure. The magnitude of these risks requires that model
developers more carefully and systematically weigh risks against
benefits when making open-sourcing decisions for highly capable
foundation models than for present-day foundation models.

Perhaps in the future we will use AI models to guard against the risks
and harms presented by the misuse of, and accidents caused by, other AI
models, allowing us to safely deploy AI models with increasingly
powerful capabilities. However, such solutions are currently technically
under-developed, and there are substantial challenges to effectively
deploying defensive solutions for AI at a societal level and at scale
\cite{shevlane2020}. We
therefore focus on forthcoming models that may take us into a zone of
high risk against which we do not yet have sufficient social or
technological resilience.

In section~\ref{sec3:risks-of-open-sourcing-foundation-models} we discuss many risks that foundation models at the
frontier of today's capabilities currently present. Arguably, these
capabilities do not yet surpass a critical threshold of capability for
the most extreme risks. However, we are seeing some dangerous
capabilities emerge, and this trend is likely to continue as models
become increasingly capable and as it becomes easier and requires less
expertise and compute resources for users to deploy and fine-tune these
models.\footnote{According to \emph{Anthropic's Responsible Scaling
  Policy} \cite{anthropic2023a},
  current cutting-edge foundation model capabilities are at AI Safety
  Level 2 (ASL-2). Anthropic defines ASL-2 models as those ``\emph{that
  do not yet pose a risk of catastrophe, but do exhibit early signs of
  the necessary capabilities required for catastrophic harms. For
  example, ASL-2 models may (in absence of safeguards) (a) provide
  information related to catastrophic misuse, but not in a way that
  significantly elevates risk compared to existing sources of knowledge
  such as search engines, or (b) provide information about catastrophic
  misuse cases that cannot be easily found in another way, but is
  inconsistent or unreliable enough to not yet present a significantly
  elevated risk of actual harm.}'' Given current indications from ASL-2
  models, it is prudent to expect that ALS-3 models (see footnote 8)
  will begin to emerge in the near future, and developers and
  policymakers should prepare accordingly.} Recently, after extensive
testing of their large language model, Claude, by biosecurity experts,
Anthropic reported that ``unmitigated LLMs could accelerate a bad
actor's efforts to misuse biology relative to solely having internet
access, and enable them to accomplish tasks they could not without an
LLM.'' They note that these effects, while ``likely small today'', are
on the near-term horizon and could materialize ``in the next two to
three years, rather than five or more''
\cite{anthropic2023b}.

Our general recommendation is that it is prudent to assume that the next
generation of foundation models could exhibit a sufficiently high level
of general-purpose capability to actualize specific extreme risks.
Developers and policymakers should therefore implement measures now to
guide responsible model research decisions in anticipation of more
highly capable models.

These recommendations are driven by the fast pace of AI progress, the
immense challenge of verifying the safety of AI systems, and our ongoing
struggle to effectively prevent harms from even current-day systems on a
technical and social level. It is difficult to predict when more extreme
risks may arise. The level of risk that a model presents is intimately
tied to model capability, and it is hard to know when a critical line of
capability has been or will likely be passed to pose extreme risks. In
the past, model capabilities often have arisen unexpectedly or have been
discovered only after model deployment
\cite{wei2022}.

\paragraph{AI models do not need to be \emph{general-purpose} to pose a
risk.} Finally, it is worth noting that high-risk AI models do not
necessarily need to be general-purpose in nature like foundation models,
nor must they be at the frontier of current capabilities to pose the
risks described above. For example, Urbina et al.
\cite{urbina2022} demonstrated
that standard, narrow AI tools used within the pharmaceutical industry
can be repurposed to assist with the design of chemical weapons. There
are also more pressing concerns that AI systems might soon present
extreme biological risks
\cite{helena2023}. So while
outside the remit of this paper, care should similarly be taken in the
open-sourcing of narrow AI models that could, for example, be used to
aid in chemical or biological weapons development.

\subsection{Open-Source AI: Definition and
Disanalogy}\label{open-source-ai-definition-and-disanalogy}

``Open-source'' is a term borrowed from open-source software (OSS). In
the context of open-source software, ``open-source'' was defined in 1998
as a ``social contract'' (and later a certification) describing software
designed to be publicly accessible---meaning anyone can view, use,
modify, and distribute the source-code---and that is released under an
open-source license. An open-source license must meet ten core criteria,
including free source code access, permission for derived works, and no
discrimination against which fields or groups may use the software \cite{dibona1999,github}.

With the release of AI models like LLaMA, LLaMA2, Dolly, StableLM the
term ``open-source'' has become disjointed from open-source license
requirements
\cite{fanelli2023}. Some
developers use ``open-source'' merely to mean that their model is
available for download, while the license may still disallow certain use
cases and distribution. For example, while Meta refers to LLaMA-2 as an
open-source model, the LLaMA-2 license caveat is that the model cannot
be used commercially by downstream developers with over 700 million
monthly users, and the outputs cannot be used to train other large
language models. Strictly speaking, LLaMA2 is therefore not open-source
according to the traditional OSS definition
\cite{maffulli2023}, and the
marketing of it as such has been criticized as false and misleading by
the Open Source Initiative
\cite{maffulli2023}.\footnote{Indeed,
  there are likely economic, strategic, and reputational benefits for a
  company to `open-source' a model in this way
  \cite{graywidder2023}.
  Open-source innovation building on publicly available architectures
  can easily be reincorporated into the model developer's downstream
  products. ``Openness'' also has a reputationally positive connotation.
  ``Openwashing'' is a term that describes companies who spin an
  appearance of open-source and open-licensing for marketing purposes,
  while continuing proprietary practices
  \cite{finley2011}.}

\parbox{\linewidth}{\paragraph{However, in this paper we set licensing considerations aside, as
we are concerned with the risks and benefits of public model
accessibility.} From an AI risk perspective, even where more restrictive
licenses such as RAIL (Responsible AI License) include clauses that
restrict certain use cases
\cite{RAIL}, license
breaches are difficult to track and enforce when models are feely and
publicly available for download
\cite{widder2022}. License
breach will also not be of great concern for malicious actors intending
to cause significant harm. Accordingly, and in line with increasing
common parlance, we use the term open-source only to refer to models
that are publicly accessible at no cost.\protect\footnote{We acknowledge that it
  may be best for open-source terminology to be reserved for software
  and AI models that do carry specific licensing permissions. In which
  case, open-source should not be used to refer to models like LLaMa 2
  and StableLM that are free to download but have restrictive licenses.
  A term like ``public access'' might be preferable. We set this
  discussion aside for the time being. Regardless of what term is being
  used, we are concerned with models that are being publicly shared
  irrespective of associated licensing permissions. For the time being
  we use the term ``open-source'' to reflect its use (appropriate or
  not) by AI developers.}}

\paragraph{Licensing aside, the open-source software concept---referring
only to ``free and publicly downloadable source code''---does not
translate directly to AI due to differences in how AI systems are built \cite{fanelli2023,sijbrandij2023}.}
For AI systems, ``source code'' can refer to either
or both of the inference code and the training code which can be shared
independently. AI systems also have additional system components beyond
source code, such as model weights and training data, all of which can
be shared or kept private independent of the source code and of each
other.

Experts disagree on precisely which model components need to be shared
for an AI model to be considered open-source. Rather, the term is being
used to encapsulate a variety of system access options ranging on a
spectrum from what Irene Solaiman
\cite{solaiman2023a} calls
non-gated downloadable to fully open models. For \emph{fully open
models}, training and inference code, weights, and all other model
components and available documentation are made public (e.g., GPT-J
\cite{gpt-j}). For
\emph{non-gated downloadable models}, key model components are publicly
available for download while others are withheld. The available
components generally include some combination of training code
(minimally model architecture), model weights, and training
data.\footnote{For \emph{gated downloadable models,} in contrast,
  privileged download access is granted only to specific actors.}

Table~\ref{table1} presents a useful reference list of standard model components
and definitions. See Appendix~\ref{appendixA} for a more detailed breakdown.

\renewcommand{\arraystretch}{1.65}
\begin{center}
\vspace*{-2mm}
\resizebox{\linewidth}{!}{\begin{tabular}{|>{\centering\arraybackslash}m{0.12\linewidth}|m{0.85\linewidth}|}
\hline
\multicolumn{2}{|c|}{\parbox{0.98\linewidth}{\captionof{table}{Useful definitions of commonly-shared AI model components\rule[-1mm]{0pt}{6mm}\label{table1}}}}\\\hline
\bfseries\small Term & \multicolumn{1}{c|}{\bfseries\small Definition} \\ \hline
\textit{Model \mbox{architecture}} & The code that specifies the structure and design of an AI model, including the types of layers, the connections between them, and any additional components or features that need to be incorporated. It also specifies the types of inputs and outputs to the model, how input data are processed, and how learning happens in the model. \\ \hline
\textit{Model weights} & The variables or numerical values used to specify how the input (e.g., text describing an image) is transformed into the output (e.g., the image itself). These are iteratively updated during model training to improve the model’s performance on the tasks for which it is trained. \\ \hline
\textit{Inference code} & The code that, given the model weights and architecture, implements the trained model. In other words, it runs the AI model and allows it to perform tasks (like writing, classifying images and playing games). \\ \hline
\textit{Training code} & The code that defines the model architecture and implements the algorithms used to optimize the model weights during training. The training algorithms iteratively update the model weights to improve the AI model’s performance on the training tasks. \\ \hline
\end{tabular}}
\end{center}

\paragraph{The more model components that are publicly released, the easier
it is for other actors to reproduce, modify, and use the model.} For
example, access to model architecture and trained weights (e.g.,
StabilityAI's Stable Diffusion
\cite{stabilityai}), when
combined with inference code, is sufficient for anyone to use a
pre-trained model to perform tasks. Inference code can be easily written
by downstream developers or even generated by large language models such
as ChatGPT. It also does not need to match the original inference code
used by the model developer to run the model. Access to model weights
also allows downstream developers to fine-tune and optimize model
performance for specific tasks and applications.

Releasing other useful parts of the training code makes it much easier
for other actors to reproduce and use the trained model. For instance,
providing the optimal hyperparameters would make a pre-trained OS AI
model more capable (and possibly dangerous), and releasing the code used
to clean, label and load the training data would reduce the burden on
actors trying to reproduce model weights.

Sometimes, an AI developer will release the training and inference code
for a model, but not the trained model weights (e.g., Meta's LLaMA
\cite{metaai2023a} before the
weights were leaked).\footnote{Furthermore, we should expect model
  weight leaks to be frequent. Weights are contained in relatively small
  files (usually less than 256 GB) that can be easily and untraceably
  shared. Meta, for instance, chose to restrict access to the weights of
  its large language model LLaMa to researchers on a case-by-case basis,
  but a week later the weights were leaked and are now available
  publicly on the internet
  \cite{vincent2023}. If
  weights for a trainable open-source model are leaked, the public
  functionally has access to a pre-trained open-source model.} In such
cases, actors with sufficient computing resources and data access could
train the model and, with some inference code, run it.\footnote{Note
  that if the model weights were not made publicly available, external
  actors who trained a trainable OS model may discover a set of model
  weights distinct from those discovered by the original developer who
  released the model. Using a different set of weights, however, does
  not preclude a model from performing equally well as (or perhaps even
  better than) a model using the original weights.} However, at the
moment, few actors (realistically, only large technology companies, state-level actors, or well-funded start-ups) have the computing resources available to train
highly capable foundation models that represent the frontier of model
performance.\footnote{Training frontier foundation
  models costs \$10--100 million in compute costs and is projected to
  increase to \$1--10 billion in coming years
  \cite{cottier2023}. However,
  the cost to train a model that matches the performance of a previous
  state-of-the-art system has fallen rapidly. For instance, training
  GPT-3, the most powerful foundation model available in June 2020, was
  estimated to cost at least \$4.6 million
  \cite{li2020}, but by
  September 2022 an equivalently powerful model was theoretically
  available for \$450,000
  \cite{venigalla2022}. This is
  due to both advances in AI chip technology and the discovery of more
  efficient AI algorithms
  \cite{sevilla2022,erdil2023,hsieh2023}.}

\vspace{3mm}
\begin{minipage}{\linewidth}\bfseries
Therefore, in this paper, when we refer to open-source
foundation models, we mean models for which at least model architecture
and trained weights are publicly available unless otherwise specified.
\end{minipage}
\vspace{2mm}

Box 1 describes the need for further work defining open-source gradients beyond the definition we give here; releasing different (combinations of) model components in addition to trained weights and training code enables different downstream activities.  

\section{Risks of Open-Sourcing Foundation
Models}\label{sec3:risks-of-open-sourcing-foundation-models}

Due to their vast application space and pace of development, foundation
models have potential for broad and significant benefit and harm.
Accordingly, open-sourcing these models poses some substantial risks
which we present in two categories: malicious use (3.1) and
proliferation of unresolved flaws (3.2).

These harms are intensified by the fact that once a decision has been
made to open-source, there is no ``undo'' function. A published
model cannot be rolled back if major safety issues emerge or if
malicious actors find an AI tool to be particularly useful for scamming,
hacking, deceptive influence, or acts of terror. Methods exist that
allow even partially open-sourced models (e.g., code with some or no
other model components) to be replicated and shared in full
\cite{goldman2023a}.

\newpage
\vspace*{\fill}
\begin{tcolorbox}[enhanced jigsaw,width=\linewidth,boxrule=1pt,sharp corners,pad at break*=1mm,colbacktitle=white,colback=white,colframe=black,coltitle=black,toptitle=1mm,bottomtitle=1mm,fonttitle=\bfseries\small\centering,title=Box 1: Further research is needed to define open-source gradients,before float=\vspace{1cm},after float=\vspace{1cm}]

\subsection*{Gradient of System Access}

The idea that models are either released open-source or maintained closed-source presents a false dichotomy; there are a variety of model release options ranging from fully closed to fully open model \cite{sijbrandij2023,sastry2021,liang2022a}. 

    \vspace{1em}
    \includegraphics[width=\textwidth]{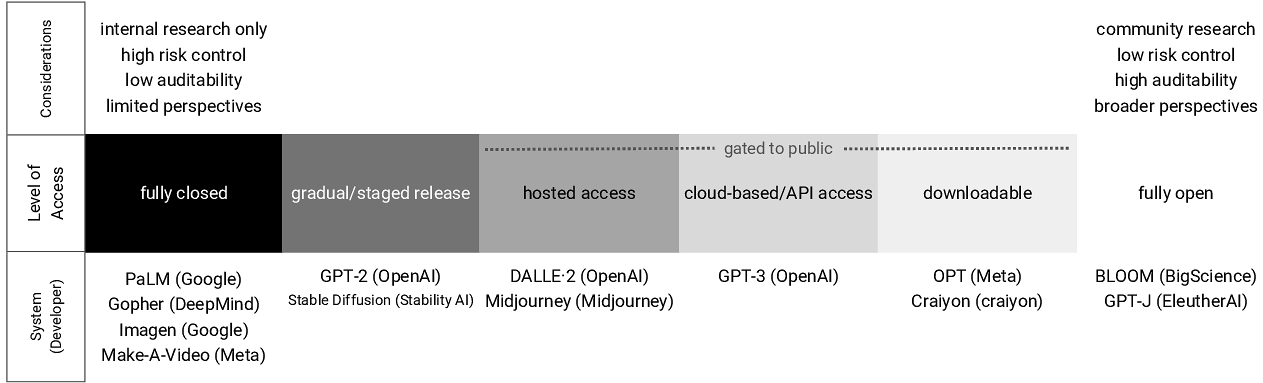}
    \begin{center}
    \begin{minipage}[t]{0.7\linewidth}
    \centering\small
    “Considerations and Systems Along the Gradient of System Access” [figure reproduced from Solaiman \cite{solaiman2023a}]
    \end{minipage}
    \vspace{1em}
    \end{center}

What is generally referred to as ``open-source'' model release spans the
two system access categories on the far right of Irene Solaiman's
\cite{solaiman2023a} gradient:
\emph{Downloadable} (specifically non-gated downloadable---meaning that
anyone is free to download the available components) and \emph{Fully
Open.}

\subsection*{Gradient of Open-Source Access}

For \emph{fully-open} models, source code, weights, training data, and
all other model components and available documentation are made public.
However, in the \emph{non-gated downloadable} category---in which some
components are publicly downloadable (usually including weights and
architecture) while others are withheld---there is room for further
specification. Importantly, the precise benefits and risks of
open-sourcing are determined by the specific combinations of model
components and documentation that are made publicly available.

\subsection*{Precise Definitions for Precise Standards}

Near-term investment in a project is needed to investigate and
articulate what activities are made possible by access to different
(combinations of) model components. This information will be key to
constructing effective and fine-grained model release standards that are
not overly burdensome, and to ensure open-source values are protected
and benefits enjoyed where safe.\\[-1pt]

We make a start in Appendix~\ref{appendixA}, though it is a much larger and more
involved project than we can do justice here, and it is a project on
which members of open-source communities should be centrally involved.
The Open Source Initiative recently launched one such initiative to
define what machine learning systems will be characterized as
open-source
\cite{maffulli2023a}.
\end{tcolorbox}
\vspace*{\fill}\mbox{}
\clearpage

\subsection{Malicious Use}\label{sec3.1:malicious-use}

Open-source publication increases foundation models' vulnerability to
misuse. Given access to the model's weights and architecture, any actor
with the requisite technical background\footnote{Knowledge equivalent to
  that from a graduate-level machine learning course would be sufficient
  to perform fine-tuning, but additional experience in training models
  would likely be useful in addressing the myriad of issues that
  sometimes come up, like divergence and memory issues. Depending on the
  malicious use case, it may be more or less difficult to source the
  required data set.} can write their own inference code---or modify
available inference code---to run the model without safety filters. They
can also fine-tune the model to enhance the model's dangerous
capabilities or introduce new ones.

There are several ways in which open-source publication can facilitate
misuse:

Firstly, open-sourcing a model allows actors to run the model using new
or modified inference code that lacks any content safety filters
included in the original code. Stable Diffusion's safety filter, for
example, can be removed by deleting a single line of inference
code.\footnote{This observation comes from personal correspondence with
  several technical researchers. We do not provide further details on
  specific technical flaws since we believe it would be irresponsible to
  do so. Please see Rando et al.
  \cite{rando2022} on
  red-teaming the Stable Diffusion safety filter for related
  information.} This is possible because such filters are implemented
post-hoc, appending additional processes to the model's inference code,
rather than fundamentally changing the behavior of the model itself.
With content safety filters removed, there is nothing to prevent users
from presenting the models with unsafe requests or to prevent the model
from yielding unsafe outputs.

Secondly, the ability to fine-tune an open-source model without
restrictions enables the modification of models specifically for
malicious purposes. Fine-tuning that occurs through an API can be
monitored; for example, the API owner can inspect the contents of the
fine-tuning data set. Without such monitoring, fine-tuning could involve
the reintroduction of potentially dangerous capabilities that were
initially removed by developers pre-release through their own
fine-tuning. Fine-tuning can also lead models to become even more
dangerous than they were before safety measures were applied. However,
increasing a model’s dangerous capabilities by fine-tuning would be more difficult than removing certain kinds of
post-hoc safeguards like filters; fine-tuning requires the curation of a
dataset to promote those dangerous capabilities, as well as requiring
the necessary compute and technical expertise to successfully fine-tune
the model.

Thirdly, access to model weights can aid adversarial actors in
effectively jailbreaking system safeguards (including for copies of
the system that have not been modified). Traditional jailbreaks use
clever prompt engineering to override safety controls in order to elicit
dangerous behavior from a model (e.g., getting a large language model
(LLMs) to provide instructions for building a bomb by asking it to write
a movie script in which one character describes how to build a bomb).
Creative prompting only requires model query access. However,
researchers recently discovered a method of adversarial attack in which
the network weights of open-source LLMs aided researchers in optimizing
the automatic and unlimited production of ``adversarial suffixes'',
sequences of characters that, when appended to a query, will reliably
cause the model to obey commands even if it produces harmful content
\cite{zou2023}. Notably,
this method, which was developed using open-source models Vicuna-7B and
Meta's LLaMA-2, is transferable; it also works against other LLMs such
as GPT-4 (OpenAI), Bard (Google), and Claude (Anthropic), indicating
that open-sourcing one model can expose the vulnerabilities of others.

The above methods have the potential of reducing, if not entirely
nullifying, the measures taken by developers to limit the misuse
potential of their models. These measures would be much more difficult
to bypass in cases where the model weights and training code are not
openly released, and where user interaction with the model is
facilitated through an API. Fine-tuning, in particular, can also lead
models to be more dangerous than they might have been originally.

\subsubsection[Varieties of Malicious Use]{Varieties of Malicious Use\protect\footnote{To be clear, open-sourcing is not to blame for the malicious use of AI. Foundation models are a
  dual use technology, and where the technology is built by malicious
  actors or where effective safety restrictions are not in-place for
  models accessible via API, misuse can occur. Open-sourcing risks the
  diffusion of potentially dangerous capabilities to malicious actors
  and lowers barriers against misuse.}} Potential epistemic, social and
political consequences of foundation model misuse include the following \cite{anderljung2023a,brundage2018}.

\begin{itemize}
\item
  \textbf{Influence operations.} There is a wealth of existing research
  theorizing AI's utility in automating, or otherwise scaling, political
  or ideological influence campaigns through the production and targeted
  dissemination of false or misleading information
  \cite{bommasani2022,brundage2018,weidinger2021,goldstein2023}. There is concern about multimodal foundation
  models being used to create interactive deepfakes of politicians or constructing
  and catering detailed and seemingly verifiable false histories
  \cite{horvitz2022}. A recent
  experiment demonstrated the potential for AI-based influence
  operations when the LLM-based system, CounterCloud, was deployed to
  autonomously identify political articles, to generate and publish
  counter-narratives, and then to direct internet traffic by writing
  tweets and building fake journalist profiles to create a veneer of
  authenticity
  \cite{banias2023}.

Concerns about AI being used to manipulate public views, undermine
trust, drive polarization, or otherwise shape community epistemics have
led some scholars to speculate that \emph{`whoever controls language
models controls politics'}
\cite{bajohr2023}.

\item
  \textbf{Surveillance and population control.} AI advances the means of
  states to monitor and control their populations through immersive data
  collection, such as facial and voice recognition
  \cite{almeida2022}, the
  nascent practice of affect recognition
  \cite{kaklauskas2022}, and
  predictive policing
  \cite{ferguson2017}. AI also
  allows automating and thus ever more cheaply analyzing unprecedented
  amounts of data \cite{shevlane2023}. Authoritarian governments may be most likely
  to make use of AI to monitor and control their populations or to
  suppress subpopulations
  \cite{xu2021,kendall-taylor2020},
  but and? other types of governments are employing AI enabled
  surveillance capabilities as well. Nascent AI surveillance
  technologies are spreading globally and in countries with political
  systems ranging from closed autocracies to advanced democracies
  \cite{crawford2019,feldstein2019}.
\item
  \textbf{Scamming and spear phishing.} Malicious actors can use AI to
  fraudulently pose as a trusted individual for the purpose of theft or
  extraction of sensitive information
  \cite{gupta2018}. For
  example, large language models have been shown to be proficient in
  generating convincing spear phishing emails, targeted at specific
  individuals, at negligible cost
  \cite{hazell2023}.

Evidence from online forums also indicates that malicious AI tools and
the use of ``jailbreaks'' to produce sensitive
information and harmful content are proliferating amongst cyber
criminals \cite{kelley2023}.
High profile scams using generative AI have also been observed, with one
report detailing how \$35million was stolen from a Japanese firm by
scammers who used AI voice cloning tools to pose as a company executive
to employees
\cite{brewster2021}.

\item
  \textbf{Cyber attacks.} Foundation models have applications for both
  cybersecurity and cyber warfare
  \cite{centerforsecurityandemergingtechnology2020,horvitz2022a}. Early demonstrations show that LLMs' current coding
  abilities can already find direct application in the development of
  malware and the design of cyber attacks
  \cite{shimony23}. With
  improved accessibility and system capability, the pace of customized
  malware production may increase as could the variability of the
  malware generated. This poses a threat to the production of viable
  defense mechanisms. Especially in the near term, there is some
  evidence that AI generated malware can evade current detection systems
  designed for less variable, human-written programs
  \cite{fritsch2022,stoecklin2018,li2019}.

Ultimately, information gained from cyberattacks might be used to steal
identities, or to gather personal information used to mount more
sophisticated and targeted influence operations and spear phishing
attacks. Cyberattacks could also be used to target government agencies
or critical infrastructure such as electrical grids
\cite{garcia2017}, financial
infrastructures, and weapons controls.

\item
  \textbf{Biological and chemical weapons development.} Finally, current
  foundation models have shown nascent capabilities in aiding and
  automating scientific research, especially when augmented with
  external specialized tools and databases
  \cite{boiko2023,bran2023}. Foundation models may therefore reduce the human expertise
  required to carry-out dual-use scientific research, such as
  gain-of-function research in virology, or the synthesis of dangerous
  chemical compounds or biological pathogens
  \cite{sandbrink2023,soice2023}. For example, pre-release model evaluation of GPT-4 showed
  that the model could re-engineer known harmful biochemical compounds
  \cite{OpenAI_2023b}, and
  red-teaming on Anthropic's Claude 2 identified significant potential
  for biosecurity risks
  \cite{anthropic2023b,gerrit2023}.

Specialized AI tools used within these domains can also be easily
modified for the purpose of designing potent novel toxins
\cite{urbina2022}. Integrating
narrow tools with a foundation model could increase risk further: During
pre-deployment evaluation of GPT-4, a red-teamer was able to use the
language model to generate the chemical formula for a novel, unpatented
molecule and order it to the red-teamer's house
\cite{OpenAI_2023b}. Law-makers
in the United States are beginning to take this biosecurity threat
seriously, with bipartisan legislation---the Artificial Intelligence and
Biosecurity Risk Assessment Act---being proposed that would monitor and
study the potential threats of generative and open-source AI models
being used ``intentionally or unintentionally to develop novel
pathogens, viruses, bioweapons, or chemical weapons''
\cite{markey2023}.

\end{itemize}

\subsubsection{Ease of Malicious Use} One factor that potentially mitigates
the misuse of open-source foundation models is that the pool of actors
with the requisite talent and compute resources to download, run and,
when necessary, modify highly capable models effectively is relatively
small. Nevertheless, there are still several reasons to be concerned.

First, there is an increasing number of individuals who have the skills
to train, use, and fine-tune AI models as illustrated by growing
computer science PhD enrollment as well as ballooning attendance at AI
conferences
\cite{Maslej2023a}. This is
supplemented by an increasing number of tutorials and guides available
online to use and fine-tune AI systems.

Second, running a pre-trained AI model at a small scale requires only a
small amount of compute---far less compute than training does. We
estimate the largest Llama 2 model (Llama-2-70B) costs between \$1.7
million and \$3.4 million to train,\footnote{Meta reported using
  1,720,320 A100 GPU-hours to train Llama-2-70B
  \cite{touvron2023}. A single
  consumer A100 GPU can be rented privately for \$1.99/hour (e.g. from
  RunPod \cite{runpod2023}.
  Our range assumes that Meta's cost was between \$1 and \$2 per hour.}
while the inference costs for Llama-2-70B are estimated to be between
0.2 and 6 cents per 750-word prompt
\cite{aman2023} and \$4 per
hour of GPU time.\footnote{Since the Llama-2-70B model is about 129GB,
  it requires 2 80GB A100 GPUs to store, each of which can be rented for
  about \$2/hour (e.g. from RunPod
  \cite{runpod2023}).} While
the compute requirement becomes large when running models at a very
large scale (that is, performing many inferences),\footnote{Both
  training and inference processes are typically more economical when
  run on centralized high-performance computing (HPC) systems optimized
  for AI workloads housed within data centers. While a single training
  run demands more compute than a single inference, the majority of
  compute for AI systems is not being used for training runs. As with
  most infrastructure, the operating costs will eventually be larger
  than the upfront cost. As the final product of AI systems, inferences
  are triggered by a multitude of daily actions, ranging from chatbot
  interactions and Google searches to commands to virtual personal
  assistants like Siri or Alexa.

  Consider image generation: the cumulative compute used for generating
  images via a generative AI model has now likely surpassed the initial
  training compute for the most popular generative systems by orders of
  magnitude. The key difference between development and deployment lies
  in timeframe and independence. In inference, the computational
  resources can be distributed across multiple copies of the trained
  model across multiple compute infrastructures over a longer time
  duration. Whereas, in training, the computational resources are
  required over a smaller time frame within one closed system, usually
  one compute cluster.} large-scale runs may not be required for
impactful misuses of a model. It is conceivable that only a few
inferences may be needed in certain domains for models to be dangerous
(e.g., a malicious actor may only need to find one critical
vulnerability to disrupt critical infrastructure).

Third, while the overall cost of training frontier models is increasing
\cite{cottier2023},\footnote{See
  Footnote 9.} algorithmic progress focuses heavily on reducing demands
on compute resource, both for training\footnote{For example, Meta's
  recently released I-JEPA (Image Joint Embedding Predictive
  Architecture) offers a non-generative approach for self-supervised
  learning that does not rely on hand-crafted data-augmentations, and
  requires significantly fewer GPU hours to train for a better
  performing model
  \cite{assran2023,metaai2023b}.}
and for fine-tuning
\cite{hu2021}. This,
combined with the decreasing cost of compute (measured in FLOP/s per
\$)\cite{hobbhahn2022}, means
that while initial model development and training may remain
prohibitively expensive for many actors, we should not expect compute
accessibility to always act as a strong limiting factor for fine-tuning
existing open-source foundation models. Targeted fine-tuning of a
pre-trained model to create dangerous models would remain much less
expensive than building a model from scratch.

\subsubsection{Offense-Defense Balance} Another argument against the threat of
malicious use posed by open-sourcing is that while open-sourcing may
increase model vulnerability to exploitation by malicious actors, it
does more to help developers identify those vulnerabilities before
malicious actors do and to support development of tools to guard against
model exploitation and harms
\cite{zellers2019}. In other
words, in the offense-defense balance---a term referring to the
``relative ease of carrying out and defending against attacks''
\cite{jervis1978,garfinkel2019}---it has been argued that open-sourcing favors defense.

This is often true in the context of software development; open-sourcing
software and disclosing software vulnerabilities often facilitate
defensive activities more than they empower malicious actors to
offensively identify and exploit system vulnerabilities. However, the
same might not be safely assumed for open-source AI, especially for
larger and more highly capable models
\cite{shevlane2020}. Shevlane
and Dafoe \cite{shevlane2020}
explain that when a given publication (e.g., publication of software, AI
models, or of research in biology or nuclear physic etc.) is potentially
helpful for both people seeking to misuse a technology and those seeking
to prevent misuse, whether offensive or defensive activities are favored
depends on several factors:

\begin{itemize}
\item
  \textbf{Counterfactual possession.} How likely would a would-be attacker
  or defender be able to acquire the relevant knowledge without
  publication? If counterfactual possession by the attacker or defender
  is probable, then the impact of publication on their respective
  offensive and defensive activities is less.
\item
  \textbf{Absorption and application capacity.} A publication only
  benefits attackers and defenders to the extent that they can absorb
  and apply the knowledge toward their desired ends. This depends on
  how much knowledge is disclosed, how the knowledge is presented, and the
  attentiveness and comprehension of the recipients.
\item
  \textbf{Resources for solution finding.} For defenders, given
  publication, how many additional actors will help develop defenses?
  Impact of publication is greater if many people are likely to
  contribute to defensive applications.
\item
  \textbf{Availability of effective solutions.} Are vulnerability patches
  easy to implement, or will developing solutions be a more complicated
  and time intensive endeavor? The positive effects of publication
  decrease the more difficult vulnerabilities are to address.
\item
  \textbf{Difficulty/cost of propagating solutions.} Even where defensive
  solutions exist, if they are difficult to propagate then the impact is
  less.
\end{itemize}

For software development, the offense-defense balance of open-source
publication often comes out in favor of defense. Software
vulnerabilities are easy to find, so counterfactual possession by
attackers is likely, and software patches are relatively easy to make,
usually fully resolve the vulnerability, and are easily rolled out
through automatic updates.

However, in the context of AI research, Shevlane and Dafoe offer the
tentative conclusion that as AI models grow in capability and
complexity, open-source publication will likely skew the balance towards
offense. As discussed at the start of this section, attacker knowledge
of vulnerabilities and their ability to exploit those vulnerabilities is
greatly increased by open-source publication. For some vulnerabilities,
researching solutions is time consuming and resource intensive (See
Section~\ref{sec4.2:accelerate-beneficial-ai-progress}). Solutions developed also tend not to be perfect fixes.
This is for a variety of reasons: (i) given our current lack of
understanding of how advanced AI systems work internally, it may be
difficult to identify the source of risk or failure; (ii) certain risks,
such as bias and discrimination, may be learned from the training data,
and it could be impossible to ``remove'' all bias from training data
\cite{ferrara2023}; (iii)
reducing misuse of AI systems may require changes to social systems
beyond changes to technical ones
\cite{shevlane2020}; (iv) the
structure of AI systems introduces new sources of failure specific to AI
that are resistant to quick fixes (e.g., the stochastic nature of large
language models may make it difficult to eliminate all negative outputs,
and the inability to distinguish prompt injections from ``regular''
inputs may make it difficult to defend against such attacks)
\cite{kassab2022}. Finally,
it is difficult to ensure improvements to open-source models are
implemented by downstream users and developers which can result in
widespread proliferation of unresolved model flaws. We address this
topic in Section~\ref{sec3.2:risks-from-the-proliferation-of-unresolved-model-flaws}.

The conclusion that the offense-defense balance skews towards offense
when open-sourcing AI remains tentative because the offense-defense
balance is influenced by a myriad of factors making it difficult to
reliably predict outcomes. The balance will vary with each model,
application space, and combination of released model components. In
addition, we may develop measures in the future that build our defensive
capabilities. Nonetheless, the general notion holds; open-sourcing AI
leans towards offense more so than open-sourcing software. AI developers
should therefore think critically about the potential for, and potential
protections against, misuse before every model release decision.

\subsection{Risks from the Proliferation of Unresolved Model
Flaws}\label{sec3.2:risks-from-the-proliferation-of-unresolved-model-flaws}

Excitement about foundation models stems from the large number of
potential downstream capability modifications and applications. These
can include applications involving malicious intent and misuse, but more
frequently will involve well-intentioned commercial, scientific, and
personal applications of foundation models. If they have the necessary
resources and model access (via open-source or sufficient API access),
downstream individuals, AI labs, and other industry and government
actors can:

\begin{enumerate}

\item
  Employ foundation models to new tasks that were not previously subject
  to risk assessments due to the general capabilities of these models.
\item
  Fine-tune or otherwise alter open-sourced foundation models to enable
  specialized or additional (narrow and general) capabilities.
\item
  Combine foundation models with other AI models, tools, and services,
  such as the internet or other APIs, to create a system of AI models
  which can have new narrow and general capabilities.\footnote{For
    example, ChemCrow is a large language model that integrates 17
    expert-designed computational chemistry tools to accomplish tasks
    across organic synthesis, drug discovery, and materials design. The
    developers note that ChemCrow aids expert chemists and lowers
    barriers for non-experts which can foster scientific advancement but
    could also pose significant risk of misuse
    \cite{bran2023}. Also
    see Boiko, MacKnight, \& Gomes
    \cite{boiko2023} on
    combining large language models.} For example, AutoGPT is an
  open-source app that integrates with GPT-3.5 and GPT-4. While GPT-3.5
  and GPT-4 can respond one prompt at a time, AutoGPT handles follow-ups
  to an initial prompt. This allows users to ask AutoGPT autonomously to
  complete higher-level goals that require iteratively responding to and
  generating new prompts \cite{wiggers2023,auto-gpt2023}.
\end{enumerate}

\paragraph{In all three cases, the risks, flaws, system vulnerabilities,
and unresolved safety issues of the initial foundation model propagate
downstream.} For instance, biased and discriminatory behavior,
vulnerabilities to prompt injection
\cite{bagdasaryan2023} and
adversarial attacks
\cite{zou2023}, autonomous
self-proliferation abilities
\cite{kinniment2023}, or other
dangerous capabilities could quickly proliferate if not caught and fixed
before being integrated into downstream products and applications.

The fact that the models can be applied to new contexts (1), but also
adapted (2 and 3) to unlock new narrow and general capabilities, also
means that further, difficult to predict risks and harms could emerge.
Consequently, it is not certain that the safeguards put in place by the
foundation model developer will continue to be effective if downstream
developers fine-tune, alter, and combine AI models. This means that not
only will existing model flaws proliferate, but previously fixed  flaws and new flaws may also arise.

If (1), (2), and (3) are enabled via structured API access (e.g.,
OpenAI's davinci-002 and GPT-3.5 can be fine-tuned via API
\cite{openaia}, then
developer monitoring of API use may go some way towards mitigating the
proliferation harms described above. There is no such recourse, however,
if a model is made open-source. \textbf{Once a model is open-sourced,
there are no take-backs if harms ensue.}

When risks and vulnerabilities are proliferated there is no way of
ensuring that when a fix is rolled out (assuming a fix is possible - see
end of 3.1) that it is adopted or integrated effectively by downstream
AI developers and users. Even in the context of traditional open-source
software, software flaws are proliferated
\cite{ponta2020} as
downstream developers and users more often than not fail to implement
patches and version updates, even where the open-source license requires
they do so \cite{synopsyseditorialteam2023}.
Very often consumers are unaware that their systems are running on
out-of-date software or that vulnerability patches are available. Other
times an updated software version will not integrate well with other
software packages and existing infrastructure. We should expect the same
challenges to undermine the maintenance of open-source foundation
models, though a given foundation model will likely be applied to a much
wider range of applications than a piece of software.

There are also different incentives influencing decisions to implement
updates for traditional software than for foundation models. For
traditional software, patches and version updates improve system
performance and functionality and resolve vulnerabilities that could
cause harm to the user. It is to the user's benefit to implement
software updates when feasible. In comparison, for increasingly capable
foundation models, safety patches and updates often aim to reduce system
functionality, disallowing certain activities that were possible with
previous versions. If downstream developers and users wish to retain
those functionalities (e.g., to be able to produce nude art with an
image generator), they are incentivized not to update versions and, in
some cases, not to disclose the existence of potential risks and system
vulnerabilities.

\parbox{\linewidth}{Due to the potential of proliferating risks and model flaws from highly
capable foundation models, developers need to consider model release
decisions carefully. Developers of highly capable foundation models must
be cognizant of the potential downstream harms of their models (harms
which they would be powerless to backtrack) and carefully consider
alternative methods by which open-source benefits might be pursued but
at significantly less risk
\cite{whittlestone2020}. We discuss
alternatives further in Section~\ref{sec4:benefits-of-open-sourcing-foundation-models-and-alternative-methods-for-achieving-them}. Clear legislation is also needed to
hold developers and controllers of AI systems liable for the impacts of
their systems.}

\section{Benefits of Open-Sourcing Foundation Models and Alternative
Methods for Achieving
Them}\label{sec4:benefits-of-open-sourcing-foundation-models-and-alternative-methods-for-achieving-them}

In this section we analyze three key benefits of open-source software:
facilitating external evaluation \eqref{sec4.1:external-model-evaluation}, accelerating beneficial progress
\eqref{sec4.2:accelerate-beneficial-ai-progress}, and distributing control over technological development and
benefits \eqref{sec4.3:distribute-control-over-ai}. For each, we first present the benefit, then evaluate
the benefit in the context of highly capable foundation models, and
finally consider other strategies that might contribute to the same
goals. A summary table is provided at the start of each subsection.

\subsection{External Model Evaluation}\label{sec4.1:external-model-evaluation}

\begin{table}[ht!]
\setlength{\tabcolsep}{3pt}
%\caption{Section summary: Open-sourcing as a mechanism for enabling external model evaluation}
\resizebox{\linewidth}{!}{\begin{tabular}{|>{\arraybackslash\centering}m{0.165\linewidth}|m{0.81\linewidth}|}
\hline
\multicolumn{2}{|c|}{\parbox{\linewidth}{\captionof{table}{Section summary: Open-sourcing as a mechanism for enabling external model evaluation\rule[-3mm]{0pt}{6mm}}}}\\\hline
\small\textbf{The argument for open-source AI} & Open-sourcing enables independent model evaluations of projects by wider communities of developers. Tapping into the wider AI community helps to catch bugs, biases, and safety issues that may otherwise go unnoticed, ultimately leading to better performing and safer AI products. \\ \hline
\small\textbf{Evaluation of benefit} & 
\begin{itemize}[after=\vspace{-0.95\baselineskip}]
    \item Open-sourcing is most useful for evaluating complex safety issues and less useful for identifying discrete bugs. There may also be suitable alternatives to open-sourcing that achieve these same benefits with fewer risks.
\end{itemize} \\ \hline
\small\textbf{Alternative methods} & \begin{itemize}[itemsep=-2pt,after=\vspace{-1.3\baselineskip}]
\item Grant privileged model access to trusted (independently selected) third-party auditors via gated-download or research API.  
\item Establish a community of (independently selected) red-team professionals to stress-test models pre-release.
\item Explore social impacts and safety issues through incremental, staged release of models.
\item Employ safety bounties to incentivize wide public involvement in reporting new behaviors and safety issues.\end{itemize}\vspace{-\baselineskip}\\  \hline
\end{tabular}}
\end{table}
\newpage

\subsubsection{The Argument for
Open-Source}\label{the-argument-for-open-source}

A clear benefit for open-source software development is that
open-sourcing facilitates independent evaluations of projects by wider
communities of developers and many more people than a single developer
would be able to employ internally to check for bugs and safety issues.
This means a more diverse pool of expertise can be tapped, with a low
barrier to entry for individuals to contribute, whose skill to identify
and solve problems is enhanced by increased access to relevant
materials. So in the case of highly capable foundation models, it is
reasonable to expect that open-sourcing would leverage the same talent
multiplier as with OSS. Tapping into the wider AI community would enable
audit and analysis of foundation models and any model components (e.g.,
training data, weights, documentation) by interested parties helping to
catch bugs, biases, and safety issues that may otherwise go unnoticed.
Such external oversight would help hold AI developers to account for the
quality and consequences of their products at a team and an industry
level, and ultimately lead to better performing and safer AI products.

\subsubsection{\texorpdfstring{Evaluating the Benefit for Foundation
Models
}{Evaluating the Benefit for Foundation Models }}\label{evaluating-the-benefit-for-foundation-models}

In this section we consider the benefits of open-sourcing for enabling
external model evaluations according to two classes of model issues: (1)
discrete bugs, and (2) complex safety challenges.

\paragraph{Discrete Bugs.} Discrete bugs such as interface glitches, data
exposures and authentication issues are self-contained flaws that are
relatively simple to fix. Once discovered, discrete bugs can be easy and relatively low
cost for model developers to fix in-house. But bug spotting certainly
benefits from additional eyes, and there are alternative methods to open-sourcing
that attempt to facilitate more widespread participation in model review. For example, AI developers can
set up community reporting systems as they are encountered and even
incentivize engagement via bug bounties like that employed by OpenAI
\cite{bugcrowd}. That being
said, conscious steps need to be taken to ensure that the benefits of
open-sourcing can be replicated: active efforts need to be made to
engage the attention of a diverse set of experts and it remains
difficult to mimic open-sourcing here in all respects. For example, not
having full access to materials will impede individuals in their ability
to find bugs.

A further advantage of open-sourcing is that it allows downstream
developers to patch such issues on their own and to pass those patches
back to the developer for integration into future model versions.

\paragraph{Complex Safety Challenges.} Increasingly capable foundation
models are bringing with them an array of new behaviors and safety
challenges that arise unpredictably and are not well-understood by
developers \cite{bowman2023}.
For example, emergent abilities are unexpected and unintended features
or behaviors that arise in AI models as they become more advanced. These
abilities are not observed in smaller precursors and are not explicitly
programmed by developers
\cite{wei2022}.
``Capability overhang'' is a concept that further describes how these
emergent abilities can be latent within a system only to emerge
unexpectedly when elicited, for example, by clever prompt engineering or
integration with other software. Sometimes new capabilities continue to
be elicited many months after model release
\cite{sastry2021}.

Drawing input from a large pool of contributors will be instrumental to
exploring this evolving space of unknown unknowns; what do new safety
issues look like and, if not immediately evident, how are they
triggered? Furthermore, because some model behaviors will only emerge
with downstream modification of model weights, model evaluators will
need to be able to experiment with model fine-tuning to test a variety of
possible model versions.

Open-sourcing provides the necessary access to model weights and
parameters for attempting to elicit new behaviors from models for safety
evaluation (although it simultaneously allows malicious attempts to
elicit new dangerous behaviors and avenues of misuse). For models that
are not open-sourced, fine-tuning might also be facilitated via APIs
that allow users to manipulate model weights and parameters (e.g.
OpenAI's davinci-002 and GPT-3.5
\cite{openaia}. However,
some APIs may introduce additional limitations on fine-tuning. For
example, API controllers could attempt to limit the format or content of
data used to fine-tune a model, limit access to weights and parameters
(e.g. provide access to weights and parameters of base-line models but
not to fine-tuned model versions), or limit the amount of fine-tuning
that can be done. These limitations might be in place to protect the
developer's commercial interests or to reduce risk of misuse.

Safety research, such as alignment and interpretability research which
aim to understand and resolve complex safety issues, also require
varying degrees of model access. We will discuss the benefits of
open-sourcing for promoting safety research in section~\ref{sec4.2:accelerate-beneficial-ai-progress}.

\subsubsection{Other Ways to Enable External
Evaluation}\label{sec4.1.3:other-ways-to-enable-external-evaluation}

There are some alternatives to open-sourcing that can facilitate the
identification and evaluation of bugs and safety issues, with less risk
than open model release.

\paragraph{Staged-Release Impact Testing.} AI developers can conduct
staged-release impact testing to gather observational data about how a
model is likely to be (mis)used and modified if open-sourced.
Staged-released impact testing is a process by which incrementally
larger versions of a model are released behind API
\cite{solaiman2019,shevlane2022a}.
Each stage of release allows time to observe how the model is used, to
study its social impacts, and to implement any patches or new safety
measures before the next, more powerful version is released (if it is
deemed safe to do so).

If many safety measures need to be implemented between stages to
mitigate harms, this is a solid indication that open-sourcing will lead
to malicious use because, once open-sourced, those measures could be
easily circumvented.

Conducting staged release impact testing allows AI developers to be more
comfortable with open-sourcing their models, assuming no other
significant issues emerge in model evaluation and risk assessment
process. However, this can come at a cost to the developer by allowing
competitors to capture market share in the meantime if such processes
are not implemented for the industry at large through regulation. In
addition, any benefits from the model are also delayed from reaching the
relevant communities that could benefit from them.

\paragraph{External Audits \& Red-teaming.} In addition to staged-release
impact testing, developers can grant privileged model access to trusted
third-party auditors. These are external actors (government departments,
private expert organizations, or some combination thereof) tasked with
evaluating the safety and security of foundation models prior to model
release or assessing and verifying the model evaluation measures
employed by AI labs.

Though they are in early stages of development, external auditing has
been proposed as a key institutional mechanism for facilitating
trustworthy AI development
\cite{brundage2020,raji2020,mokander2023,khlaaf2022}.
One early example is the Alignment Research Center's (ARC) pre-release
evaluation of GPT-4 for dangerous capabilities
\cite{arcevals2023}.

The ARC evaluations largely involved red-teaming GPT-4. Red-teaming is
an evaluation method that stress-tests models to discover how and where
safety concerns arise. The aim is to identify potentially dangerous
model properties (e.g., manipulative or power-seeking behavior),
security flaws (e.g., jailbreaks), and possible misuse applications.
Stress-testing requires that red-teams are able to prompt models to
elicit new and dangerous behavior which can be facilitated with model
query access---that is, being able prompt models and receive outputs
without open-source access to model code and weights.

Where model weight access is needed to experiment with fine-turning,
access might be granted to identified individuals or research groups via
gated download or API. For \textbf{gated download} developers make
models (minimally weights and training code) available for specific
actors to download and run on their own hardware.\footnote{For further
  discussion on gradients of model release, including gated and
  non-gated downloadable models, see Box 1 and
  \cite{solaiman2023a}} The risk
with gated download is that model leaks could result in the
dissemination of potentially dangerous models. Download recipients would
need to be vetted carefully. Another option is for developers to provide
fine-tuning access via API. However, as mentioned above, some developers
may choose to implement limitations on fine-tuning in order to prevent
misuse or model reproduction. For this reason, Bucknall et al.
\cite{bucknallForthcoming} recommend
the design and implementation of \textbf{`research APIs'} whereby more
flexible fine-tuning permissions are granted to trusted researchers,
red-teams, and auditors depending on their access needs.

Red-teams such as those employed by OpenAI
\cite{openai2023d,openai2022,murgia2023} and Anthropic
\cite{anthropic2023b,ganguli2022}
are increasingly common, though best practices are still being
developed. Model evaluation is a nascent field. This makes it difficult
to evaluate the skill of potential auditors. Moving forward, standards
will need to be developed and implemented to ensure the quality and
consistency of third-party audits
\cite{costanza-chock2022} as numerous
governments and private actors move to occupy a growing AI assurance
sector \cite{centrefordataethicsandinnovation2021}.
Mechanisms will also be needed to ensure developers provide sufficient
model access to auditors and respond to audit findings. For instance,
audit reports should be published publicly or shared with a government
overseer while regulatory requirements ensure labs respond and disclose
their efforts and results. Governments should consider establishing
mandatory auditing regimes for large and potentially dangerous
foundation models to minimize the risk of model developers only granting
access to favored auditors, who might be less likely to expose failure
modes that are potentially embarrassing or inconvenient for the
developer.

Much work is needed developing new model evaluation techniques and
establishing best practice. Some evaluation processes may benefit from
leveraging foundation model capabilities
\cite{perez2022} as well as
input from wider AI developer communities. Decisions about how and by
whom models are audited are currently entirely at the discretion of
individual developers. Without standardized risk assessment procedures a
lab could choose an ``easy'' or ``friendly'' auditor.

Another possibility Brundage et al.
\cite{brundage2020} suggests,
is to extend red-teams to elicit input from a wider community of
\textbf{`red-team professionals'}. Such a community would be composed of
members from the wider AI community as well as security professionals,
and representatives from high-risk domains to which foundation models
might be put to use. This would help distribute red-teaming costs for
labs less-inclined to form internal red-teams, and the community of
red-team professionals would benefit from greater insight to common
attack vectors and useful red-teaming strategies shared within the
community. But again, risks arise by allowing AI developers to choose
red-teamers on their own, including capture of the safety evaluation
process and a potential narrowing of focus and values by not ensuring an
optimally diverse and comprehensive set of experts. Further best
practices and regulatory mechanisms need to be put in place to make sure
red-teaming can provide effective safety evaluations of AI models.

\paragraph{Bug Bounties and Safety Bounties.} Safety bounty programs have
been proposed as another method of tapping into a wider global community
to help identify and surface new safety and alignment issues in large
foundation models
\cite{levermore2023}. Bounty
``hunters'' are not pre-vetted as with selected red-teams.

Analogous to bug bounty programs commonly used in cybersecurity, safety
bounty programs would offer financial and reputational rewards to
members of the public who discover and responsibly report new safety
failures, such as novel jailbreaks, or capabilities beyond those found
in internal tests. As with red-teaming, bounty ``hunters'' can do this
by interacting with systems behind an API. However, it is as yet unclear
to what extent this impedes the ability of external testers to surface
and probe safety issues.

An early safety bounty trial by OpenAI for ChatGPT incentivized over
1500 submissions, with limited publicization and \$20,000 of API prizes
in total \cite{openai2022a}.
While OpenAI noted that the submissions seemed to yield few new
discoveries beyond the safety issues that internal red-teams had already
noticed, the exercise produced insight into the most common routes of
attack and lessons for future public engagement
\cite{levermore2023}.

Safety bounty programs can also be leveraged to identify promising
talent. Bounty hunters who submit multiple helpful tips could be
contacted and employed to perform more extensive system testing, and be
granted deeper levels of system access after appropriate vetting. In
cybersecurity, some bug bounty hunters earn payouts totaling over \$1
million for their work, and go on to work for large firms
\cite{hackerone2022,zhao2015}.

\clearpage

\subsection{Accelerate (beneficial) AI Progress }\label{sec4.2:accelerate-beneficial-ai-progress}

%\begin{table}[ht]
%\caption{Section summary: Open-sourcing as a mechanism for accelerating AI progress}
%\setlength{\tabcolsep}{5pt}\resizebox{\linewidth}{!}{
\begin{tabular}{|>{\centering\arraybackslash}m{0.15\linewidth}|m{0.79\linewidth}|}\hline
\multicolumn{2}{|c|}{\parbox{0.9\linewidth}{\captionof{table}{Section summary: Open-sourcing as a mechanism for accelerating AI progress\rule[-1mm]{0pt}{7mm}}}}\\\hline
\small\textbf{The argument for open-source AI} & Open-sourcing allows more people to
contribute to AI development processes and enables large-scale
collaborative efforts. The idea is that more expertise, more diverse
perspectives, and simply more human creativity and hours put into AI
development will drive innovation in new and useful downstream
integrations, advance AI safety research, and help push forward the
boundaries of AI capability. \\\hline
\small\textbf{Evaluation \mbox{of Benefit}} & 
\textit{Integration Progress}
\begin{itemize}[itemsep=1pt]

\item
  Open-sourcing is most helpful for integration progress. Model access
  allows more people to tinker, innovate, and optimize for integration
  with new downstream applications.
\end{itemize}

\textit{Capability Progress}

\begin{itemize}[itemsep=1pt]
\item
  Open-sourcing is less beneficial for capability progress than for
  integration progress.
\item
  The benefit is limited by bottlenecks in the talent, compute, and data
  resources needed for contributing to cutting-edge AI capability
  research.
\end{itemize}

\textit{Safety Progress}

\begin{itemize}[after=\vspace{-0.82\baselineskip},itemsep=1pt]
\item
  Academic safety research is often curtailed by insufficient access to
  highly capable models.
\item
  The benefit of open-source might be reduced by insufficient
  computation infrastructure outside of leading AI labs for running
  highly capable models.
\end{itemize}
 \\\hline
\small\textbf{Alternative methods for driving AI progress} &
\textit{Integration Progress}

\begin{itemize}[itemsep=1pt]
\item
  Use plugins for exploration of new applications.
\item
  Provide gated access {[}i.e. full access restricted to identified
  third parties{]} coupled with Know-Your-Customer Requirements.
\end{itemize}

\textit{Capability Progress / Safety Progress}

\begin{itemize}[after=\vspace{-0.82\baselineskip},itemsep=1pt]
\item
  Provide privileged model access to identified AI research groups,
  possibly via structured access research APIs.
\item
  Seek and organize collaborations with trusted parties and provide
  gated download access to collaborators.
\item
  Establish a multistakeholder governing body to mediate research access
  to protect against favoritism and to facilitate independent academic
  research.
\item
  Build incentive structures like large rewards programs for major
  scientific discoveries (e.g., protein folding) or pro-social advances
  (e.g. health and equity applications) using AI and for AI safety
  breakthroughs (e.g., interpretability).
\item
  Commit a certain percentage of profits or research hours towards AI
  safety projects.
\end{itemize}\\\hline
\end{tabular}
%}

\subsubsection{The Argument for
Open-Source}\label{the-argument-for-open-source-1}

Another argument for open-sourcing AI is that doing so helps to accelerate
progress that pushes the boundaries of AI capability, advances AI safety
research, and drives innovation of new downstream applications and
integrations. The idea is that open-sourcing allows more people to
contribute to AI development processes. It allows downstream developers
to optimize and perfect existing models instead of having to start from
scratch for each new application, and it enables large-scale
collaborative efforts. Furthermore, progress created by the wider AI
community will benefit from more diverse perspectives and insights,
which will ultimately help develop AI aligned to unique community needs
and cultural preferences. In addition, open-source efforts may be more
likely to focus on pro-social applications of AI, and be less influenced by the
financial and commercial incentives than industry AI developers.

These are benefits widely enjoyed by open-source software communities.
Linus Torvalds' open-source release of the Linux kernel, in particular,
showed how taking advantage of community-wide co-creation allows OSS
tools to be developed and released quickly, maintained cheaply, and
customized for individual needs without compromising quality. For cloud
computing especially, these benefits allowed the Linux operating system
to directly compete with Windows and MacOS, commercial systems backed by
significantly more resources such as specialized knowledge, corporate
information-sharing infrastructure, performance accountability
mechanisms, and marketing and legal support
\cite{dardaman2023,teamnuggets2018}.

It follows that we might expect AI progress to benefit similarly.

\subsubsection{Evaluating the Benefit for Foundation
Models}\label{evaluating-the-benefit-for-foundation-models-1}

In this section we evaluate the influence of open-source model sharing
on driving beneficial foundation model progress. To focus the
conversation we differentiate between three kinds of AI progress: (1)
integration progress, (2) safety progress, and (3) capability progress.

\paragraph{(1) Integration progress.} Integration progress is about the
discovery of new applications and integrations for foundation models to
serve a greater variety of needs---i.e. how a model can be applied to
new tasks and integrated with other applications. For example, ChatGPT
embedded with Duolingo has made for an effective language tutoring and
practice tool
\cite{marr2023}.

Of the three forms of progress, integration progress benefits most from
open-source. Open-sourcing models and model components gives more people
access to tinker and innovate. But perhaps more importantly, passing on
a model with all life-cycle documentation to downstream developers
enables those developers to optimize the model's performance by
fine-tuning its training and to infinitely test and evaluate the model
when integrated into the final product --- as Alex Engler writes, there
is ``simply too much at stake for downstream developers to use AI
systems they do not fully understand''
\cite{engler2022a}.

Indeed, recent breakthroughs in fine-tuning---specifically Low Rank
Adaptation
(LoRA)\cite{dettmers2023}---were
driven by open-source communities out of necessity for reducing costs
and compute requirements. It is a process by which the performance of
smaller models can be significantly improved by optimizing model weights
using the outputs of more high-capable models as training
data.\footnote{We classify fine-tuning as a form of integration progress
  instead of capability progress because the impressive performance of
  fine-turned models bootstraps on the capabilities of existing models.
  Pushing the frontier of AI capability still requires significant
  talent and compute at a scale only found in large, well-resourced labs
  \cite{gudibande2023}.}

\paragraph{(2) Safety Progress.} Safety progress refers to advances made in
AI safety research. AI Safety research works to improve AI safety by
identifying causes of unintended and harmful behavior, aligning AI
behavior with human values, improving model interpretability and
robustness, and otherwise developing tools to ensure AI systems work
safely and reliably
\cite{centerforsecurityandemergingtechnology2021,hendrycks2022}.

Current safety research is often limited by insufficient access to
large, cutting-edge models and relevant information such as their
architecture and training processes
\cite{bucknallForthcoming}.
Open-sourcing does alleviate these restrictions but is not necessary for
all safety research.

Different areas of safety research require different kinds of model
access. For example, evaluation and benchmarking research aims to
develop and test methods to assess the capabilities and safety of AI
systems. Often the ability to sample from a model via an API will be
sufficient for this research, as current approaches are based on
observing a model's output in response to a given
prompt.

In contrast, research areas such as alignment and interpretability
require more comprehensive access. Alignment research, which aims to
help AI systems better reflect user preferences and values, typically
requires researchers to be able to modify a model through fine-tuning,
including through the use of reinforcement learning. Like model
sampling, fine-tuning might also be facilitated through an API (e.g.
\cite{openaia}. However,
some experts express concern that current interfaces often do not
provide enough information about underlying models for them to draw
meaningful conclusions from their research.\footnote{For example, when
  attempting to evaluate the effect of instruct fine-tuning across
  multiple models, Wei et al.
  \cite{wei2023} write:
  \emph{``We do not compare InstructGPT against GPT-3 models in this
  experiment because we cannot determine if the only difference between
  these model families is instruction tuning (e.g., we do not even know
  if the base models are the same).''} Bucknall et al.
  \cite{bucknallForthcoming} discuss
  this and other examples from literature and expert interviews that
  elucidate the limitations many APIs pose for researchers.}
Interpretability research further requires that researchers can directly
modify model internals such as learned parameters and activation
patterns. Full (or nearly full) model access is needed for
interpretability research. That said, current interpretability research
is not limited by access to large models because interpretability
techniques are not mature enough to be ``computationally doable'' in the
largest models. In other words, we have a way to go before open-sourcing
our most capable models is a significant benefit to interpretability
research.

Even where comprehensive model access is crucial to a research agenda,
other factors can reduce the benefits of open-sourcing highly capable
models. For example, safety researchers external to private labs
sometimes lack sufficient computational infrastructure to run highly
capable foundation models
\cite{bucknallForthcoming}. Yet, some
research agendas, such as those studying emergent capabilities, require
access to the largest models at the bleeding edge of development;
smaller models that can be run on local hardware do not reliably exhibit
the emergent capabilities under investigation, even when fine-tuned on
the outputs of larger models.

\paragraph{(3) Capability progress.} Finally, capability progress
describes advancement in frontier AI research toward developing more
powerful and capable systems (i.e. working towards AGI).

The extent to which open-source contributions can drive progress on AI
frontier capabilities may be limited by access to compute and data
resources, as well as the distribution of talent.

Few AI actors have the requisite financial, compute, high-quality data
and talent resources to operate at the cutting edge of AI research and
development. Training new foundation models costs \$10-100 million in
compute costs and is projected to increase to \$1-10 billion in coming
years \cite{cottier2023}; the
stock of high-quality data used to train large language models (such as
books) currently freely available on the internet may be depleted in a
few years, requiring potentially costly new sources of data, innovations
in data efficiency
\cite{villalobos2022}, or
expensive human feedback data; and AI talent is most heavily
concentrated in high-paying positions at leading AI labs, primarily
based in the United States, while smaller labs struggle to fill
positions \cite{macropolo2023}.

Open-sourcing large pre-trained models does allow less-well-resourced
actors such as academic labs and open-source developers to study and
innovate on these existing models. These communities can make technical
and conceptual innovation and refinements within open-source
environments that generate knowledge that can be incorporated to advance
the AI capability frontier. If a high variance of research and
development strategies are employed by open-source communities, their
contributions may be particularly valuable for advancing state of the
art AI.

Furthermore, open-source model sharing also facilitates talent
development. More people being able to interact with pre-trained cutting
edge-models may, over time, lead to a larger and more diverse AI talent
pool for government regulators, AI labs, universities, and auditing
institutions to draw from. On a longer time scale this could have a
positive effect on capability progress (and safety progress) by
increasing the talent pool.

Realistically, however, the advancements that push the capability
frontier will nearly exclusively take place at frontier labs in
leading nations. In these locations, in-house expertise can draw upon
open-source innovations and top talent to run giant training runs using
huge compute, data, and engineering resources not available to the open
source community. (See Section~\ref{sec4.3:distribute-control-over-ai} for discussion on distributing AI
development away from big tech.)

\paragraph{Furthermore, the desirability of accelerating capability
progress is presently hotly debated.} This is due to concerns over risks
as well as benefits of more advanced models, in addition to the
governance challenge of preparing appropriate regulation and oversight
for such a rapidly advancing technology
\cite{laion.ai2023,jeffries2023,grace2022,futureoflifeinstitute2023}.
Accordingly, ``Accelerating AI capability progress'', to the extent that
open-sourcing does drive capability progress, should only be considered
an open-source benefit if the effect of open-sourcing is to drive
beneficial progress disproportionately to increasing risk and severity
of harm.

Toward beneficial AI progress, one benefit of open-sourcing is that it
puts AI tools in the hands of safety researchers, e.g., in academia, who
would otherwise not have access to the cutting edge models. We expand on
this point shortly under ``Safety Progress''. Open-sourcing also
increases opportunity for external scrutiny.

However, open-sourcing frontier models might also drive progress in
undesirable directions. One example of this is the potential effect of
open-source model sharing on the offense-defense balance; open-sourcing
may empower malicious actors to offensively identify and exploit system
vulnerabilities to a greater extent than it facilitates defensive
activities to protect against malicious use (See Section~\ref{sec3.1:malicious-use} for further
details).

\vspace{5mm}
\begin{tcolorbox}[boxrule=1pt,enhanced jigsaw, sharp corners,pad at break*=1mm,colbacktitle=white,colback=white,colframe=black,coltitle=black,toptitle=1mm,bottomtitle=1mm,fonttitle=\bfseries\small\centering,title=Box 2: Strategies for driving safety progress alongside model
sharing]

Alongside alternative model sharing strategies, there are also other
activities that can be employed to help safety progress. These are not
alternatives to model sharing, but are worthwhile considerations if
accelerating safety progress is the desired outcome.

\subsection*{Large rewards programs}

Progress might be accelerated in crucial AI safety domains by building
new incentive structures, for instance, large rewards programs on the
scale of millions or billions of dollars to reward major AI safety
breakthroughs (e.g., in model interpretability). The goal is to make
safety progress, like capability progress, a financially lucrative
endeavor.

\subsection*{Committing profits to safety research}

Safety progress could also be prioritized by orchestrating agreements
between frontier AI labs to commit a certain percentage of profits or
research hours towards AI safety projects. This would reduce incentives
for labs to cut corners on safety research and help remedy the large
mismatch in resources currently dedicated to capability progress versus
safety progress by major labs.

\subsection*{International institutions and collaborations for AI Safety}

Finally, in the long term we may benefit greatly from establishing
international institutions and collaboration to promote AI safety
\cite{ho2023}. For
instance, there is budding interest in establishing global collaboration
on advancing AI safety research akin to CERN or ITER\footnotemark
\cite{ho2023,marcus2023}.
Such a project could funnel significant resources towards AI safety
research, enable open and secure sharing of insights between leading
nations, and reduce the burden of cost (financial and opportunity costs)
associated with dedicating significant resources to AI safety research.

There is a risk that collaborative AI safety research would facilitate
the diffusion of dual use technologies and disincentivize leading labs
from conducting their own safety research. It is therefore imperative
that any such project be coupled with efforts to involve safety
researchers from leading labs (e.g., by offering dual appointment or
advisory positions) and to implement careful membership restrictions and
information security measures
\cite{ho2023}.
\end{tcolorbox}
\footnotetext{The
  International Thermonuclear Experimental Reactor (ITER) is an
  international nuclear fusion research and engineering megaproject
  aimed at creating energy through nuclear fusion. https://www.iter.org/}
\clearpage

\subsubsection{\texorpdfstring{Other Ways to Drive (Beneficial) Progress
}{Other Ways to Drive (Beneficial) Progress }}\label{other-ways-to-drive-beneficial-progress}

There are a variety of methods that might be employed to help pursue
open-source objectives. These methods do not necessarily cover all losses from
not open-sourcing, but they do not suffer the same risks as
open-sourcing and can be used in combination.

Toward \textbf{integration progress}, for example, new integrations and
applications can be explored and implemented through the development of
\textbf{``plugins''} allowing a model to integrate with other services
\cite{openaib}. The plugin
could be submitted to the developer or a third-party auditor before
publication. This option provides a mechanism for new integrations and
applications to be reviewed and approved before being shipped while
still tapping into public creativity and representation of interests and
needs.

In so far as model access allows downstream developers to more
thoroughly understand and test the performance and safety of their
integrations, labs could also provide identified downstream developers
with privileged access to requested model components via \textbf{gated
download}. One policy recommendation is that labs are held to a
\textbf{``know-your-customer'' requirement} whereby labs must vet and
keep a record of potential model recipients (e.g., proposed use, past
activities, funding source, etc.)
\cite{anderljung2023,schuett2023}.
Additionally, technical safety measures such as applying a unique
fingerprint to each copy of the model's weights should be applied when
feasible \cite{yu2022}.

As discussed above, the benefit of open-sourcing for \textbf{safety
progress} and \textbf{capability progress} is dampened by limited talent
and compute resources external to major labs. There are, however, other
means of driving both forward.

As mentioned in 4.1.3, developers might provide \textbf{privileged model
access} to AI safety research groups, possibly via structured access
research APIs. While not yet fully realized, there is hope that suitably
comprehensive researcher access to closed models can also be provided
through structured access approaches
\cite{shevlane2022}, such as
specialized \textbf{researcher API} access
\cite{bucknallForthcoming}. Such
solutions could be used in addition to existing social and legal
mechanisms for ensuring information security, such as researcher NDAs,
thereby potentially providing more comprehensive security guarantees
than either approach could in isolation.

For the purpose of propelling capability progress, labs could also
actively \textbf{seek collaborations} with trusted parties and provide
gated download access to collaborators. This is similar, for example, to
how OpenAI partnered with research institutions during the staged
release of GPT-2, providing access to models for carrying out research
into biases and methods for detecting GPT-2-generated text
\cite{solaiman2019}. As before,
any time gated download access is provided, it should be backed by
know-your-customer investigation and documentation requirements, and any
applicable technical safety measures. Selectively providing model
weights to only those researchers whose work requires them would also
help reduce the risk of leaks.

There is a challenge, however, regarding the decision as to which actors are provided
privileged model access (gated downloadable or via research API) to
conduct external evaluation and research or for collaborations. Where
labs are inundated with an unmanageable number of requests for research
access, favoritism and in-group model evaluations may emerge out of
necessity. Labs are also likely to prioritize external collaborators who
they believe will support their market interests. One possible solution
could be to \textbf{establish a multistakeholder governance body or
system for mediating researcher access} to highly capable foundation
models. For example, within the UK, we might imagine the recently
established Frontier AI Taskforce taking on such a role.

Such a body could also determine the degree of access provided to
external researchers (if through research API). This is important for
preventing ``independence by permissions'' whereby academic
collaborators are able to conduct high-quality independent research, but
research directions are ultimately determined by the access permissions
given by the developer
\cite{wagner2023}. For
cutting-edge models especially, researchers may not know which access
permissions they need to request, and the incentives are not clear for
developers to reveal everything they know (or suspect) about their
proprietary models.
\newpage

\subsection{Distribute Control Over
AI}\label{sec4.3:distribute-control-over-ai}

\begin{center}
%\caption{Section summary: Open-sourcing as a mechanism for distributing control over AI}
\setlength{\tabcolsep}{6pt}
\resizebox{\linewidth}{!}{\begin{tabular}{|>{\centering\arraybackslash}m{0.18\linewidth}|m{0.84\linewidth}|}\hline
\multicolumn{2}{|l|}{\parbox{0.98\linewidth}{\captionof{table}{Section summary: Open-sourcing as a mechanism for distributing control over AI\rule[-1mm]{0pt}{7mm}}}\hspace*{-5mm}}\\\hline
\small\textbf{The argument for open-source AI} &
Open-sourcing foundation models will help distribute influence over the
future of AI away from major labs by empowering smaller groups and
independent developers. The idea is that open-sourcing ``democratizes
AI'', giving more people influence over how AI is developed, optimized,
and used, and promotes the representation of more diverse interests and
needs in the direction of the field.\\\hline
\begin{tabular}{c}\small\bfseries
Evaluation \\[-7pt]\small\bfseries of Benefit\end{tabular} & 
\begin{itemize}[after=\vspace{-0.9\baselineskip},itemsep=-1pt]
\item
  Open-sourcing helps distribute control over downstream integration
  progress to open-source communities.
\item
  The effect of open-source on distributing influence over capability
  and safety progress is reduced by concentration of compute, data, and
  talent resources needed to influence frontier AI capability progress
  in large, well-resourced labs.
\item
  Open-sourcing large and highly capable models can also help amplify
  the original developer's influence over AI ecosystems; downstream
  innovations building on open-sourced models are easily integrated back
  into the developers' products, and the open-source communities become
  go-to hiring pools already familiar with the company's
  tools and models.
\item
  Open-sourcing is a tool that can aid the democratization of AI. But AI
  democratization is a multifaceted and proactive project to distribute
  influence over highly capable AI systems---how they are used,
  distributed, developed, and regulated---to wider communities of
  stakeholders and impacted populations. Open-sourcing alone cannot
  fulfill the goal of AI democratization.
\end{itemize}\\\hline
\small\textbf{Alternative methods for distributing control over AI} &
\begin{itemize}[after=\vspace{-0.9\baselineskip},itemsep=-1pt]
\item
  Implement participatory or representative deliberative processes to
  democratically inform high-impact decisions about AI development, use,
  and governance, including decisions about model access.
\item
  Institutionalize democratic structures (e.g., via democratically
  selected boards or by requiring the use of such deliberative processes
  for all decisions on particular topics) within large labs to dissipate
  control away from unilateral decision-makers.
\item
  Support appropriate regulatory intervention to developer behaviors and
  to guard against regulatory capture.
\end{itemize}\\\hline
\end{tabular}}
\end{center}
\vspace{3mm}
%\newpage

\subsubsection{The Argument for
Open-Source}\label{the-argument-for-open-source-2}

A commonly cited argument for open-sourcing foundation models is that
doing so will help distribute influence over the future of AI away from
major labs and to the wider AI community
\cite{howard2023,laion.ai2023a}.

There are very good reasons for wanting to distribute influence over AI.
There are economic implications; if open-sourcing foundation models
enables downstream developers to independently innovate and capitalize
on a lucrative technology, this could help to ensure that the huge value AI
promises to produce does not accrue only to a handful of tech giants.

There are also social and political implications; major AI labs are
unelected entities that primarily serve their own and shareholder
interests. The idea is that distributing influence over AI development
processes prevents private labs from exercising too much control over
numerous aspects of public life that emerging AI capabilities promise to
transform. As Emad Mostaque explains Stability AI's decision to
open-source Stable Diffusion, ``We trust people, and we trust the
community, as opposed to having a centralized, unelected entity
controlling the most powerful technology in the world''
\cite{scalevirtualevents2022}.

Overall, the idea is that open-sourcing ``democratizes AI'', giving more
people influence over how AI is developed and used, and promoting the
representation of more diverse interests and needs in the direction of
the field.

\subsubsection{Evaluating the Benefit for Foundation
Models}\label{evaluating-the-benefit-for-foundation-models-2}

Historically, open-source software development has had a noteworthy
influence-distributing effect. For instance, the open-source Linux
kernel now underpins numerous operatings systems (e.g., Ubuntu, Fedora,
Debian) that offer competitive and highly-utilized alternatives to
Windows and MacOS. We caution, however, that this effect should not be
expected to translate perfectly to the context of open-source foundation
models.

AI democratization is a multifaceted project. Open-sourcing certainly
contributes to AI democratization, though for some aspects of AI
democratization the effect is marginal. All aspects of AI
democratization benefit from investment in other proactive activities
aimed at distributing influence over AI and AI impacts. We briefly
review four aspects of AI democratization originally outlined in \cite{seger2023} and comment
on the extent to which open-source model sharing contributes to each.

\begin{itemize}[left=5.5mm, font=\bfseries]
    \item[(1)] \textbf{Democratization of AI development}
\end{itemize}

The democratization of AI development is about helping a wider range of
people contribute to AI design and development processes. Of the four
forms of AI democratization, open-sourcing promotes the democratization
of development most, and most directly. Open-sourcing places models in
the hands of large communities of open-source developers who can
continue to examine and modify the model. Open-sourcing also supports
self-learning and education among open-source developers, allowing them
to keep up with advances in model design and safety research and to
continue participating in AI development as techniques evolve.

There are, however, some ways in which the effect of open-source on the
democratization of AI development is limited.

First, especially with respect to highly-capable models, open-source
development activities may be increasingly limited by resource
accessibility. Participating at the cutting-edge of AI research and
development requires significant financial, compute, talent, and
high-quality data resources, and few actors outside of major labs and
government actors have these requisite resources (See Section~\ref{sec4.2:accelerate-beneficial-ai-progress}). As
Widder et al.
\cite{graywidder2023} write,
``\emph{even maximalist varieties of `open' AI don't democratize or
extend access to the resources needed to build AI from scratch---during
which highly significant `editorial' decisions are made.''} Accordingly,
toward the goals of facilitating wider and more diverse participation in
driving AI development, the benefit of open-sourcing is limited.

Second, open-sourcing can help leading AI developers to further entrench
their control over AI ecosystems and value production
\cite{patel2023,engler2021}.
While a near term, first-order effect is that downstream developers gain
influence over model application and integration progress, a
longer-term, second-order effect of open-sourcing large foundation
models is to feed back value and influence to the original developer.
Open-sourcing grants wider AI communities access to a technology that
they can fine-tune and customize to a variety of new applications.
However, these downstream innovations which build on top of the original
open-sourced model architecture, are then easily integrated back into
the original developer's own products and ecosystems. Open-source
communities also become go-to hiring pools already familiar with the
company's tools and models.

Third, the wider AI community, including open-source communities, are
relatively homogenous in terms of economic, cultural, gender, and
geographic grouping
\cite{macropolo2023,maslej2023}.
Open-source communities are often better than tech companies at building
diverse and inclusive spaces, and they put significant effort into
engaging with the broader world.\footnote{For example, the open-source
  AI research organization EleutherAI \cite{eleutherai2023} and the
  open-source collective BigScience
  \cite{bigscience} have
  teams spanning four or more continents and have projects focusing on
  increasing access to NLP technologies for people who speak
  non-dominant languages. Similarly, Cohere is running a program to
  collect fine-tuning data in hundreds of languages
  \cite{kayid2022}, and
  LAION is the only organization, at time of writing, to be training
  massively multilingual CLIP models \cite{beaumont2022,ilharco2021}.} However, something is still lost conflating the
distribution of power to open-source communities and the distribution of
power to communities generally. There is a risk that by missing this
nuance we exaggerate the benefits of open-sourcing alone and underplay
the need for other mechanisms for promoting the democratization of AI.
In addition to model sharing, democratizing AI development requires the
provision of educational and upskilling opportunities and technical
support infrastructure (e.g., high bandwidth network access and cloud
compute services) to encourage and enable wider and more diverse
participation in AI development processes.

\begin{itemize}[left=5.5mm, font=\bfseries]
    \item[(2)] \textbf{Democratization of AI use}
\end{itemize}

The democratization of AI use is about enabling a wide range of people
to use and benefit from AI applications. Open-sourcing allows downstream
developers to tailor models to serve diverse needs. For most people,
using an AI system also requires the provision of intuitive interfaces
to facilitate human-AI interaction without extensive training or
technical knowhow. Open-source communities can help develop these
interfaces.

However, one thing to consider is that benefiting from the use of an AI
system does not always require that everyone be able to use the AI
system. Especially for highly-capable and potentially high-risk systems,
a designated user could employ the system for the benefit of the
community. For example, a drug discovery system which could be
maliciously used to discover new toxins, could be used in a controlled,
limited-access setting while resulting pharmaceuticals are
``democratized'' in the sense that they are made accessible to anyone in
need.

\begin{itemize}[left=5.5mm, font=\bfseries]
    \item[(3)] \textbf{Democratization of AI profits}
\end{itemize}

The democratization of AI profits is about facilitating the broad and
equitable distribution of value accrued to organizations that build and
control advanced AI capabilities. Subgoals of profit democratization
include: smoothing economic transition in case of massive growth of the
AI industry, easing financial burden of job loss to automation,
preventing a widening economic divide between AI leading and lagging
nations, and acknowledging through compensation the human labor and
creativity that goes into producing and catering the data upon which
highly lucrative AI capabilities are built.

Open-sourcing helps democratize profits in two ways. First, by
open-sourcing their models, rather than charging for access, companies
will tend to capture less of the wealth produced by these models; users
can employ the models to generate profits (e.g. through increased
productivity) without having to pay some portion back to the developer.
Second, open-sourcing helps democratize profits insofar as it allows a
more widespread array of downstream developers to iterate upon AI models
and place competitive pressure on large labs; open-sourcing can make it
more difficult for large labs to build profitable downstream applications of
their models, since they will need to compete with open-source developer communities that are building competing applications.

However, the effect of open-source on distributing profits from
highly-capable AI will likely be limited in a couple respects. First,
open-source community participation in the development of cutting-edge
models will be curbed by inadequate access to necessary compute and
financial resources (Section~\ref{sec4.2:accelerate-beneficial-ai-progress}), thus limiting the competitive
pressure open-source developers can put on well-resourced large labs.
Second, as discussed earlier in this section, open-sourcing frontier
systems can also be financially advantageous to large companies in the
long run as they can use downstream developers as a free labor source,
easily feeding their best contributions and insights back into the
company's own products.

Additional proactive measures are needed to help pursue the goals of
profit democratization. These might include implementation of a profit
redistribution scheme such as taxation and redistribution by the state
\cite{altman2021,miller2021},
lab commitments to a windfall clause whereby developers obligate
themselves to donate windfall profits (measured as ``a substantial
fraction of the world's total economic output'') for redistribution
\cite{okeefe2020}, and
mechanisms for compensating content creators for the data on which
generative AI models are trained, for instance, through the creation of
licensed data sets
\cite{bigcode2020,vincent2022}.

\begin{itemize}[left=5.5mm, font=\bfseries]
    \item[(4)] \textbf{Democratization of AI governance}
\end{itemize}

Finally, the democratization of AI governance is about distributing
influence over decisions about AI to a wider community of stakeholders
and impacted populations. AI governance decisions involve balancing AI
related risks and benefits to determine how and by whom AI is used,
distributed, developed, and regulated.

Of the four forms of AI democratization, open-sourcing has the least
impact on distributing influence over AI governance decisions.
Open-sourcing distributes influence over AI governance decisions away
from major labs insofar as it enables wider AI research and development
communities to participate in, and therefore direct, AI development
processes. However, open-sourcing does little to gain influence over AI
governance decisions for the public more broadly. In this respect,
democratizing AI governance involves applying democratic processes
directly to high-impact decisions made by AI developers, subjugating
labs to regulation by democratic governments, or some combination
thereof. We expand on these possibilities shortly in 4.3.3.

\paragraph{Overall, open-sourcing AI should not be conflated with
democratizing AI}. Open-sourcing is but one option for sharing models
and model components; model sharing is but one mechanism for
democratizing AI development; and the democratization of AI development
is but one dimension of distributing influence and control over the
future of AI. Indeed, the decision to open-source is itself a
consequential decision over which influence can and likely should be
distributed away from private labs.

\subsubsection{Other Ways to Reduce Corporate or Autocratic
Control}\label{sec4.3.3:other-ways-to-reduce-corporate-or-autocratic-control}

A comprehensive approach will be needed to counteract the centralisation
of power in AI companies as AI systems become more capable and therefore
confer more political and economic power. This section presents options
for distributing influence over AI via the democratization of AI
governance. It is not an exhaustive list, but it illustrates that there are
a wide variety of methods that can be used to decentralize power and to
better facilitate representation of diverse stakeholder interests and
needs in decisions about how and by whom AI is developed, used,
distributed, and regulated.

\paragraph{Public participation and deliberation.} AI labs and policymakers
could institute participatory and deliberative democratic processes to
guide decision-making about complex issues in AI
\cite{seger2023}. For
example, \textbf{participatory platforms} such as Pol.is
\cite{polis2023} might be
used to solicit and synthesize public input into complex normative
decisions about AI at low cost. Alternatively, \textbf{representative
deliberations}, such as citizens assemblies, can convene representative
microcosms of impacted populations (or even global populations) selected
by sortition (i.e. stratified sampling) to tackle AI governance
questions \cite{coy2023,thecollectiveintelligenceproject2023a}.

Such efforts by large tech companies are not unprecedented. Meta, for
example, has quietly run a set of national and transnational pilots
\cite{costa2022,ovadya2023}
to navigate their `complex normative challenges' and have since scaled
up to a near-global deliberative process
\cite{harris2022}. Twitter
had also planned to pilot such processes before its acquisition
\cite{ovadya2023a}, and OpenAI
recently has launched a ``democratic inputs to AI'' grant program to
experiment with setting up democratic processes for deciding what rules
AI systems should follow within legal bounds
\cite{zaremba2023}.

\paragraph{Institutional structure.} Instead of, or in addition to,
directly eliciting public input to inform key decisions, another option
is for AI labs to introduce organizational structures that are more
democratic in nature. These structures would help maintain transparency
of internal practices and to dissipate control away from unilateral
decision-makers in such a way that better reflects stakeholder
interests. Relevant stakeholders importantly include public communities
whose lives are impacted by emerging AI capabilities.

For example, AI labs can \textbf{incorporate as Public Benefit
Corporations} (PBC).\footnote{There is increased momentum toward this
  now, as two leading AI organizations, Anthropic and Inflection AI, are
  both PBC's.} A PBC is a for-profit corporation intended to produce
public benefits and to operate in a responsible and sustainable manner.
Incorporating as a PBC does not necessitate public involvement, but it
does give a corporation clearer legal standing to make decisions about
institutional structure that aim to maximize public benefit, even if
that might conflict with maximizing shareholder interests.

For more direct public control, a \emph{golden share} (a nominal share
which is able to outvote all other shares) could be held by a
\emph{perpetual purpose trust} (a non-charitable trust established for
the benefit of a purpose) governed by a committee that is a
representative sample of the public selected by sortition or elected by
stakeholders.

Alternatively, AI labs could \textbf{implement democratically selected
oversight boards.} Such a board might, for instance, be composed of
representatives from the public selected by sortition or, perhaps a more
palatable option, a sortition body is used to ``elect'' board members
from among a nominated list. ``Nominators'' could be members of
government (e.g., state governors), and perhaps two to three board
members are committed to `voting' on issues as determined by a
democratic process (e.g., public polling or citizen assembly, whichever
is appropriate to the situation).

\paragraph{Regulation by democratic governments.} Finally, of course, labs
can encourage government regulation that restricts their behavior and
capacity for independent decision-making where the potential for
significant societal impact is high. For example, governments could
require authorization for large foundation model release and institute
multistakeholder committees to mediate research access to highly capable
models. Regulatory interventions should be developed in response to
deliberative processes involving developers, open source communities,
academia, and civil society to reflect diverse stakeholder interests and
to guard against regulatory capture by AI industry. In this way
appropriate government regulation can help systematically reduce
unilateral control over AI by leading private labs.

\section{Recommendations}\label{sec5:recommendations}

We conclude this paper with five high-level recommendations for AI
developers, standard setting bodies, and governments for working towards
safe and responsible model sharing decisions. These recommendations are necessarily incomplete and preliminary because best practices for open-sourcing highly capable models will be highly context-dependent and require input from numerous parties. We
look forward to further development of these recommendations in future
work.

Table~\ref{table5} summarizes the recommendations. Each recommendation is explained
in more detail below.

\vspace{4mm}
\begin{tabular}{|m{0.965\linewidth}|}\hline
\multicolumn{1}{|c|}{\parbox{0.95\linewidth}{\captionof{table}{Recommendations for working towards responsible model-sharing\rule[-0.5mm]{0pt}{4.5mm}}\label{table5}}}\\[-4pt]\hline
\begin{itemize}[left=2pt,after=\vspace{-0.9\baselineskip}]
\item[1.]
  \textbf{Developers and governments should recognise that some highly
  capable models will be too risky to open-source, at least initially.}
  These models may become safe to open-source in the future as societal
  resilience to AI risk increases and improved safety mechanisms are
  developed.
\item[2.] \textbf{Decisions about open-sourcing highly capable foundation models should be informed by rigorous risk assessments.} In addition to evaluating models for dangerous capabilities and immediate misuse
  applications, risk assessments must consider how a model might be
  fine-tuned or otherwise amended to facilitate misuse.

\item[3.]
  \textbf{Developers should consider alternatives to open-source release
  that capture some of the same {[}distributive, democratic, and
  societal{]} benefits, without creating as much risk.} Some promising
  alternatives include gradual or ``staged'' model release, model access
  for researchers and auditors, and democratic oversight of AI
  development and governance decisions.
\item[4.]
  \textbf{Developers, standards setting bodies, and open-source
  communities should engage in collaborative and multi-stakeholder
  efforts to define fine-grained standards for when model components
  should be released.} These standards should be based on an
  understanding of the risks posed by releasing (different combinations
  of) model components.
  
\item[5.]
  \textbf{Governments should exercise oversight of open source AI models and enforce safety measures when stakes are sufficiently high.} AI
  developers may not voluntarily adopt risk assessment and model sharing
  standards. Governments will need to enforce such measures through
  options such as liability law and regulation (e.g. via licensing
  requirements, fines, or penalties). Governments will also need to build the
  capacity to enforce such oversight mechanisms effectively.
\end{itemize}\\\hline
\end{tabular}

\newpage

\begin{itemize}[left=4.5mm, font=\bfseries]
    \item[1.] \textbf{Developers and governments should recognise that some highly
  capable models will be too dangerous to open-source, at least
  initially.}
\end{itemize}

If models are determined to pose significant threats, and those risks
are determined to outweigh the potential benefits of open-sourcing, then
those models should not be open-sourced. Such models may include those
that can materially assist development of biological and chemical
weapons \cite{sandbrink2023,soice2023}, enable successful cyberattacks against critical national
infrastructure
\cite{centerforsecurityandemergingtechnology2020}, or
facilitate highly-effective manipulation and persuasion
\cite{goldstein2023}.\footnote{Note
  that we do not claim that existing models are already too risky. We
  also do not make any predictions about how risky the next generation
  of models will be. Our claim is that developers need to assess the
  risks and be willing to not open-source a model if the risks outweigh
  the benefits.}

This is not to say that a given highly capable model should \emph{never}
be open-sourced. Expected model impacts are likely to change with
increasing societal resilience and development of new defensive
techniques. However, model developers should consider a default policy
of pursuing release through alternative methods rather than open-source
if they find that a model poses significant threats, and that the
benefits of open-sourcing do not outweigh the risks of doing so.

\begin{itemize}[left=4.5mm, font=\bfseries]
    \item[2.] \textbf{Decisions about open-sourcing highly capable foundation models should be informed by rigorous risk assessments.}
\end{itemize}

In the past, the benefits of open-sourcing seem to have clearly
outweighed the risks. However, we are not confident that this will
continue to be the case in the future for highly capable foundation
models (Section~3). It is therefore important to carefully assess
potential risks and benefits before open-sourcing the model, especially
since the decision to open-source a model is irreversible. The need to
conduct risk assessments prior to model release seems to be generally
accepted \cite{anderljung2023,thewhitehouse2023,schuett2023a}.

The National Institute of Standards and Technology (NIST) provides
guidance for how to conduct such an assessment
\cite{tabassi2023} which might
be applied to inform open-sourcing decisions. Some scholars have
suggested ways in which the NIST AI Risk Management Framework could be
adapted to general-purpose AI systems
\cite{centerforlong-termcybersecurity2023} and
catastrophic risks
\cite{barrett2023}. In the
future, we think that developers of highly capable foundation models
will need to combine qualitative and quantitative approaches. They may
need to conduct deterministic safety assessment as well as probabilistic
risk assessments, as is common in the nuclear industry
\cite{international2001iaea}.

Since risks associated with certain model capabilities are particularly
concerning, risk assessments should be informed by evaluations of
dangerous model capabilities
\cite{shevlane2023}. Both
internal \cite{openai2023d,anthropic2023c} and external model evaluations should be conducted. External
assessments can take many different shapes, such as model evaluations
\cite{kinniment2023,arcevals2023},
model audits
\cite{mokander2023,raji2019,raji2022}, red-teaming
\cite{ganguli2022,perez2022},
or researcher access via API \cite{bucknallForthcoming}.

Developers intending to open-source a model that is likely to be highly capable
should conduct more involved risk assessments than they would have
otherwise. Firstly, the risk assessment should be more thorough to have
the decision be as well-informed as possible, given the irreversibility
of decisions to open-source. Methods such as additional red teaming,
internal testing, and staged release approaches should be pursued.

Secondly, risk assessments ahead of open-sourcing decisions need to
assess how the model can be amended to facilitate misuse. The risk
assessment must consider the ease with which safeguards can be removed
and ``uncensored'' versions of the model can be distributed. Often,
safeguards will be so easy to remove that it is better to avoid the
model having the worrying capability altogether (Section~\ref{sec3:risks-of-open-sourcing-foundation-models}). For
example, while Stable Diffusion 1.0 had a safety filter, it was easy to
disable \cite{rando2022}. In
future releases, Stability AI therefore opted to remove inappropriate
content from the training data instead
\cite{stabilityai2022}.

Risk assessments should also consider the extent to which risks can be
exacerbated by malicious actors fine-tuning or otherwise amending the
model to elicit or develop more dangerous capabilities (Section~\ref{sec3:risks-of-open-sourcing-foundation-models}). It
is difficult to anticipate how the model is going to be fine-tuned. It
is therefore crucial that red-teamers have fine-tuning access to the
model ahead of release.

Thirdly, risk assessments should consider factors external to the model.
The social impacts of a model (e.g., on democratic processes) are
difficult to forecast and necessitate consideration of how the model
will interact with other tools and outside institutions, cultures and
material conditions
\cite{solaiman2023}.

Finally, for red-teaming, model evaluations and other external safety
assessments to be effective, AI developers need to elicit participation
from a diverse and comprehensive set of experts. Only by harnessing a
varied set of viewpoints and expertise can we ensure a broad spectrum of
potential risks are adequately identified and evaluated.

\begin{itemize}[left=4.5mm, font=\bfseries]
    \item[3.] \textbf{Developers should consider alternatives to open-source release as possibilities for working towards distributive, democratic, and
  societal advancement goals with less risk.}
\end{itemize}

Before open-sourcing a highly capable foundation model, developers
should first clarify goals---reflecting on why specifically they want to
open-source a model---and then consider alternatives that may reach
those goals at lower risk.

With respect to alternative model-sharing strategies, some options may
offer some of the same benefits as open-sourcing, but unlike
open-sourcing, still allow developers to adjust their deployment
strategy after release. The idea that models are either fully open or
fully closed is a false dichotomy. As discussed in Section~\ref{sec4:benefits-of-open-sourcing-foundation-models-and-alternative-methods-for-achieving-them}, there are
numerous options for gated, API, or hosted access in between which allow
for varying degrees of model probing and researchability
\cite{solaiman2023a,bucknallForthcoming},
and there are proposed frameworks to help navigate these options
\cite{liang2022a,whittlestone2020}.

Developers could also deploy the model in stages (staged-release) and
gather observational data about how a model is likely to be (mis)used
and modified if open-sourced (Section~\ref{sec4.1.3:other-ways-to-enable-external-evaluation}). Finally, developers could
employ proactive efforts to pursue desired benefits, such as by
implementing democratic processes to distribute influence over
development and release decisions (Section~\ref{sec4.3.3:other-ways-to-reduce-corporate-or-autocratic-control}).

\begin{itemize}[left=4.5mm, font=\bfseries]
    \item[4.] \textbf{A collaborative and multi-stakeholder effort is needed to
  define fine-grained standards for when model components should be
  released.}
\end{itemize}

Standard-setting organizations or industry bodies should develop
model-sharing standards that provide guidance relating to decisions
about whether, and if so how, to open-source highly capable foundation
models. Such a standard would contribute to more consistent industry
practices and could be an important step towards regulation. There are a
wide range of model-sharing options, even within the currently
ill-defined category of ``open-source'' systems (see Box 1).

Model-sharing standards should both support safe model distribution and
protect open-source practices and benefits. To achieve both, these standards
must be fine-grained and built on a well-researched understanding of the
extent to which access to different (combinations of) model components
enable unrestricted model use, reproduction, and modification.

We make a start at breaking down and defining the numerous model components
that can be independently shared in Appendix~\ref{appendixA}. It is, however, a much
larger project than we can do justice to here, and it is a project on
which members of open-source communities should be centrally involved. A
clear understanding of activities enabled by access to various model
components can then be used to inform model-sharing standards that are
well-tailored to their purpose, that are not overly burdensome, that
prevent distribution of dangerous capabilities, and that do not
unnecessarily undermine open-source benefits.

Technical experts, open-source communities, policymakers, and civil
society all need to be involved in this process. There are several
actors who could develop such standards. Although standard-setting
organizations like NIST
\cite{tabassi2023} and ISO/IEC
\cite{iso2023} have
published standards for AI, they do not seem to have engaged with
questions around open-sourcing foundation models specifically. The
Partnership on AI (PAI) has a working group on foundation models
\cite{partnershiponaistaff2023} and they
have published similar guidelines for publishing research in the past
\cite{partnershiponaistaff2021}. The Open
Source Initiative recently started a working group to define what makes
an AI system ``open source''
\cite{maffulli2023a}. Another
body that could contribute industry expertise is the recently-announced
Frontier Model Forum
\cite{microsoft2023}, however
current participants have generally not open-sourced their most advanced
foundation models.

\newpage

\begin{itemize}[left=4.5mm, font=\bfseries]
    \item[5.] \textbf{Government should exercise oversight and enforcement where
  stakes are sufficiently high.}
\end{itemize}

AI developers may not voluntarily adopt the risk assessment and model
sharing standards described above, and government involvement will
likely be needed. Without such involvement, developers may not be
sufficiently incentivised to voluntarily conduct thorough risk
assessments ahead of model release, to appropriately act on those
results, to provide sufficient external access to their models, or put
in place appropriate safeguards. For instance, AI developers may
preferentially choose ``friendly'' external assessors who share similar
concerns around certain types of risk, or whose financial incentives
undermine their ability to provide an independent assessment.

To mitigate such potentialities, governments should increase oversight
capacity and set up mechanisms for enforcing rigorous risk assessments
and responsible model release in sufficiently high-stakes contexts.
Governments need to ensure that oversight is rigorous and independent,
supported by a diverse and comprehensive set of independent advisors,
and investigates a wide range of AI risks. Similarly, enforcement
mechanisms need to guard against the risk of regulatory capture.

There are multiple options governments could consider in terms of
enforcement, such as:

\paragraph{Liability.} Developers could be held liable for harms caused by
their models that could have been reasonably foreseen\footnote{See
  \cite{americanlawinstitute1965} § 4
  (Duty) and § 6 (Tortious Conduct) (1965), and § 901 on the general
  principle of liability (1979); See
  \cite{americanlawinstitute1998} on
  products liability.} or avoided through an exercise of due
care.\footnote{See
  \cite{goldberg2001} on the
  legal concept of negligence.} While courts will ultimately have to
decide liability on a case-by-case basis, there are strong incentives
for developers to demonstrate due care, by, for example, conducting
thorough risk assessments and model evaluations, implementing adequate
precautionary measures, refraining from or reducing high-risk
activity,\footnote{See
  \cite[p. 61]{landes1987}} and
maintaining their ability to limit harms that occur post-release.
Existing tort law already covers unjustifiably risky acts and omissions,
via negligence for failing to exercise due care (including to prevent
foreseeable criminal conduct by a third parties\footnote{See
  \cite{americanlawinstitute1965} §§ 302A-B
  (1965); Restatement (Third) of Torts: General Principles § 17
  (Discussion Draft April 5, 1999) ("The conduct of a defendant can lack
  reasonable care insofar as it can foreseeably combine with or bring
  about the improper conduct of . . . a third party."); \emph{see,
  e.g.}, \emph{Hamilton v. Accu-Tek}, 62 F. Supp. 2d 802, 825 (E.D.N.Y.
  1999), 222 F. 3d 36 (2d Cir. 2000), 95 N.Y.2d 878 (N.Y. 2000) (Holding
  that gun manufacturers had a duty ``to take reasonable steps available
  . . . to reduce the possibility that {[}their products would{]} fall
  into the hands of those likely to misuse them'' and thus could be held
  legally responsible under New York negligence law for criminal
  shootings resulting from failures to ``minimize the risk'' through
  their distribution and marketing choices).}), products liability for
defective designs, and strict liability for abnormally dangerous
activities.\footnote{See
  \cite{americanlawinstitute1965} § 520
  (1977).} A critical task will be to clarify the application of these
doctrines to open-sourcing highly capable foundation models
\cite{hacker2023}. Where the
application of existing liability regimes fails to address significant
risks, new statutory duties and liability laws may need to be developed.

\paragraph{Regulation.} Governments could legally require developers of
highly capable foundation models to conduct pre-deployment risk
assessments, report potentially dangerous capabilities discovered during
model evaluation, and provide model access pre-deployment to government
auditors. Regulations may also specify under which conditions models may
be open-sourced
\cite{anderljung2023}. They could
also encourage or mandate that significant model deployments are
preceded by notifications to relevant parts of government
\cite{mulani2023}. Such
requirements could be enforced by administrative enforcement measures,
both before model deployments (e.g., via a licensing regime) as well as
after (e.g., via fines and penalties)
\cite{anderljung2023}.

It is worth noting that liability and regulation each have their
strengths and weaknesses. While liability is generally less onerous and
more flexible, enforcing liability rules might be difficult (e.g.,
because of causation and attribution problems, especially when a
malicious actor intervenes) and it is not possible to enforce liability
rules ahead of model deployments. Regulation is the only way to enforce
compliance before a model is open-sourced. However, regulation typically
leads to higher compliance costs and there are risks of regulatory
capture. In general, liability should be seen as a complement to, rather
than a substitute for, regulation
\cite{anderljung2023}. Since the
right mix of policies will be highly context-specific, we do not make
any further recommendations.

Policy interventions on open-sourcing are delicate because of the
obvious benefits of open-sourcing and because for-profit companies might
use safety concerns as an excuse to gain a competitive advantage. These
concerns should be taken seriously, and further research is needed to
understand the risks, benefits, and legal feasibility of different
policy options. However, policy interventions still seem necessary
because open-sourcing highly capable foundation models might essentially
democratize the ability to cause significant harm and because the
decision to open-source a model is irreversible
\cite{anderljung2023b}. We think
the current debate around the issue
\cite{henshall2023} is healthy
and necessary to strike the right balance between open-source risks and benefits. In this paper, we have
advocated for a risk-based approach that could be summarized as
\textbf{``make open-source decisions with care''}.

\section{Conclusion}\label{conclusion}

Open-sourcing offers clear advantages including enabling external
oversight, accelerating progress, and decentralizing control over a
potentially transformative technology. To date, open-source practice has
provided substantial net benefits for most software and AI development
processes, distributing influence over the direction of technological
innovation and facilitating the development of products well-tailored to
diverse user needs.

However, as AI research progresses and capabilities improve,
open-sourcing also presents a growing potential for misuse and
unintended consequences. Open-sourcing increases the risk of
proliferation of model flaws downstream. With access to model weights
and code, malicious actors can also more easily bypass safety measures
and modify models or fine-tune models to display dangerous capabilities.
Some of the most worrying potentialities involve the use of highly
capable foundation models to build new biological and chemical weapons,
to mount cyberattacks against critical infrastructures and institutions,
and to execute highly-effective political influence operations.

For some highly capable foundation models these risks may come to
outweigh open-source benefits. In such cases, developers and regulators
should acknowledge that the model should not be open-sourced, at least
initially. These models may become safe to open-source in the future as
societal resilience to AI risk increases and improved safety mechanisms
are developed.

Model release decisions should therefore be responsive to comprehensive
risk assessments and a fine-grained understanding of what activities are
enabled by freely sharing different combinations of model components.
These decisions should also take into account how alternative model
sharing options (e.g. staged release, gated access, and research API)
might further some of the same goals as open-sourcing. Alternative
proactive measures to organize secure collaborations, and to encourage
and enable wider involvement in AI development, evaluation, and
governance processes might also be employed. Open-sourcing is but one
option for sharing models, and model sharing is but one mechanism for
facilitating wider community contributions to AI evaluation,
development, and control.

Overall, openness, transparency, accessibility, and wider community
input are key to facilitating a future for beneficial AI. The goal of
this paper is therefore not to argue that foundation model development
should be kept behind closed doors. Model sharing, including
open-sourcing, remains a valuable practice in most cases. Rather, we
submit that decisions to open-source increasingly capable models must be
considered with great care. Comprehensive risk assessments and careful
consideration of alternative methods for pursuing open-source objectives
are minimum first steps.

\newpage
\setcode{utf8}
\addcontentsline{toc}{section}{References}
\printbibliography

@online{thecollectiveintelligenceproject2023,
  title = {Introducing the {{Collective Intelligence Project Solving}} the {{Transformative Technology Trilemma}} through {{Governance R}}\&{{D}}},
  author = {{The Collective Intelligence Project}},
  date = {2023},
  url = {https://cip.org/whitepaper},
  urldate = {2023-09-23}
}

@article{hoffmann2022,
  title = {Training {{Compute-Optimal Large Language Models}}},
  author = {Hoffmann, Jordan and Borgeaud, Sebastian and Mensch, Arthur and Buchatskaya, Elena and Cai, Trevor and Rutherford, Eliza and Casas, Diego de Las and Hendricks, Lisa Anne and Welbl, Johannes and Clark, Aidan and Hennigan, Tom and Noland, Eric and Millican, Katie and family=Driessche, given=George, prefix=van den, useprefix=false and Damoc, Bogdan and Guy, Aurelia and Osindero, Simon and Simonyan, Karen and Elsen, Erich and Rae, Jack W. and Vinyals, Oriol and Sifre, Laurent},
  date = {2022-03-29},
  eprint = {2203.15556},
  eprinttype = {arxiv},
  eprintclass = {cs},
  doi = {10.48550/arXiv.2203.15556},
  pubstate = {preprint},
  journaltitle = {}
}

@online{openai,
  title = {{{GPT-4}} Is {{OpenAI}}’s Most Advanced System, Producing Safer and More Useful Responses},
  author = {OpenAI},
  url = {https://openai.com/gpt-4},
  urldate = {2023-09-23},
  langid = {american}
}

@online{anthropic2023,
  title = {Claude 2},
  author = {Anthropic},
  date = {2023-07-11},
  url = {https://www.anthropic.com/index/claude-2},
  urldate = {2023-09-24},
  langid = {english},
  organization = {{Anthropic}}
}

@online{brockman2023,
  title = {Introducing {{ChatGPT}} and {{Whisper APIs}}},
  author = {Brockman, Greg and Eleti, Atty and Georges, Elie and Jang, Joanne and Kilpatrick, Logan and Lim, Rachel and Miller, Luke and Pokrass, Michelle},
  date = {2023-03-01},
  url = {https://openai.com/blog/introducing-chatgpt-and-whisper-apis},
  urldate = {2023-09-24},
  langid = {american}
}

@online{goldman2023,
  title = {Hugging {{Face}}, {{GitHub}} and More Unite to Defend Open Source in {{EU AI}} Legislation},
  author = {Goldman, Sharon},
  date = {2023-07-26T07:00:00+00:00},
  url = {https://venturebeat.com/ai/hugging-face-github-and-more-unite-to-defend-open-source-in-eu-ai-legislation/},
  urldate = {2023-09-24},
  langid = {american},
  organization = {{VentureBeat}}
}

@report{creativecommons2023,
  title = {Supporting {{Open Source}} and {{Open Science}} in the {{EU AI Act}}},
  author = {{Creative Commons} and Eleuther.ai and GitHub and {Hugging Face} and LAION and {Open Future}},
  date = {2023},
  url = {https://huggingface.co/blog/assets/eu_ai_act_oss/supporting_OS_in_the_AIAct.pdf}
}

@article{assran2023,
  title = {Self-{{Supervised Learning}} from {{Images}} with a {{Joint-Embedding Predictive Architecture}}},
  author = {Assran, Mahmoud and Duval, Quentin and Misra, Ishan and Bojanowski, Piotr and Vincent, Pascal and Rabbat, Michael and LeCun, Yann and Ballas, Nicolas},
  date = {2023-04-13},
  eprint = {2301.08243},
  eprinttype = {arxiv},
  eprintclass = {cs, eess},
  doi = {10.48550/arXiv.2301.08243},
  urldate = {2023-09-24},
  pubstate = {preprint},
  journaltitle = {}
}

@online{metaai2023,
  title = {Introducing {{Llama}} 2: {{The}} next Generation of Our Open Source Large Language Model},
  author = {{Meta AI}},
  date = {2023},
  url = {https://ai.meta.com/llama-project},
  urldate = {2023-09-24},
  langid = {english},
  organization = {{Meta AI}}
}

@misc{inskeep2023,
  title = {Meta Leans on 'wisdom of Crowds' in {{AI}} Model Release},
  author = {Inskeep, Steve and Hampton, Olivia},
  date = {2023-07-19},
  journaltitle = {Morning Edition},
  publisher = {{NPR}},
  url = {https://www.npr.org/2023/07/19/1188543421/metas-nick-clegg-on-the-companys-decision-to-offer-ai-tech-as-open-source-softwa},
  urldate = {2023-09-24},
  langid = {english},
  entrysubtype = {podcastepisode},
  author+an:role = {=hosts}
}

@article{milmo2023,
  entrysubtype = {newspaper},
  title = {Nick {{Clegg}} Defends Release of Open-Source {{AI}} Model by {{Meta}}},
  author = {Milmo, Dan},
  date = {2023-07-19T11:52:14},
  journaltitle = {The Guardian},
  url = {https://www.theguardian.com/technology/2023/jul/19/nick-clegg-defends-release-open-source-ai-model-meta-facebook},
  urldate = {2023-09-24},
  journalsubtitle = {Technology},
  langid = {british}
}

@inproceedings{langenkamp2022,
  title = {How {{Open Source Machine Learning Software Shapes AI}}},
  booktitle = {Proceedings of the 2022 {{AAAI}}/{{ACM Conference}} on {{AI}}, {{Ethics}}, and {{Society}}},
  author = {Langenkamp, Max and Yue, Daniel N.},
  date = {2022-07-26},
  pages = {385--395},
  publisher = {{ACM}},
  location = {{Oxford United Kingdom}},
  doi = {10.1145/3514094.3534167},
  urldate = {2023-09-24},
  eventtitle = {{{AIES}} '22: {{AAAI}}/{{ACM Conference}} on {{AI}}, {{Ethics}}, and {{Society}}},
  isbn = {978-1-4503-9247-1},
  langid = {english}
}

@report{engler2021,
  type = {AI Governance Report},
  title = {How Open-Source Software Shapes {{AI}} Policy},
  author = {Engler, Alex},
  date = {2021-08-10},
  institution = {{Brookings}},
  url = {https://www.brookings.edu/articles/how-open-source-software-shapes-ai-policy/},
  urldate = {2023-09-24},
  langid = {american}
}

@online{engler2022,
  title = {The {{EU}}’s Attempt to Regulate Open-Source {{AI}} Is Counterproductive},
  author = {Engler, Alex},
  date = {2022-08-24},
  url = {https://www.brookings.edu/articles/the-eus-attempt-to-regulate-open-source-ai-is-counterproductive/},
  urldate = {2023-09-24},
  langid = {american},
  organization = {{Brookings}}
}

@online{zwetsloot2019,
  title = {Thinking {{About Risks From AI}}: {{Accidents}}, {{Misuse}} and {{Structure}}},
  shorttitle = {Thinking {{About Risks From AI}}},
  author = {Zwetsloot, Remco and Dafoe, Allan},
  date = {2019-02-11},
  url = {https://www.lawfaremedia.org/article/thinking-about-risks-ai-accidents-misuse-and-structure},
  urldate = {2023-09-24},
  langid = {english},
  organization = {{Default}}
}

@article{shevlane2022,
  title = {Structured Access: An Emerging Paradigm for Safe {{AI}} Deployment},
  shorttitle = {Structured Access},
  author = {Shevlane, Toby},
  date = {2022-04-11},
  eprint = {2201.05159},
  eprinttype = {arxiv},
  eprintclass = {cs},
  doi = {10.48550/arXiv.2201.05159},
  urldate = {2023-09-24},
  pubstate = {preprint},
  journaltitle = {}
}

@article{bommasani2022,
  title = {On the {{Opportunities}} and {{Risks}} of {{Foundation Models}}},
  author = {Bommasani, Rishi and Hudson, Drew A. and Adeli, Ehsan and Altman, Russ and Arora, Simran and family=Arx, given=Sydney, prefix=von, useprefix=true and Bernstein, Michael S. and Bohg, Jeannette and Bosselut, Antoine and Brunskill, Emma and Brynjolfsson, Erik and Buch, Shyamal and Card, Dallas and Castellon, Rodrigo and Chatterji, Niladri and Chen, Annie and Creel, Kathleen and Davis, Jared Quincy and Demszky, Dora and Donahue, Chris and Doumbouya, Moussa and Durmus, Esin and Ermon, Stefano and Etchemendy, John and Ethayarajh, Kawin and Fei-Fei, Li and Finn, Chelsea and Gale, Trevor and Gillespie, Lauren and Goel, Karan and Goodman, Noah and Grossman, Shelby and Guha, Neel and Hashimoto, Tatsunori and Henderson, Peter and Hewitt, John and Ho, Daniel E. and Hong, Jenny and Hsu, Kyle and Huang, Jing and Icard, Thomas and Jain, Saahil and Jurafsky, Dan and Kalluri, Pratyusha and Karamcheti, Siddharth and Keeling, Geoff and Khani, Fereshte and Khattab, Omar and Koh, Pang Wei and Krass, Mark and Krishna, Ranjay and Kuditipudi, Rohith and Kumar, Ananya and Ladhak, Faisal and Lee, Mina and Lee, Tony and Leskovec, Jure and Levent, Isabelle and Li, Xiang Lisa and Li, Xuechen and Ma, Tengyu and Malik, Ali and Manning, Christopher D. and Mirchandani, Suvir and Mitchell, Eric and Munyikwa, Zanele and Nair, Suraj and Narayan, Avanika and Narayanan, Deepak and Newman, Ben and Nie, Allen and Niebles, Juan Carlos and Nilforoshan, Hamed and Nyarko, Julian and Ogut, Giray and Orr, Laurel and Papadimitriou, Isabel and Park, Joon Sung and Piech, Chris and Portelance, Eva and Potts, Christopher and Raghunathan, Aditi and Reich, Rob and Ren, Hongyu and Rong, Frieda and Roohani, Yusuf and Ruiz, Camilo and Ryan, Jack and Ré, Christopher and Sadigh, Dorsa and Sagawa, Shiori and Santhanam, Keshav and Shih, Andy and Srinivasan, Krishnan and Tamkin, Alex and Taori, Rohan and Thomas, Armin W. and Tramèr, Florian and Wang, Rose E. and Wang, William and Wu, Bohan and Wu, Jiajun and Wu, Yuhuai and Xie, Sang Michael and Yasunaga, Michihiro and You, Jiaxuan and Zaharia, Matei and Zhang, Michael and Zhang, Tianyi and Zhang, Xikun and Zhang, Yuhui and Zheng, Lucia and Zhou, Kaitlyn and Liang, Percy},
  date = {2022-07-12},
  eprint = {2108.07258},
  eprinttype = {arxiv},
  eprintclass = {cs},
  doi = {10.48550/arXiv.2108.07258},
  urldate = {2023-09-24},
  pubstate = {preprint},
  journaltitle = {}
}

@report{jones2023,
  title = {Explainer: {{What}} Is a Foundation Model?},
  shorttitle = {Explainer},
  author = {Jones, Elliot},
  date = {2023-07-17},
  institution = {{Ada Lovelace Institute}},
  url = {https://www.adalovelaceinstitute.org/resource/foundation-models-explainer/},
  urldate = {2023-09-24},
  langid = {british}
}

@article{shea2023,
  title = {Use of {{GPT-4}} to {{Analyze Medical Records}} of {{Patients With Extensive Investigations}} and {{Delayed Diagnosis}}},
  author = {Shea, Yat-Fung and Lee, Cynthia Min Yao and Ip, Whitney Chin Tung and Luk, Dik Wai Anderson and Wong, Stephanie Sze Wing},
  date = {2023-08-14},
  journaltitle = {JAMA Network Open},
  shortjournal = {JAMA Netw Open},
  volume = {6},
  number = {8},
  pages = {e2325000},
  issn = {2574-3805},
  doi = {10.1001/jamanetworkopen.2023.25000},
  urldate = {2023-09-24},
  langid = {english}
}

@online{openai2023,
  title = {Be {{My Eyes}}: {{Be My Eyes}} Uses {{GPT-4}} to Transform Visual Accessibility},
  author = {OpenAI},
  date = {2023-03-14},
  url = {https://openai.com/customer-stories/be-my-eyes},
  urldate = {2023-09-24},
  langid = {american}
}

@online{openai2023a,
  title = {Viable: {{Viable}} Uses {{GPT-4}} to Analyze Qualitative Data at a Revolutionary Scale with Unparalleled Accuracy},
  author = {OpenAI},
  date = {2023-07-07},
  url = {https://openai.com/customer-stories/viable},
  urldate = {2023-09-24}
}

@online{openai2023b,
  title = {Inworld {{AI}}: {{Using GPT-3}} to Create the next Generation of {{AI-powered}} Characters},
  author = {OpenAI},
  date = {2023-01-01},
  url = {https://openai.com/customer-stories/inworld-ai},
  urldate = {2023-09-24}
}

@online{altmann2023,
  title = {{{GPT-4 Chatbot}} for {{Customer Service}} | {{The New ChatGPT Beta Chatbot}} in {{Test}}},
  author = {Altmann, Yasmin},
  date = {2023-03-27},
  url = {https://omq.ai/blog/gpt-4-chatbot-in-customer-service-beta-chatbot/},
  urldate = {2023-09-24},
  langid = {english},
  organization = {{OMQ Blog}}
}

@online{marr2023,
  title = {The {{Amazing Ways Duolingo Is Using AI And GPT-4}}},
  author = {Marr, Bernard},
  date = {2023-04-28},
  url = {https://www.forbes.com/sites/bernardmarr/2023/04/28/the-amazing-ways-duolingo-is-using-ai-and-gpt-4/},
  urldate = {2023-09-24},
  langid = {english},
  organization = {{Forbes}}
}

@online{openai2023c,
  title = {Stripe: {{Stripe}} Leverages {{GPT-4}} to Streamline User Experience and Combat Fraud},
  author = {OpenAI},
  date = {2023-03-14},
  url = {https://openai.com/customer-stories/stripe},
  urldate = {2023-09-24},
  langid = {american}
}

@online{harvey.ai,
  title = {Harvey: {{Unprecedented}} Legal {{AI}}},
  author = {Harvey.ai},
  url = {https://www.harvey.ai/},
  urldate = {2023-09-24},
  langid = {english}
}

@article{rombach2022,
  title = {High-{{Resolution Image Synthesis}} with {{Latent Diffusion Models}}},
  author = {Rombach, Robin and Blattmann, Andreas and Lorenz, Dominik and Esser, Patrick and Ommer, Björn},
  date = {2022-04-13},
  eprint = {2112.10752},
  eprinttype = {arxiv},
  eprintclass = {cs},
  doi = {10.48550/arXiv.2112.10752},
  urldate = {2023-09-24},
  pubstate = {preprint},
  journaltitle = {}
}

@article{ramesh2022,
  title = {Hierarchical {{Text-Conditional Image Generation}} with {{CLIP Latents}}},
  author = {Ramesh, Aditya and Dhariwal, Prafulla and Nichol, Alex and Chu, Casey and Chen, Mark},
  date = {2022-04-12},
  eprint = {2204.06125},
  eprinttype = {arxiv},
  eprintclass = {cs},
  doi = {10.48550/arXiv.2204.06125},
  urldate = {2023-09-24},
  pubstate = {preprint},
  journaltitle = {}
}

@article{openai2023d,
  title = {{{GPT-4 Technical Report}}},
  author = {OpenAI},
  date = {2023-03-27},
  eprint = {2303.08774},
  eprinttype = {arxiv},
  eprintclass = {cs},
  doi = {10.48550/arXiv.2303.08774},
  urldate = {2023-09-24},
  pubstate = {preprint},
  journaltitle = {}
}

@online{mehdi2023,
  title = {Furthering Our {{AI}} Ambitions – {{Announcing Bing Chat Enterprise}} and {{Microsoft}} 365 {{Copilot}} Pricing},
  author = {Mehdi, Yusuf and Spataro, Jared},
  date = {2023-07-18T15:30:09+00:00},
  url = {https://blogs.microsoft.com/blog/2023/07/18/furthering-our-ai-ambitions-announcing-bing-chat-enterprise-and-microsoft-365-copilot-pricing/},
  urldate = {2023-09-24},
  langid = {american},
  organization = {{Official Microsoft Blog}}
}

@online{vincent2023,
  title = {Meta’s Powerful~{{AI}}~Language Model Has Leaked Online — What Happens Now? - {{The Verge}}},
  author = {Vincent, James},
  date = {2023-03-08},
  url = {https://www.theverge.com/2023/3/8/23629362/meta-ai-language-model-llama-leak-online-misuse},
  urldate = {2023-09-24},
  organization = {{The Verge}}
}

@online{fries2022,
  title = {How {{Foundation Models Can Advance AI}} in {{Healthcare}}},
  author = {Fries, Jason and Steinberg, Ethan and Fleming, Scott and Wornow, Michael and Xu, Yizhe and Morse, Keith and Dash, Dev and Shah, Nigam},
  date = {2022-12-15},
  url = {https://hai.stanford.edu/news/how-foundation-models-can-advance-ai-healthcare},
  urldate = {2023-09-24},
  langid = {english},
  organization = {{Stanford HAI}}
}

@online{marr2023a,
  title = {Digital {{Twins}}, {{Generative AI}}, {{And The Metaverse}}},
  author = {Marr, Bernard},
  date = {2023-05-23},
  url = {https://www.forbes.com/sites/bernardmarr/2023/05/23/digital-twins-generative-ai-and-the-metaverse/},
  urldate = {2023-09-24},
  langid = {english},
  organization = {{Forbes}}
}

@article{milmo2023a,
  entrysubtype = {newspaper},
  title = {Paedophiles Using Open Source {{AI}} to Create Child Sexual Abuse Content, Says Watchdog},
  author = {Milmo, Dan},
  date = {2023-09-13T10:46:28},
  journaltitle = {The Guardian},
  url = {https://www.theguardian.com/society/2023/sep/12/paedophiles-using-open-source-ai-to-create-child-sexual-abuse-content-says-watchdog},
  urldate = {2023-09-24},
  journalsubtitle = {Society},
  langid = {british}
}

@inproceedings{horvitz2022,
  title = {On the {{Horizon}}: {{Interactive}} and {{Compositional Deepfakes}}},
  shorttitle = {On the {{Horizon}}},
  booktitle = {{{ICMI}} '22: {{Proceedings}} of the 2022 {{International Conference}} on {{Multimodal Interaction}}},
  author = {Horvitz, Eric},
  date = {2022-11-07},
  pages = {653--661},
  publisher = {{ACM}},
  location = {{Bengaluru India}},
  doi = {10.1145/3536221.3558175},
  urldate = {2023-09-24},
  isbn = {978-1-4503-9390-4},
  langid = {english}
}

@article{verma2023,
  entrysubtype = {newspaper},
  title = {They Thought Loved Ones Were Calling for Help. {{It}} Was an {{AI}} Scam.},
  author = {Verma, Pranshu},
  date = {2023-03-10},
  journaltitle = {Washington Post},
  url = {https://www.washingtonpost.com/technology/2023/03/05/ai-voice-scam/},
  urldate = {2023-09-24},
  langid = {american}
}

@online{brewster2021,
  title = {Fraudsters {{Cloned Company Director}}’s {{Voice In}} \$35 {{Million Heist}}, {{Police Find}}},
  author = {Brewster, Thomas},
  date = {2021-10-14},
  url = {https://www.forbes.com/sites/thomasbrewster/2021/10/14/huge-bank-fraud-uses-deep-fake-voice-tech-to-steal-millions/},
  urldate = {2023-09-24},
  langid = {english},
  organization = {{Forbes}}
}

@inproceedings{weidinger2022,
  title = {Taxonomy of {{Risks}} Posed by {{Language Models}}},
  booktitle = {2022 {{ACM Conference}} on {{Fairness}}, {{Accountability}}, and {{Transparency}}},
  author = {Weidinger, Laura and Uesato, Jonathan and Rauh, Maribeth and Griffin, Conor and Huang, Po-Sen and Mellor, John and Glaese, Amelia and Cheng, Myra and Balle, Borja and Kasirzadeh, Atoosa and Biles, Courtney and Brown, Sasha and Kenton, Zac and Hawkins, Will and Stepleton, Tom and Birhane, Abeba and Hendricks, Lisa Anne and Rimell, Laura and Isaac, William and Haas, Julia and Legassick, Sean and Irving, Geoffrey and Gabriel, Iason},
  date = {2022-06-21},
  pages = {214--229},
  publisher = {{ACM}},
  location = {{Seoul Republic of Korea}},
  doi = {10.1145/3531146.3533088},
  urldate = {2023-09-24},
  eventtitle = {{{FAccT}} '22: 2022 {{ACM Conference}} on {{Fairness}}, {{Accountability}}, and {{Transparency}}},
  isbn = {978-1-4503-9352-2},
  langid = {english}
}

@article{solaiman2023,
  title = {Evaluating the {{Social Impact}} of {{Generative AI Systems}} in {{Systems}} and {{Society}}},
  author = {Solaiman, Irene and Talat, Zeerak and Agnew, William and Ahmad, Lama and Baker, Dylan and Blodgett, Su Lin and Daumé III, Hal and Dodge, Jesse and Evans, Ellie and Hooker, Sara and Jernite, Yacine and Luccioni, Alexandra Sasha and Lusoli, Alberto and Mitchell, Margaret and Newman, Jessica and Png, Marie-Therese and Strait, Andrew and Vassilev, Apostol},
  date = {2023-06-12},
  eprint = {2306.05949},
  eprinttype = {arxiv},
  eprintclass = {cs},
  doi = {10.48550/arXiv.2306.05949},
  urldate = {2023-09-24},
  pubstate = {preprint},
  journaltitle = {}
}

@article{shelby2023,
  title = {Sociotechnical {{Harms}} of {{Algorithmic Systems}}: {{Scoping}} a {{Taxonomy}} for {{Harm Reduction}}},
  shorttitle = {Sociotechnical {{Harms}} of {{Algorithmic Systems}}},
  author = {Shelby, Renee and Rismani, Shalaleh and Henne, Kathryn and family=Moon, given=Ajung, given-i={{Aj}} and Rostamzadeh, Negar and Nicholas, Paul and Yilla, N'Mah and Gallegos, Jess and Smart, Andrew and Garcia, Emilio and Virk, Gurleen},
  date = {2023-07-18},
  eprint = {2210.05791},
  eprinttype = {arxiv},
  eprintclass = {cs},
  doi = {10.48550/arXiv.2210.05791},
  urldate = {2023-09-24},
  pubstate = {preprint},
  journaltitle = {}
}

@book{crawford2021,
  title = {Atlas of {{AI}}: Power, Politics, and the Planetary Costs of Artificial Intelligence},
  shorttitle = {Atlas of {{AI}}},
  author = {Crawford, Kate},
  date = {2021},
  publisher = {{Yale University Press}},
  location = {{New Haven London}},
  isbn = {978-0-300-26463-0},
  langid = {english},
  pagetotal = {327}
}

@book{gray2019,
  title = {Ghost Work: How to Stop {{Silicon Valley}} from Building a New Global Underclass},
  shorttitle = {Ghost Work},
  author = {Gray, Mary L. and Suri, Siddharth},
  date = {2019},
  publisher = {{Houghton Mifflin Harcourt}},
  location = {{Boston}},
  isbn = {978-1-328-56628-7},
  pagetotal = {1}
}

@article{li2023,
  title = {Making {{AI Less}} "{{Thirsty}}": {{Uncovering}} and {{Addressing}} the {{Secret Water Footprint}} of {{AI Models}}},
  shorttitle = {Making {{AI Less}} "{{Thirsty}}"},
  author = {Li, Pengfei and Yang, Jianyi and Islam, Mohammad A. and Ren, Shaolei},
  date = {2023-04-06},
  eprint = {2304.03271},
  eprinttype = {arxiv},
  eprintclass = {cs},
  doi = {10.48550/arXiv.2304.03271},
  urldate = {2023-09-24},
  pubstate = {preprint},
  journaltitle = {}
}

@article{strubell2019,
  title = {Energy and {{Policy Considerations}} for {{Deep Learning}} in {{NLP}}},
  author = {Strubell, Emma and Ganesh, Ananya and McCallum, Andrew},
  date = {2019-06-05},
  eprint = {1906.02243},
  eprinttype = {arxiv},
  eprintclass = {cs},
  doi = {10.48550/arXiv.1906.02243},
  urldate = {2023-09-24},
  pubstate = {preprint},
  journaltitle = {}
}

@article{patterson2021,
  title = {Carbon {{Emissions}} and {{Large Neural Network Training}}},
  author = {Patterson, David and Gonzalez, Joseph and Le, Quoc and Liang, Chen and Munguia, Lluis-Miquel and Rothchild, Daniel and So, David and Texier, Maud and Dean, Jeff},
  date = {2021-04-23},
  eprint = {2104.10350},
  eprinttype = {arxiv},
  eprintclass = {cs},
  doi = {10.48550/arXiv.2104.10350},
  urldate = {2023-09-24},
  pubstate = {preprint},
  journaltitle = {}
}

@article{liang2022,
  title = {Holistic {{Evaluation}} of {{Language Models}}},
  author = {Liang, Percy and Bommasani, Rishi and Lee, Tony and Tsipras, Dimitris and Soylu, Dilara and Yasunaga, Michihiro and Zhang, Yian and Narayanan, Deepak and Wu, Yuhuai and Kumar, Ananya and Newman, Benjamin and Yuan, Binhang and Yan, Bobby and Zhang, Ce and Cosgrove, Christian and Manning, Christopher D. and Ré, Christopher and Acosta-Navas, Diana and Hudson, Drew A. and Zelikman, Eric and Durmus, Esin and Ladhak, Faisal and Rong, Frieda and Ren, Hongyu and Yao, Huaxiu and Wang, Jue and Santhanam, Keshav and Orr, Laurel and Zheng, Lucia and Yuksekgonul, Mert and Suzgun, Mirac and Kim, Nathan and Guha, Neel and Chatterji, Niladri and Khattab, Omar and Henderson, Peter and Huang, Qian and Chi, Ryan and Xie, Sang Michael and Santurkar, Shibani and Ganguli, Surya and Hashimoto, Tatsunori and Icard, Thomas and Zhang, Tianyi and Chaudhary, Vishrav and Wang, William and Li, Xuechen and Mai, Yifan and Zhang, Yuhui and Koreeda, Yuta},
  date = {2022-11-16},
  eprint = {2211.09110},
  eprinttype = {arxiv},
  eprintclass = {cs},
  doi = {10.48550/arXiv.2211.09110},
  urldate = {2023-09-24},
  pubstate = {preprint},
  journaltitle = {}
}

@article{hendrycks2021,
  title = {Measuring {{Massive Multitask Language Understanding}}},
  author = {Hendrycks, Dan and Burns, Collin and Basart, Steven and Zou, Andy and Mazeika, Mantas and Song, Dawn and Steinhardt, Jacob},
  date = {2021-01-12},
  eprint = {2009.03300},
  eprinttype = {arxiv},
  eprintclass = {cs},
  doi = {10.48550/arXiv.2009.03300},
  urldate = {2023-09-24},
  pubstate = {preprint},
  journaltitle = {}
}

@article{shevlane2023,
  title = {Model Evaluation for Extreme Risks},
  author = {Shevlane, Toby and Farquhar, Sebastian and Garfinkel, Ben and Phuong, Mary and Whittlestone, Jess and Leung, Jade and Kokotajlo, Daniel and Marchal, Nahema and Anderljung, Markus and Kolt, Noam and Ho, Lewis and Siddarth, Divya and Avin, Shahar and Hawkins, Will and Kim, Been and Gabriel, Iason and Bolina, Vijay and Clark, Jack and Bengio, Yoshua and Christiano, Paul and Dafoe, Allan},
  date = {2023-05-24},
  eprint = {2305.15324},
  eprinttype = {arxiv},
  eprintclass = {cs},
  doi = {10.48550/arXiv.2305.15324},
  urldate = {2023-09-24},
  pubstate = {preprint},
  journaltitle = {}
}

@report{anthropic2023a,
  title = {Anthropic's {{Responsible Scaling Policy}}, {{Version}} 1.0},
  author = {Anthropic},
  date = {2023-09-19},
  institution = {{Anthropic}},
  url = {https://www.anthropic.com/index/anthropics-responsible-scaling-policy},
  urldate = {2023-09-24},
  langid = {english}
}

@article{sandbrink2023,
  title = {Artificial Intelligence and Biological Misuse: {{Differentiating}} Risks of Language Models and Biological Design Tools},
  shorttitle = {Artificial Intelligence and Biological Misuse},
  author = {Sandbrink, Jonas B.},
  date = {2023-08-12},
  eprint = {2306.13952},
  eprinttype = {arxiv},
  eprintclass = {cs},
  doi = {10.48550/arXiv.2306.13952},
  urldate = {2023-09-24},
  pubstate = {preprint},
  journaltitle = {}
}

@article{mirsky2021,
  title = {The {{Threat}} of {{Offensive AI}} to {{Organizations}}},
  author = {Mirsky, Yisroel and Demontis, Ambra and Kotak, Jaidip and Shankar, Ram and Gelei, Deng and Yang, Liu and Zhang, Xiangyu and Lee, Wenke and Elovici, Yuval and Biggio, Battista},
  date = {2021-06-29},
  eprint = {2106.15764},
  eprinttype = {arxiv},
  eprintclass = {cs},
  doi = {10.48550/arXiv.2106.15764},
  urldate = {2023-09-24},
  pubstate = {preprint},
  journaltitle = {}
}

@report{centerforsecurityandemergingtechnology2020,
  title = {A {{National Security Research Agenda}} for {{Cybersecurity}} and {{Artificial Intelligence}}},
  author = {{Center for Security and Emerging Technology} and Buchanan, Ben},
  date = {2020-05},
  institution = {{Center for Security and Emerging Technology}},
  doi = {10.51593/2020CA001},
  urldate = {2023-09-24}
}

@article{anderljung2023,
  title = {Frontier {{AI Regulation}}: {{Managing Emerging Risks}} to {{Public Safety}}},
  shorttitle = {Frontier {{AI Regulation}}},
  author = {Anderljung, Markus and Barnhart, Joslyn and Korinek, Anton and Leung, Jade and O'Keefe, Cullen and Whittlestone, Jess and Avin, Shahar and Brundage, Miles and Bullock, Justin and Cass-Beggs, Duncan and Chang, Ben and Collins, Tantum and Fist, Tim and Hadfield, Gillian and Hayes, Alan and Ho, Lewis and Hooker, Sara and Horvitz, Eric and Kolt, Noam and Schuett, Jonas and Shavit, Yonadav and Siddarth, Divya and Trager, Robert and Wolf, Kevin},
  date = {2023-09-04},
  eprint = {2307.03718},
  eprinttype = {arxiv},
  eprintclass = {cs},
  doi = {10.48550/arXiv.2307.03718},
  urldate = {2023-09-24},
  pubstate = {preprint},
  journaltitle = {}
}

@report{kinniment2023,
  title = {Evaluating Language-Model Agents on Realistic Autonomous Tasks},
  author = {Kinniment, Megan and Jun Koba Sato, Lucas and Du, Haoxing and Goodrich, Brian and Hasin, Max and Chan, Lawrence and Harold Miles, Luke and Lin, Tao R. and Wijk, Hjalmar and Burget, Joel and Ho, Aaron and Barnes, Elizabeth and Christiano, Paul},
  date = {2023-07},
  institution = {{Alignment Research Center}},
  url = {https://evals.alignment.org/Evaluating_LMAs_Realistic_Tasks.pdf}
}

@article{shevlane2020,
  title = {The {{Offense-Defense Balance}} of {{Scientific Knowledge}}: {{Does Publishing AI Research Reduce Misuse}}?},
  shorttitle = {The {{Offense-Defense Balance}} of {{Scientific Knowledge}}},
  author = {Shevlane, Toby and Dafoe, Allan},
  date = {2020-01-09},
  eprint = {2001.00463},
  eprinttype = {arxiv},
  eprintclass = {cs},
  doi = {10.48550/arXiv.2001.00463},
  urldate = {2023-09-24},
  pubstate = {preprint},
  journaltitle = {}
}

@online{anthropic2023b,
  title = {Frontier {{Threats Red Teaming}} for {{AI Safety}}},
  author = {Anthropic},
  date = {2023-07-26},
  url = {https://www.anthropic.com/index/frontier-threats-red-teaming-for-ai-safety},
  urldate = {2023-09-24},
  langid = {english},
  organization = {{Anthropic}}
}

@article{wei2022,
  title = {Emergent {{Abilities}} of {{Large Language Models}}},
  author = {Wei, Jason and Tay, Yi and Bommasani, Rishi and Raffel, Colin and Zoph, Barret and Borgeaud, Sebastian and Yogatama, Dani and Bosma, Maarten and Zhou, Denny and Metzler, Donald and Chi, Ed H. and Hashimoto, Tatsunori and Vinyals, Oriol and Liang, Percy and Dean, Jeff and Fedus, William},
  date = {2022-10-26},
  eprint = {2206.07682},
  eprinttype = {arxiv},
  eprintclass = {cs},
  doi = {10.48550/arXiv.2206.07682},
  urldate = {2023-09-24},
  pubstate = {preprint},
  journaltitle = {}
}

@article{urbina2022,
  title = {Dual Use of Artificial-Intelligence-Powered Drug Discovery},
  author = {Urbina, Fabio and Lentzos, Filippa and Invernizzi, Cédric and Ekins, Sean},
  date = {2022-03-07},
  journaltitle = {Nature Machine Intelligence},
  shortjournal = {Nat Mach Intell},
  volume = {4},
  number = {3},
  pages = {189--191},
  issn = {2522-5839},
  doi = {10.1038/s42256-022-00465-9},
  urldate = {2023-09-24},
  langid = {english}
}

@online{helena2023,
  title = {Biosecurity in the {{Age}} of {{AI}}},
  author = {HELENA},
  date = {2023},
  url = {https://www.helenabiosecurity.org},
  urldate = {2023-09-24},
  langid = {english}
}

@book{dibona1999,
  title = {Open Sources: Voices from the Open Source Revolution},
  shorttitle = {Open Sources},
  editor = {DiBona, Chris and Ockman, Sam and Stone, Mark},
  date = {1999},
  edition = {1st ed},
  publisher = {{O'Reilly}},
  location = {{Beijing ; Sebastopol, CA}},
  isbn = {978-1-56592-582-3},
  pagetotal = {272}
}

@online{github,
  title = {Licenses},
  author = {Github},
  url = {https://choosealicense.com/licenses/},
  urldate = {2023-09-24},
  langid = {english},
}

@online{fanelli2023,
  title = {{{LLaMA2}} Isn't "{{Open Source}}"---and Why It Doesn't Matter},
  author = {Fanelli, Alessio},
  date = {2023-07-19},
  url = {https://www.alessiofanelli.com/blog/llama2-isnt-open-source},
  urldate = {2023-09-24},
  organization = {{Alessio Fanelli's blog}}
}

@online{maffulli2023,
  title = {Meta’s {{LLaMa}} 2 License Is Not {{Open Source}}},
  author = {Maffulli, Stefano},
  date = {2023-07-20T20:45:00+00:00},
  url = {https://blog.opensource.org/metas-llama-2-license-is-not-open-source/},
  urldate = {2023-09-24},
  langid = {american},
  organization = {{Voices of Open Source}}
}

@article{graywidder2023,
  title = {Open ({{For Business}}): {{Big Tech}}, {{Concentrated Power}}, and the {{Political Economy}} of {{Open AI}}},
  shorttitle = {Open ({{For Business}})},
  author = {Gray Widder, David and West, Sarah and Whittaker, Meredith},
  date = {2023},
  journaltitle = {SSRN Electronic Journal},
  shortjournal = {SSRN Journal},
  issn = {1556-5068},
  doi = {10.2139/ssrn.4543807},
  urldate = {2023-09-24},
  langid = {english}
}

@online{finley2011,
  title = {How to {{Spot Openwashing}}},
  author = {Finley, Klint},
  date = {2011-02-03T13:15:00+00:00},
  url = {https://readwrite.com/how_to_spot_openwashing/},
  urldate = {2023-09-24},
  langid = {american},
  organization = {{ReadWrite}}
}

@online{RAIL,
  title = {Responsible {{AI Licenses}}},
  author = {{Responsible AI Licenses}},
  url = {https://www.licenses.ai},
  urldate = {2023-09-24},
  langid = {american}
}

@inproceedings{widder2022,
  title = {Limits and {{Possibilities}} for “{{Ethical AI}}” in {{Open Source}}: {{A Study}} of {{Deepfakes}}},
  shorttitle = {Limits and {{Possibilities}} for “{{Ethical AI}}” in {{Open Source}}},
  booktitle = {2022 {{ACM Conference}} on {{Fairness}}, {{Accountability}}, and {{Transparency}}},
  author = {Widder, David Gray and Nafus, Dawn and Dabbish, Laura and Herbsleb, James},
  date = {2022-06-21},
  pages = {2035--2046},
  publisher = {{ACM}},
  location = {{Seoul Republic of Korea}},
  doi = {10.1145/3531146.3533779},
  urldate = {2023-09-24},
  eventtitle = {{{FAccT}} '22: 2022 {{ACM Conference}} on {{Fairness}}, {{Accountability}}, and {{Transparency}}},
  isbn = {978-1-4503-9352-2},
  langid = {english}
}

@online{sijbrandij2023,
  title = {{{AI}} Weights Are Not Open "Source"},
  author = {Sijbrandij},
  date = {2023-06-27T00:00:00+00:00},
  url = {https://opencoreventures.com/blog/2023-06-27-ai-weights-are-not-open-source/},
  urldate = {2023-09-24},
  langid = {english}
}

@article{solaiman2023a,
  title = {The {{Gradient}} of {{Generative AI Release}}: {{Methods}} and {{Considerations}}},
  shorttitle = {The {{Gradient}} of {{Generative AI Release}}},
  author = {Solaiman, Irene},
  date = {2023-02-05},
  eprint = {2302.04844},
  eprinttype = {arxiv},
  eprintclass = {cs},
  doi = {10.48550/arXiv.2302.04844},
  urldate = {2023-09-24},
  pubstate = {preprint},
  journaltitle = {}
}

@misc{gpt-j,
  author = {Wang, Ben and Komatsuzaki, Aran},
  title = {{GPT-J-6B: A 6 Billion Parameter Autoregressive Language Model}},
  howpublished = {\url{https://github.com/kingoflolz/mesh-transformer-jax}},
  year = 2021,
  month = May
}

@online{stabilityai,
  title = {Stable {{Diffusion Public Release}}},
  shorttitle = {Aug 22, 2023},
  author = {{Stability AI}},
  url = {https://stability.ai/blog/stable-diffusion-public-release},
  urldate = {2023-09-24},
  langid = {british},
  organization = {{stability.ai}}
}

@online{metaai2023a,
  title = {Introducing {{LLaMA}}: {{A}} Foundational, 65-Billion-Parameter Language Model},
  shorttitle = {Introducing {{LLaMA}}},
  author = {{Meta AI}},
  date = {2023-02-24},
  url = {https://ai.meta.com/blog/large-language-model-llama-meta-ai/},
  urldate = {2023-09-24},
  langid = {english}
}

@online{cottier2023,
  title = {Trends in the Dollar Training Cost of Machine Learning Systems},
  author = {Cottier, Ben},
  date = {2023-01-31T00:00:00+00:00},
  url = {https://epochai.org/blog/trends-in-the-dollar-training-cost-of-machine-learning-systems},
  urldate = {2023-09-24},
  langid = {english},
  organization = {{EPOCH}}
}

@online{li2020,
  title = {{{OpenAI}}'s {{GPT-3 Language Model}}: {{A Technical Overview}}},
  shorttitle = {{{OpenAI}}'s {{GPT-3 Language Model}}},
  author = {Li, Chuan},
  date = {2020-06-03},
  url = {https://lambdalabs.com/blog/demystifying-gpt-3},
  urldate = {2023-09-24},
  langid = {english},
  organization = {{Lambda}}
}

@online{venigalla2022,
  title = {Mosaic {{LLMs}} ({{Part}} 2): {{GPT-3}} Quality for {$<\$$}500k},
  shorttitle = {Mosaic {{LLMs}} ({{Part}} 2)},
  author = {Venigalla, Abhinav and Linden, Li},
  date = {2022-09-29},
  url = {https://www.mosaicml.com/blog/gpt-3-quality-for-500k},
  urldate = {2023-09-24},
  organization = {{Mosaic ML}}
}

@article{sevilla2022,
  title = {Compute {{Trends Across Three Eras}} of {{Machine Learning}}},
  author = {Sevilla, Jaime and Heim, Lennart and Ho, Anson and Besiroglu, Tamay and Hobbhahn, Marius and Villalobos, Pablo},
  date = {2022-03-09},
  eprint = {2202.05924},
  eprinttype = {arxiv},
  eprintclass = {cs},
  doi = {10.48550/arXiv.2202.05924},
  urldate = {2023-09-24},
  pubstate = {preprint},
  journaltitle = {}
}

@article{erdil2023,
  title = {Algorithmic Progress in Computer Vision},
  author = {Erdil, Ege and Besiroglu, Tamay},
  date = {2023-08-24},
  eprint = {2212.05153},
  eprinttype = {arxiv},
  eprintclass = {cs},
  doi = {10.48550/arXiv.2212.05153},
  urldate = {2023-09-24},
  pubstate = {preprint},
  journaltitle = {}
}

@article{hsieh2023,
  title = {Distilling {{Step-by-Step}}! {{Outperforming Larger Language Models}} with {{Less Training Data}} and {{Smaller Model Sizes}}},
  author = {Hsieh, Cheng-Yu and Li, Chun-Liang and Yeh, Chih-Kuan and Nakhost, Hootan and Fujii, Yasuhisa and Ratner, Alexander and Krishna, Ranjay and Lee, Chen-Yu and Pfister, Tomas},
  date = {2023-07-05},
  eprint = {2305.02301},
  eprinttype = {arxiv},
  eprintclass = {cs},
  doi = {10.48550/arXiv.2305.02301},
  urldate = {2023-09-24},
  pubstate = {preprint},
  journaltitle = {}
}

@online{sastry2021,
  title = {Beyond “{{Release}}” vs. “{{Not Release}}”},
  author = {Sastry, Girish},
  date = {2021},
  url = {https://crfm.stanford.edu/commentary/2021/10/18/sastry.html},
  urldate = {2023-09-24},
  organization = {{Center for Research on Foundation Models}}
}

@online{liang2022a,
  title = {The Time Is Now to Develop Community Norms for the Release of Foundation Models},
  author = {Liang, Percy and Bommasani, Rishi and Creel, Kathleen A. and Reich, Rob},
  date = {2022},
  url = {https://crfm.stanford.edu/2022/05/17/community-norms.html},
  organization = {{Center for Research on Foundation Models}}
}

@online{maffulli2023a,
  title = {Towards a Definition of "{{Open Artificial Intelligence}}": {{First}} Meeting Recap},
  shorttitle = {Towards a Definition of "{{Open Artificial Intelligence}}"},
  author = {Maffulli, Stefano},
  date = {2023-07-13T13:00:00+00:00},
  url = {https://blog.opensource.org/towards-a-definition-of-open-artificial-intelligence-first-meeting-recap/},
  urldate = {2023-09-25},
  langid = {american},
  organization = {{Voices of Open Source}}
}

@online{goldman2023a,
  title = {{{RedPajama}} Replicates {{LLaMA}} Dataset to Build Open Source, State-of-the-Art {{LLMs}}},
  author = {Goldman, Sharon},
  date = {2023-04-18T20:04:22+00:00},
  url = {https://venturebeat.com/ai/redpajama-replicates-llama-to-build-open-source-state-of-the-art-llms/},
  urldate = {2023-09-25},
  langid = {american},
  organization = {{VentureBeat}}
}

@article{rando2022,
  title = {Red-{{Teaming}} the {{Stable Diffusion Safety Filter}}},
  author = {Rando, Javier and Paleka, Daniel and Lindner, David and Heim, Lennart and Tramèr, Florian},
  date = {2022-11-10},
  eprint = {2210.04610},
  eprinttype = {arxiv},
  eprintclass = {cs},
  doi = {10.48550/arXiv.2210.04610},
  urldate = {2023-09-25},
  pubstate = {preprint},
  journaltitle = {}
}

@article{zou2023,
  title = {Universal and {{Transferable Adversarial Attacks}} on {{Aligned Language Models}}},
  author = {Zou, Andy and Wang, Zifan and Kolter, J. Zico and Fredrikson, Matt},
  date = {2023-07-27},
  eprint = {2307.15043},
  eprinttype = {arxiv},
  eprintclass = {cs},
  doi = {10.48550/arXiv.2307.15043},
  urldate = {2023-09-25},
  pubstate = {preprint},
  journaltitle = {}
}

@article{anderljung2023a,
  title = {Protecting {{Society}} from {{AI Misuse}}: {{When}} Are {{Restrictions}} on {{Capabilities Warranted}}?},
  shorttitle = {Protecting {{Society}} from {{AI Misuse}}},
  author = {Anderljung, Markus and Hazell, Julian},
  date = {2023-03-29},
  eprint = {2303.09377},
  eprinttype = {arxiv},
  eprintclass = {cs},
  doi = {10.48550/arXiv.2303.09377},
  urldate = {2023-09-25},
  pubstate = {preprint},
  journaltitle = {}
}

@article{brundage2018,
  title = {The {{Malicious Use}} of {{Artificial Intelligence}}: {{Forecasting}}, {{Prevention}}, and {{Mitigation}}},
  shorttitle = {The {{Malicious Use}} of {{Artificial Intelligence}}},
  author = {Brundage, Miles and Avin, Shahar and Clark, Jack and Toner, Helen and Eckersley, Peter and Garfinkel, Ben and Dafoe, Allan and Scharre, Paul and Zeitzoff, Thomas and Filar, Bobby and Anderson, Hyrum and Roff, Heather and Allen, Gregory C. and Steinhardt, Jacob and Flynn, Carrick and {hÉigeartaigh}, Seán Ó and Beard, Simon and Belfield, Haydn and Farquhar, Sebastian and Lyle, Clare and Crootof, Rebecca and Evans, Owain and Page, Michael and Bryson, Joanna and Yampolskiy, Roman and Amodei, Dario},
  date = {2018-02-20},
  eprint = {1802.07228},
  eprinttype = {arxiv},
  eprintclass = {cs},
  doi = {10.48550/arXiv.1802.07228},
  urldate = {2023-09-25},
  pubstate = {preprint},
  journaltitle = {}
}

@article{weidinger2021,
  title = {Ethical and Social Risks of Harm from {{Language Models}}},
  author = {Weidinger, Laura and Mellor, John and Rauh, Maribeth and Griffin, Conor and Uesato, Jonathan and Huang, Po-Sen and Cheng, Myra and Glaese, Mia and Balle, Borja and Kasirzadeh, Atoosa and Kenton, Zac and Brown, Sasha and Hawkins, Will and Stepleton, Tom and Biles, Courtney and Birhane, Abeba and Haas, Julia and Rimell, Laura and Hendricks, Lisa Anne and Isaac, William and Legassick, Sean and Irving, Geoffrey and Gabriel, Iason},
  date = {2021-12-08},
  eprint = {2112.04359},
  eprinttype = {arxiv},
  eprintclass = {cs},
  doi = {10.48550/arXiv.2112.04359},
  urldate = {2023-09-25},
  pubstate = {preprint},
  journaltitle = {}
}

@article{goldstein2023,
  title = {Generative {{Language Models}} and {{Automated Influence Operations}}: {{Emerging Threats}} and {{Potential Mitigations}}},
  shorttitle = {Generative {{Language Models}} and {{Automated Influence Operations}}},
  author = {Goldstein, Josh A. and Sastry, Girish and Musser, Micah and DiResta, Renee and Gentzel, Matthew and Sedova, Katerina},
  date = {2023-01-10},
  eprint = {2301.04246},
  eprinttype = {arxiv},
  eprintclass = {cs},
  doi = {10.48550/arXiv.2301.04246},
  urldate = {2023-09-25},
  pubstate = {preprint},
  journaltitle = {}
}

@online{banias2023,
  title = {Inside {{CounterCloud}}: {{A Fully Autonomous AI Disinformation System}}},
  author = {family=Banias, given=MJ, given-i=MJ},
  date = {2023-08-16},
  url = {https://thedebrief.org/countercloud-ai-disinformation/},
  urldate = {2023-09-25},
  organization = {{The Debrief}}
}

@online{bajohr2023,
  title = {Whoever {{Controls Language Models Controls Politics}}},
  author = {Bajohr, Hannes},
  date = {2023-04-08},
  url = {https://hannesbajohr.de/en/2023/04/08/whoever-controls-language-models-controls-politics/},
  urldate = {2023-09-25},
  langid = {american}
}

@article{almeida2022,
  title = {The Ethics of Facial Recognition Technologies, Surveillance, and Accountability in an Age of Artificial Intelligence: A Comparative Analysis of {{US}}, {{EU}}, and {{UK}} Regulatory Frameworks},
  shorttitle = {The Ethics of Facial Recognition Technologies, Surveillance, and Accountability in an Age of Artificial Intelligence},
  author = {Almeida, Denise and Shmarko, Konstantin and Lomas, Elizabeth},
  date = {2022-08},
  journaltitle = {AI and Ethics},
  shortjournal = {AI Ethics},
  volume = {2},
  number = {3},
  pages = {377--387},
  issn = {2730-5953, 2730-5961},
  doi = {10.1007/s43681-021-00077-w},
  urldate = {2023-09-25},
  langid = {english}
}

@article{kaklauskas2022,
  title = {A {{Review}} of {{AI Cloud}} and {{Edge Sensors}}, {{Methods}}, and {{Applications}} for the {{Recognition}} of {{Emotional}}, {{Affective}} and {{Physiological States}}},
  author = {Kaklauskas, Arturas and Abraham, Ajith and Ubarte, Ieva and Kliukas, Romualdas and Luksaite, Vaida and Binkyte-Veliene, Arune and Vetloviene, Ingrida and Kaklauskiene, Loreta},
  date = {2022-10-14},
  journaltitle = {Sensors},
  shortjournal = {Sensors},
  volume = {22},
  number = {20},
  pages = {7824},
  issn = {1424-8220},
  doi = {10.3390/s22207824},
  urldate = {2023-09-25},
  langid = {english}
}

@article{ferguson2017,
  title = {Policing Predictive Policing},
  author = {Ferguson, Andrew},
  date = {2017-01},
  journaltitle = {Washington University Law Review},
  volume = {94},
  number = {5},
  pages = {1109--1189}
}

@article{xu2021,
  title = {To {{Repress}} or to {{Co}}‐opt? {{Authoritarian Control}} in the {{Age}} of {{Digital Surveillance}}},
  shorttitle = {To {{Repress}} or to {{Co}}‐opt?},
  author = {Xu, Xu},
  date = {2021-04},
  journaltitle = {American Journal of Political Science},
  shortjournal = {American J Political Sci},
  volume = {65},
  number = {2},
  pages = {309--325},
  issn = {0092-5853, 1540-5907},
  doi = {10.1111/ajps.12514},
  urldate = {2023-09-25},
  langid = {english}
}

@article{kendall-taylor2020,
  entrysubtype = {magazine},
  title = {The {{Digital Dictators}}},
  author = {Kendall-Taylor, Andrea and Frantz, Erica and Wright, Joseph},
  date = {2020-02-06},
  journaltitle = {Foreign Affairs},
  volume = {99},
  number = {2},
  issn = {0015-7120},
  url = {https://www.foreignaffairs.com/articles/china/2020-02-06/digital-dictators},
  urldate = {2023-09-25},
  langid = {american}
}

@report{crawford2019,
  title = {{{AI}} Now 2019 Report},
  author = {Crawford, Kate and Dobbe, Roel and Dryer, Thoedora and Fried, Genevieve and Green, Ben and Kaziunas, Elizabeth and {kak}, Amba and Mathur, Varoon and McElroy, Erin and Nill Sánchez, Andrea and Raji, Deborah and Rankin, Lisi and Richardson, Rashida and Schultz, Jason and Myers West, Sarah and Whittaker, Meredith},
  date = {2019},
  institution = {{AI Now Institute}},
  location = {{New York}},
  url = {https://ainowinstitute.org/publication/ai-now-2019-report-2}
}

@report{feldstein2019,
  type = {Working Paper},
  title = {The Global Expansion of {{AI}} Surveillance},
  author = {Feldstein, Steven},
  date = {2019},
  institution = {{Carnegie Endowment for International Peace}},
  url = {https://carnegieendowment.org/2019/09/17/global-expansion-of-ai-surveillance-pub-79847}
}

@article{gupta2018,
  title = {The Evolution of Fraud: {{Ethical}} Implications in the Age of Large-Scale Data Breaches and Widespread Artificial Intelligence Solutions Deployment},
  author = {Gupta, Abhishek},
  date = {2018-02-02},
  journaltitle = {International Telecommunication Union Journal},
  volume = {1},
  url = {http://handle.itu.int/11.1002/pub/812a022b-en}
}

@article{hazell2023,
  title = {Large {{Language Models Can Be Used To Effectively Scale Spear Phishing Campaigns}}},
  author = {Hazell, Julian},
  date = {2023-05-12},
  eprint = {2305.06972},
  eprinttype = {arxiv},
  eprintclass = {cs},
  doi = {10.48550/arXiv.2305.06972},
  urldate = {2023-09-25},
  pubstate = {preprint},
  journaltitle = {}
}

@online{kelley2023,
  title = {{{WormGPT}} - {{The Generative AI Tool Cybercriminals Are Using}} to {{Launch BEC Attacks}}},
  author = {Kelley, Daniel},
  date = {2023-07-13T13:00:49+00:00},
  url = {https://slashnext.com/blog/wormgpt-the-generative-ai-tool-cybercriminals-are-using-to-launch-business-email-compromise-attacks/},
  urldate = {2023-09-25},
  organization = {{SlashNext}}
}

@inproceedings{horvitz2022a,
  title = {Artificial {{Intelligence}} and {{Cybersecurity}}: {{Rising Challenges}} and {{Promising Directions}}},
  author = {Horvitz, Eric},
  date = {2022-05-03},
  location = {{117th Congress}},
  url = {https://aka.ms/AAhee56},
  eventtitle = {Hearing on {{Artificial Intelligence Applications}} to {{Operations}} in {{Cyberspace}}}
}

@online{shimony23,
  title = {Chatting {{Our Way Into Creating}} a {{Polymorphic Malware}}},
  author = {Shimony, Eran and Tsarfati, Omer},
  date = {0023-01-17},
  url = {https://www.cyberark.com/resources/threat-research-blog/chatting-our-way-into-creating-a-polymorphic-malware},
  urldate = {2023-09-25},
  langid = {english},
  organization = {{CyberArk}}
}

@incollection{fritsch2022,
  title = {An {{Overview}} of {{Artificial Intelligence Used}} in {{Malware}}},
  booktitle = {Nordic {{Artificial Intelligence Research}} and {{Development}}},
  author = {Fritsch, Lothar and Jaber, Aws and Yazidi, Anis},
  editor = {Zouganeli, Evi and Yazidi, Anis and Mello, Gustavo and Lind, Pedro},
  date = {2022},
  volume = {1650},
  pages = {41--51},
  publisher = {{Springer International Publishing}},
  location = {{Cham}},
  doi = {10.1007/978-3-031-17030-0_4},
  urldate = {2023-09-25},
  langid = {english}
}

@online{stoecklin2018,
  title = {{{DeepLocker}}: {{How AI Can Power}} a {{Stealthy New Breed}} of {{Malware}}},
  author = {Stoecklin, Marc Ph. and Jang, Jiyong and Kirat, Dhilung},
  date = {2018-08-08},
  url = {https://securityintelligence.com/deeplocker-how-ai-can-power-a-stealthy-new-breed-of-malware/},
  urldate = {2023-09-25},
  organization = {{Security Intelligence}}
}

@inproceedings{li2019,
  title = {Dynamic {{Traffic Feature Camouflaging}} via {{Generative Adversarial Networks}}},
  booktitle = {2019 {{IEEE Conference}} on {{Communications}} and {{Network Security}} ({{CNS}})},
  author = {Li, Jie and Zhou, Lu and Li, Huaxin and Yan, Lu and Zhu, Haojin},
  date = {2019-06},
  pages = {268--276},
  publisher = {{IEEE}},
  location = {{Washington DC, DC, USA}},
  doi = {10.1109/CNS.2019.8802772},
  urldate = {2023-09-25},
  eventtitle = {2019 {{IEEE Conference}} on {{Communications}} and {{Network Security}} ({{CNS}})},
  isbn = {978-1-5386-7117-7}
}

@inproceedings{garcia2017,
  title = {Hey, {{My Malware Knows Physics}}! {{Attacking PLCs}} with {{Physical Model Aware Rootkit}}},
  booktitle = {Proceedings 2017 {{Network}} and {{Distributed System Security Symposium}}},
  author = {Garcia, Luis A. and Brasser, Ferdinand and Cintuglu, Mehmet H. and Sadeghi, Ahmad-Reza and Mohammed, Osama and Zonouz, Saman A.},
  date = {2017},
  publisher = {{Internet Society}},
  location = {{San Diego, CA}},
  doi = {10.14722/ndss.2017.23313},
  urldate = {2023-09-25},
  eventtitle = {Network and {{Distributed System Security Symposium}}},
  isbn = {978-1-891562-46-4},
  langid = {english}
}

@article{boiko2023,
  title = {Emergent Autonomous Scientific Research Capabilities of Large Language Models},
  author = {Boiko, Daniil A. and MacKnight, Robert and Gomes, Gabe},
  date = {2023-04-11},
  eprint = {2304.05332},
  eprinttype = {arxiv},
  eprintclass = {physics},
  doi = {10.48550/arXiv.2304.05332},
  urldate = {2023-09-25},
  pubstate = {preprint},
  journaltitle = {}
}

@article{bran2023,
  title = {{{ChemCrow}}: {{Augmenting}} Large-Language Models with Chemistry Tools},
  shorttitle = {{{ChemCrow}}},
  author = {Bran, Andres M. and Cox, Sam and White, Andrew D. and Schwaller, Philippe},
  date = {2023-06-21},
  eprint = {2304.05376},
  eprinttype = {arxiv},
  eprintclass = {physics, stat},
  doi = {10.48550/arXiv.2304.05376},
  urldate = {2023-09-25},
  pubstate = {preprint},
  journaltitle = {}
}

@article{soice2023,
  title = {Can Large Language Models Democratize Access to Dual-Use Biotechnology?},
  author = {Soice, Emily H. and Rocha, Rafael and Cordova, Kimberlee and Specter, Michael and Esvelt, Kevin M.},
  date = {2023-06-06},
  eprint = {2306.03809},
  eprinttype = {arxiv},
  eprintclass = {cs},
  doi = {10.48550/arXiv.2306.03809},
  urldate = {2023-09-25},
  pubstate = {preprint},
  journaltitle = {}
}

@unpublished{OpenAI_2023b,
  title = {{{GPT-4 System Card}}},
  author = {{OpenAI}},
  date = {2023-03-23},
  url = {https://cdn.openai.com/papers/gpt-4-system-card.pdf}
}

@article{gerrit2023,
  entrysubtype = {newspaper},
  title = {{{AI}} Leaders Warn {{Congress}} That {{AI}} Could Be Used to Create Bioweapons},
  author = {Gerrit, De Vynck},
  date = {2023-07-25},
  journaltitle = {Washington Post},
  url = {https://www.washingtonpost.com/technology/2023/07/25/ai-bengio-anthropic-senate-hearing/},
  urldate = {2023-09-25},
  langid = {american}
}

@misc{markey2023,
  title = {Text - {{S}}.2399 - 118th {{Congress}} (2023-2024): {{Artificial Intelligence}} and {{Biosecurity Risk Assessment Act}}},
  shorttitle = {Text - {{S}}.2399 - 118th {{Congress}} (2023-2024)},
  author = {Markey {[D-MA]}, Edward J. },
  date = {2023-07-19},
  url = {https://www.congress.gov/bill/118th-congress/senate-bill/2399/text},
  urldate = {2023-09-25},
  langid = {english}
}

@incollection{Maslej2023a,
  title = {Chapter 5: {{Education}}},
  booktitle = {The {{AI Index}} 2023 {{Annual Report}}},
  author = {Nestor Maslej and Loredana Fattorini and Erik Brynjolfsson and John Etchemendy and Katrina Ligett and Terah Lyons and James Manyika and Helen Ngo and Juan Carlos Niebles and Vanessa Parli and Yoav Shoham and Russell Wald and Jack Clark and Raymond Perrault},
  date = {2023-04},
  publisher = {{Institute for Human-Centered AI, Stanford University}},
  location = {{Stanford, CA}},
  url = {https://aiindex.stanford.edu/wp-content/uploads/2023/04/HAI_AI-Index-Report-2023_CHAPTER_5.pdf}
}

@article{touvron2023,
  title = {Llama 2: {{Open Foundation}} and {{Fine-Tuned Chat Models}}},
  shorttitle = {Llama 2},
  author = {Touvron, Hugo and Martin, Louis and Stone, Kevin and Albert, Peter and Almahairi, Amjad and Babaei, Yasmine and Bashlykov, Nikolay and Batra, Soumya and Bhargava, Prajjwal and Bhosale, Shruti and Bikel, Dan and Blecher, Lukas and Ferrer, Cristian Canton and Chen, Moya and Cucurull, Guillem and Esiobu, David and Fernandes, Jude and Fu, Jeremy and Fu, Wenyin and Fuller, Brian and Gao, Cynthia and Goswami, Vedanuj and Goyal, Naman and Hartshorn, Anthony and Hosseini, Saghar and Hou, Rui and Inan, Hakan and Kardas, Marcin and Kerkez, Viktor and Khabsa, Madian and Kloumann, Isabel and Korenev, Artem and Koura, Punit Singh and Lachaux, Marie-Anne and Lavril, Thibaut and Lee, Jenya and Liskovich, Diana and Lu, Yinghai and Mao, Yuning and Martinet, Xavier and Mihaylov, Todor and Mishra, Pushkar and Molybog, Igor and Nie, Yixin and Poulton, Andrew and Reizenstein, Jeremy and Rungta, Rashi and Saladi, Kalyan and Schelten, Alan and Silva, Ruan and Smith, Eric Michael and Subramanian, Ranjan and Tan, Xiaoqing Ellen and Tang, Binh and Taylor, Ross and Williams, Adina and Kuan, Jian Xiang and Xu, Puxin and Yan, Zheng and Zarov, Iliyan and Zhang, Yuchen and Fan, Angela and Kambadur, Melanie and Narang, Sharan and Rodriguez, Aurelien and Stojnic, Robert and Edunov, Sergey and Scialom, Thomas},
  date = {2023-07-19},
  eprint = {2307.09288},
  eprinttype = {arxiv},
  eprintclass = {cs},
  doi = {10.48550/arXiv.2307.09288},
  urldate = {2023-09-25},
  pubstate = {preprint},
  journaltitle = {}
}

@online{runpod2023,
  title = {{{GPU Instance Pricing}}},
  author = {RunPod},
  date = {2023},
  url = {https://www.runpod.io/gpu-instance/pricing},
  urldate = {2023-09-25}
}

@online{aman2023,
  title = {Why {{GPT-3}}.5 Is (Mostly) Cheaper than {{Llama}} 2},
  author = {Aman},
  date = {2023-07-20},
  url = {https://www.cursor.so/blog/llama-inference},
  urldate = {2023-09-25},
  langid = {english},
  organization = {{Cursor}}
}

@online{metaai2023b,
  title = {I-{{JEPA}}: {{The}} First {{AI}} Model Based on {{Yann LeCun}}’s Vision for More Human-like {{AI}}},
  shorttitle = {I-{{JEPA}}},
  author = {Meta AI},
  date = {2023-06-13},
  url = {https://ai.meta.com/blog/yann-lecun-ai-model-i-jepa/},
  urldate = {2023-09-25},
  langid = {english},
  organization = {{Meta AI}}
}

@article{hu2021,
  title = {{{LoRA}}: {{Low-Rank Adaptation}} of {{Large Language Models}}},
  shorttitle = {{{LoRA}}},
  author = {Hu, Edward J. and Shen, Yelong and Wallis, Phillip and Allen-Zhu, Zeyuan and Li, Yuanzhi and Wang, Shean and Wang, Lu and Chen, Weizhu},
  date = {2021-10-16},
  eprint = {2106.09685},
  eprinttype = {arxiv},
  eprintclass = {cs},
  doi = {10.48550/arXiv.2106.09685},
  urldate = {2023-09-25},
  pubstate = {preprint},
  journaltitle = {}
}

@online{hobbhahn2022,
  title = {Trends in {{GPU}} Price-Performance},
  author = {Hobbhahn, Marius},
  date = {2022-06-27T00:00:00+00:00},
  url = {https://epochai.org/blog/trends-in-gpu-price-performance},
  urldate = {2023-09-25},
  langid = {english},
  organization = {{EPOCH}}
}

@online{zellers2019,
  title = {Why {{We Released Grover}}},
  author = {Zellers, Rowan},
  date = {2019-07-15T23:19:03},
  url = {https://thegradient.pub/why-we-released-grover/},
  urldate = {2023-09-25},
  langid = {english},
  organization = {{The Gradient}}
}

@article{jervis1978,
  title = {Cooperation under the {{Security Dilemma}}},
  author = {Jervis, Robert},
  date = {1978-01},
  journaltitle = {World Politics},
  shortjournal = {World Pol.},
  volume = {30},
  number = {2},
  pages = {167--214},
  doi = {10.2307/2009958},
  urldate = {2023-09-25},
  langid = {english}
}

@article{garfinkel2019,
  title = {How Does the Offense-Defense Balance Scale?},
  author = {Garfinkel, Ben and Dafoe, Allan},
  date = {2019-09-19},
  journaltitle = {Journal of Strategic Studies},
  shortjournal = {Journal of Strategic Studies},
  volume = {42},
  number = {6},
  pages = {736--763},
  doi = {10.1080/01402390.2019.1631810},
  urldate = {2023-09-25},
  langid = {english}
}

@article{ferrara2023,
  title = {Should {{ChatGPT}} Be {{Biased}}? {{Challenges}} and {{Risks}} of {{Bias}} in {{Large Language Models}}},
  shorttitle = {Should {{ChatGPT}} Be {{Biased}}?},
  author = {Ferrara, Emilio},
  date = {2023-04-18},
  eprint = {2304.03738},
  eprinttype = {arxiv},
  eprintclass = {cs},
  doi = {10.48550/arXiv.2304.03738},
  urldate = {2023-09-25},
  pubstate = {preprint},
  journaltitle = {}
}

@inproceedings{kassab2022,
  title = {Investigating {{Bugs}} in {{AI-Infused Systems}}: {{Analysis}} and {{Proposed Taxonomy}}},
  shorttitle = {Investigating {{Bugs}} in {{AI-Infused Systems}}},
  booktitle = {2022 {{IEEE International Symposium}} on {{Software Reliability Engineering Workshops}} ({{ISSREW}})},
  author = {Kassab, Mohamad and DeFranco, Joanna and Laplante, Phillip},
  date = {2022-10},
  pages = {365--370},
  publisher = {{IEEE}},
  location = {{Charlotte, NC, USA}},
  doi = {10.1109/ISSREW55968.2022.00094},
  urldate = {2023-09-25},
  eventtitle = {2022 {{IEEE International Symposium}} on {{Software Reliability Engineering Workshops}} ({{ISSREW}})},
  isbn = {978-1-66547-679-9}
}

@online{wiggers2023,
  title = {What Is {{Auto-GPT}} and Why Does It Matter? | {{TechCrunch}}},
  author = {Wiggers, Kyle},
  date = {2023-04-22},
  url = {https://techcrunch.com/2023/04/22/what-is-auto-gpt-and-why-does-it-matter/?guccounter=1},
  urldate = {2023-09-25},
  organization = {{TechCrunch}}
}

@online{auto-gpt2023,
  title = {Home},
  author = {{Auto-GPT}},
  date = {2023},
  url = {https://news.agpt.co/},
  urldate = {2023-09-25},
  langid = {american},
  organization = {{The Official Auto-GPT Website}}
}

@article{bagdasaryan2023,
  title = {({{Ab}})Using {{Images}} and {{Sounds}} for {{Indirect Instruction Injection}} in {{Multi-Modal LLMs}}},
  author = {Bagdasaryan, Eugene and Hsieh, Tsung-Yin and Nassi, Ben and Shmatikov, Vitaly},
  date = {2023-07-24},
  eprint = {2307.10490},
  eprinttype = {arxiv},
  eprintclass = {cs},
  doi = {10.48550/arXiv.2307.10490},
  urldate = {2023-09-25},
  pubstate = {preprint},
  journaltitle = {}
}

@online{openaia,
  title = {Welcome to the {{OpenAI}} Platform},
  author = {OpenAI},
  url = {https://platform.openai.com},
  urldate = {2023-09-25},
  langid = {english}
}

@article{ponta2020,
  title = {Detection, Assessment and Mitigation of Vulnerabilities in Open Source Dependencies},
  author = {Ponta, Serena Elisa and Plate, Henrik and Sabetta, Antonino},
  date = {2020-09},
  journaltitle = {Empirical Software Engineering},
  shortjournal = {Empir Software Eng},
  volume = {25},
  number = {5},
  pages = {3175--3215},
  issn = {1382-3256, 1573-7616},
  doi = {10.1007/s10664-020-09830-x},
  urldate = {2023-09-25},
  langid = {english}
}

@online{synopsyseditorialteam2023,
  title = {2023 {{OSSRA}}: {{A}} Deep Dive into Open Source Trends},
  author = {{Synopsys Editorial Team}},
  date = {2023-02-21},
  url = {https://www.synopsys.com/blogs/software-security/open-source-trends-ossra-report.html},
  urldate = {2023-09-25},
  langid = {english},
  organization = {{Synopsys}}
}

@article{whittlestone2020,
  title = {The Tension between Openness and Prudence in {{AI}} Research},
  author = {Whittlestone, Jess and Ovadya, Aviv},
  date = {2020-01-13},
  eprint = {1910.01170},
  eprinttype = {arxiv},
  eprintclass = {cs},
  doi = {10.48550/arXiv.1910.01170},
  urldate = {2023-09-25},
  pubstate = {preprint},
  journaltitle = {}
}

@online{bugcrowd,
  title = {{{OpenAI}}},
  author = {Bugcrowd},
  url = {https://bugcrowd.com/openai},
  urldate = {2023-09-25},
  langid = {english}
}

@article{bowman2023,
  title = {Eight {{Things}} to {{Know}} about {{Large Language Models}}},
  author = {Bowman, Samuel R.},
  date = {2023-04-02},
  eprint = {2304.00612},
  eprinttype = {arxiv},
  eprintclass = {cs},
  doi = {10.48550/arXiv.2304.00612},
  urldate = {2023-09-25},
  pubstate = {preprint},
  journaltitle = {}
}

@article{solaiman2019,
  title = {Release {{Strategies}} and the {{Social Impacts}} of {{Language Models}}},
  author = {Solaiman, Irene and Brundage, Miles and Clark, Jack and Askell, Amanda and Herbert-Voss, Ariel and Wu, Jeff and Radford, Alec and Krueger, Gretchen and Kim, Jong Wook and Kreps, Sarah and McCain, Miles and Newhouse, Alex and Blazakis, Jason and McGuffie, Kris and Wang, Jasmine},
  date = {2019-11-12},
  eprint = {1908.09203},
  eprinttype = {arxiv},
  eprintclass = {cs},
  doi = {10.48550/arXiv.1908.09203},
  urldate = {2023-09-25},
  pubstate = {preprint},
  journaltitle = {}
}

@thesis{shevlane2022a,
  type = {phdthesis},
  title = {The {{Artefacts}} of {{Intelligence}}: {{Governing Scientists}}' {{Contribution}} to {{AI Proliferation}}},
  author = {Shevlane, Toby},
  date = {2022-04-22},
  institution = {{University of Oxford}},
  url = {https://cdn.governance.ai/Shevlane,_Artefacts_of_Intelligence.pdf},
  pagetotal = {278}
}

@article{brundage2020,
  title = {Toward {{Trustworthy AI Development}}: {{Mechanisms}} for {{Supporting Verifiable Claims}}},
  shorttitle = {Toward {{Trustworthy AI Development}}},
  author = {Brundage, Miles and Avin, Shahar and Wang, Jasmine and Belfield, Haydn and Krueger, Gretchen and Hadfield, Gillian and Khlaaf, Heidy and Yang, Jingying and Toner, Helen and Fong, Ruth and Maharaj, Tegan and Koh, Pang Wei and Hooker, Sara and Leung, Jade and Trask, Andrew and Bluemke, Emma and Lebensold, Jonathan and O'Keefe, Cullen and Koren, Mark and Ryffel, Théo and Rubinovitz, J. B. and Besiroglu, Tamay and Carugati, Federica and Clark, Jack and Eckersley, Peter and family=Haas, given=Sarah, prefix=de, useprefix=true and Johnson, Maritza and Laurie, Ben and Ingerman, Alex and Krawczuk, Igor and Askell, Amanda and Cammarota, Rosario and Lohn, Andrew and Krueger, David and Stix, Charlotte and Henderson, Peter and Graham, Logan and Prunkl, Carina and Martin, Bianca and Seger, Elizabeth and Zilberman, Noa and {hÉigeartaigh}, Seán Ó and Kroeger, Frens and Sastry, Girish and Kagan, Rebecca and Weller, Adrian and Tse, Brian and Barnes, Elizabeth and Dafoe, Allan and Scharre, Paul and Herbert-Voss, Ariel and Rasser, Martijn and Sodhani, Shagun and Flynn, Carrick and Gilbert, Thomas Krendl and Dyer, Lisa and Khan, Saif and Bengio, Yoshua and Anderljung, Markus},
  date = {2020-04-20},
  eprint = {2004.07213},
  eprinttype = {arxiv},
  eprintclass = {cs},
  doi = {10.48550/arXiv.2004.07213},
  urldate = {2023-09-25},
  pubstate = {preprint},
  journaltitle = {}
}

@article{raji2020,
  title = {Closing the {{AI Accountability Gap}}: {{Defining}} an {{End-to-End Framework}} for {{Internal Algorithmic Auditing}}},
  shorttitle = {Closing the {{AI Accountability Gap}}},
  author = {Raji, Inioluwa Deborah and Smart, Andrew and White, Rebecca N. and Mitchell, Margaret and Gebru, Timnit and Hutchinson, Ben and Smith-Loud, Jamila and Theron, Daniel and Barnes, Parker},
  date = {2020-01-03},
  eprint = {2001.00973},
  eprinttype = {arxiv},
  eprintclass = {cs},
  doi = {10.48550/arXiv.2001.00973},
  urldate = {2023-09-25},
  pubstate = {preprint},
  journaltitle = {}
}

@article{mokander2023,
  title = {Auditing Large Language Models: A Three-Layered Approach},
  shorttitle = {Auditing Large Language Models},
  author = {Mökander, Jakob and Schuett, Jonas and Kirk, Hannah Rose and Floridi, Luciano},
  date = {2023-05-30},
  journaltitle = {AI and Ethics},
  shortjournal = {AI Ethics},
  issn = {2730-5953, 2730-5961},
  doi = {10.1007/s43681-023-00289-2},
  urldate = {2023-09-25},
  langid = {english}
}

@article{khlaaf2022,
  title = {A {{Hazard Analysis Framework}} for {{Code Synthesis Large Language Models}}},
  author = {Khlaaf, Heidy and Mishkin, Pamela and Achiam, Joshua and Krueger, Gretchen and Brundage, Miles},
  date = {2022-07-25},
  eprint = {2207.14157},
  eprinttype = {arxiv},
  eprintclass = {cs},
  doi = {10.48550/arXiv.2207.14157},
  urldate = {2023-09-25},
  pubstate = {preprint},
  journaltitle = {}
}

@online{arcevals2023,
  title = {Update on {{ARC}}'s Recent Eval Efforts: More information about ARC's evaluations of GPT-4 and Claude},
  author = {{ARC Evals}},
  date = {2023-03-17},
  url = {https://evals.alignment.org/blog/2023-03-18-update-on-recent-evals/},
  urldate = {2023-09-25}
}

@unpublished{bucknallForthcoming,
  title = {Structured {{Access}} for {{Third-Party Safety Research}} on {{Frontier AI Models Investigating}} Researchers’ Model Access Requirements},
  author = {Bucknall, Ben and Trager, Robert and Shevlane, Toby},
  howpublished = {Working Paper},
  pubstate = {forthcoming},
  sortyear = 2023
}

@online{openai2022,
  title = {{{DALL}}·{{E}} 2 {{Preview}} - {{Risks}} and {{Limitations}}},
  author = {OpenAI},
  date = {2022},
  url = {https://github.com/openai/dalle-2-preview/blob/main/system-card.md},
  urldate = {2023-09-25},
  langid = {english},
  organization = {{GitHub}}
}

@article{murgia2023,
  entrysubtype = {newspaper},
  title = {{{OpenAI}}’s Red Team: The Experts Hired to ‘Break’ {{ChatGPT}}},
  shorttitle = {{{OpenAI}}’s Red Team},
  author = {Murgia, Madhumita},
  date = {2023-04-14},
  journaltitle = {Financial Times}
}

@article{ganguli2022,
  title = {Red {{Teaming Language Models}} to {{Reduce Harms}}: {{Methods}}, {{Scaling Behaviors}}, and {{Lessons Learned}}},
  shorttitle = {Red {{Teaming Language Models}} to {{Reduce Harms}}},
  author = {Ganguli, Deep and Lovitt, Liane and Kernion, Jackson and Askell, Amanda and Bai, Yuntao and Kadavath, Saurav and Mann, Ben and Perez, Ethan and Schiefer, Nicholas and Ndousse, Kamal and Jones, Andy and Bowman, Sam and Chen, Anna and Conerly, Tom and DasSarma, Nova and Drain, Dawn and Elhage, Nelson and El-Showk, Sheer and Fort, Stanislav and Hatfield-Dodds, Zac and Henighan, Tom and Hernandez, Danny and Hume, Tristan and Jacobson, Josh and Johnston, Scott and Kravec, Shauna and Olsson, Catherine and Ringer, Sam and Tran-Johnson, Eli and Amodei, Dario and Brown, Tom and Joseph, Nicholas and McCandlish, Sam and Olah, Chris and Kaplan, Jared and Clark, Jack},
  date = {2022-11-22},
  eprint = {2209.07858},
  eprinttype = {arxiv},
  eprintclass = {cs},
  doi = {10.48550/arXiv.2209.07858},
  urldate = {2023-09-25},
  pubstate = {preprint},
  journaltitle = {}
}

@inproceedings{costanza-chock2022,
  title = {Who {{Audits}} the {{Auditors}}? {{Recommendations}} from a Field Scan of the Algorithmic Auditing Ecosystem},
  shorttitle = {Who {{Audits}} the {{Auditors}}?},
  booktitle = {2022 {{ACM Conference}} on {{Fairness}}, {{Accountability}}, and {{Transparency}}},
  author = {Costanza-Chock, Sasha and Raji, Inioluwa Deborah and Buolamwini, Joy},
  date = {2022-06-21},
  pages = {1571--1583},
  publisher = {{ACM}},
  location = {{Seoul Republic of Korea}},
  doi = {10.1145/3531146.3533213},
  urldate = {2023-09-25},
  eventtitle = {{{FAccT}} '22: 2022 {{ACM Conference}} on {{Fairness}}, {{Accountability}}, and {{Transparency}}},
  isbn = {978-1-4503-9352-2},
  langid = {english}
}

@report{centrefordataethicsandinnovation2021,
  type = {Independent report},
  title = {The Roadmap to an Effective {{AI}} Assurance Ecosystem},
  author = {{Centre for Data Ethics {and} Innovation}},
  date = {2021-12-08},
  institution = {{Centre for Data Ethics and Innovation}},
  url = {https://www.gov.uk/government/publications/the-roadmap-to-an-effective-ai-assurance-ecosystem},
  urldate = {2023-09-25},
  langid = {english}
}

@article{perez2022,
  title = {Red {{Teaming Language Models}} with {{Language Models}}},
  author = {Perez, Ethan and Huang, Saffron and Song, Francis and Cai, Trevor and Ring, Roman and Aslanides, John and Glaese, Amelia and McAleese, Nat and Irving, Geoffrey},
  date = {2022-02-07},
  eprint = {2202.03286},
  eprinttype = {arxiv},
  eprintclass = {cs},
  doi = {10.48550/arXiv.2202.03286},
  urldate = {2023-09-25},
  pubstate = {preprint},
  journaltitle = {}
}

@report{levermore2023,
  title = {{{AI Safety Bounties}}},
  author = {Levermore, Patrick},
  date = {2023-08-10},
  institution = {{Rethink Priorities}},
  url = {https://rethinkpriorities.org/publications/ai-safety-bounties},
  urldate = {2023-09-25},
  langid = {canadian}
}

@misc{openai2022a,
  title = {{{ChatGPT Feedback Contest}}: {{Official Rules}}},
  author = {{OpenAI}},
  date = {2022},
  url = {https://cdn.openai.com/chatgpt/ChatGPT_Feedback_Contest_Rules.pdf},
  langid = {english}
}

@online{hackerone2022,
  title = {Hacker-{{Powered Security Report}}},
  author = {{hackerone}},
  date = {2022},
  url = {https://www.hackerone.com/resources/i/1487910-2022-hacker-powered-security-report-q4fy23/3?}
}

@inproceedings{zhao2015,
  title = {An {{Empirical Study}} of {{Web Vulnerability Discovery Ecosystems}}},
  booktitle = {Proceedings of the 22nd {{ACM SIGSAC Conference}} on {{Computer}} and {{Communications Security}}},
  author = {Zhao, Mingyi and Grossklags, Jens and Liu, Peng},
  date = {2015-10-12},
  pages = {1105--1117},
  publisher = {{ACM}},
  location = {{Denver Colorado USA}},
  doi = {10.1145/2810103.2813704},
  urldate = {2023-09-25},
  eventtitle = {{{CCS}}'15: {{The}} 22nd {{ACM Conference}} on {{Computer}} and {{Communications Security}}},
  isbn = {978-1-4503-3832-5},
  langid = {english}
}

@incollection{dardaman2023,
  title = {When Openness Fails: {{Towards}} a More Robust Governance Framework for Generative {{AI}}},
  booktitle = {Proceedings of the Sixth {{AAIA}}/{{ACM}} Conference on Artificial Intelligence, Ethics, and Society.},
  author = {Dardaman, Emily and Gupta, Abhishek},
  date = {2023},
  location = {{Montreal, Ontario, Canada}}
}

@online{teamnuggets2018,
  title = {Why {{Linux}} Runs 90 Percent of the Public Cloud Workload},
  author = {{Team Nuggets}},
  date = {2018-08-10},
  url = {https://www.cbtnuggets.com/blog/certifications/open-source/why-linux-runs-90-percent-of-the-public-cloud-workload},
  urldate = {2023-09-25},
  langid = {english},
  organization = {{CBT Nuggets}}
}

@online{engler2022a,
  title = {To {{Regulate General Purpose AI}}, {{Make}} the {{Model Move}}},
  author = {Engler, Alex},
  date = {2022-11-10T13:48:53+00:00},
  url = {https://techpolicy.press/to-regulate-general-purpose-ai-make-the-model-move/},
  urldate = {2023-09-25},
  langid = {american},
  organization = {{Tech Policy Press}}
}

@article{dettmers2023,
  title = {{{QLoRA}}: {{Efficient Finetuning}} of {{Quantized LLMs}}},
  shorttitle = {{{QLoRA}}},
  author = {Dettmers, Tim and Pagnoni, Artidoro and Holtzman, Ari and Zettlemoyer, Luke},
  date = {2023-05-23},
  eprint = {2305.14314},
  eprinttype = {arxiv},
  eprintclass = {cs},
  doi = {10.48550/arXiv.2305.14314},
  urldate = {2023-09-25},
  pubstate = {preprint},
  journaltitle = {}
}

@article{gudibande2023,
  title = {The {{False Promise}} of {{Imitating Proprietary LLMs}}},
  author = {Gudibande, Arnav and Wallace, Eric and Snell, Charlie and Geng, Xinyang and Liu, Hao and Abbeel, Pieter and Levine, Sergey and Song, Dawn},
  date = {2023-05-25},
  eprint = {2305.15717},
  eprinttype = {arxiv},
  eprintclass = {cs},
  doi = {10.48550/arXiv.2305.15717},
  urldate = {2023-09-25},
  pubstate = {preprint},
  journaltitle = {}
}

@report{centerforsecurityandemergingtechnology2021,
  title = {Key {{Concepts}} in {{AI Safety}}: {{An Overview}}},
  shorttitle = {Key {{Concepts}} in {{AI Safety}}},
  author = {{Center for Security and Emerging Technology} and Rudner, Tim and Toner, Helen},
  date = {2021-03},
  institution = {{Center for Security and Emerging Technology}},
  doi = {10.51593/20190040},
  urldate = {2023-09-25}
}

@article{hendrycks2022,
  title = {Unsolved {{Problems}} in {{ML Safety}}},
  author = {Hendrycks, Dan and Carlini, Nicholas and Schulman, John and Steinhardt, Jacob},
  date = {2022-06-16},
  eprint = {2109.13916},
  eprinttype = {arxiv},
  eprintclass = {cs},
  doi = {10.48550/arXiv.2109.13916},
  urldate = {2023-09-25},
  pubstate = {preprint},
  journaltitle = {}
}

@article{villalobos2022,
  title = {Will We Run out of Data? {{An}} Analysis of the Limits of Scaling Datasets in {{Machine Learning}}},
  shorttitle = {Will We Run out of Data?},
  author = {Villalobos, Pablo and Sevilla, Jaime and Heim, Lennart and Besiroglu, Tamay and Hobbhahn, Marius and Ho, Anson},
  date = {2022-10-25},
  eprint = {2211.04325},
  eprinttype = {arxiv},
  eprintclass = {cs},
  doi = {10.48550/arXiv.2211.04325},
  urldate = {2023-09-25},
  pubstate = {preprint},
  journaltitle = {}
}

@online{macropolo2023,
  title = {The {{Global AI Talent Tracker}}},
  author = {MacroPolo},
  date = {2023},
  url = {https://macropolo.org/digital-projects/the-global-ai-talent-tracker/},
  urldate = {2023-09-25},
  langid = {english},
  organization = {{MacroPolo}}
}

@article{wei2023,
  title = {Larger Language Models Do In-Context Learning Differently},
  author = {Wei, Jerry and Wei, Jason and Tay, Yi and Tran, Dustin and Webson, Albert and Lu, Yifeng and Chen, Xinyun and Liu, Hanxiao and Huang, Da and Zhou, Denny and Ma, Tengyu},
  date = {2023-03-08},
  eprint = {2303.03846},
  eprinttype = {arxiv},
  eprintclass = {cs},
  doi = {10.48550/arXiv.2303.03846},
  urldate = {2023-09-25},
  pubstate = {preprint},
  journaltitle = {}
}

@online{laion.ai2023,
  title = {Petition for Keeping up the Progress Tempo on {{AI}} Research While Securing Its Transparency and Safety},
  author = {LAION.ai},
  date = {2023-03-29},
  url = {https://laion.ai/blog/petition},
  urldate = {2023-09-25},
  langid = {english},
  organization = {{LAION}}
}

@online{jeffries2023,
  type = {Substack newsletter},
  title = {Let's {{Speed Up AI}}},
  author = {Jeffries, Daniel},
  date = {2023-02-04},
  url = {https://danieljeffries.substack.com/p/lets-speed-up-ai},
  urldate = {2023-09-25},
  organization = {{Future History}}
}

@online{grace2022,
  title = {Let’s Think about Slowing down {{AI}}},
  author = {Grace, Katja},
  date = {2022-12-22},
  url = {https://www.lesswrong.com/posts/uFNgRumrDTpBfQGrs/let-s-think-about-slowing-down-ai},
  urldate = {2023-09-25},
  langid = {english},
  organization = {{LESSWRONG}}
}

@online{futureoflifeinstitute2023,
  title = {Pause {{Giant AI Experiments}}: {{An Open Letter}}},
  shorttitle = {Pause {{Giant AI Experiments}}},
  author = {{Future of Life Institute}},
  date = {2023-03-22},
  url = {https://futureoflife.org/open-letter/pause-giant-ai-experiments/},
  urldate = {2023-09-25},
  langid = {american}
}

@online{openaib,
  title = {Chat {{Plugins}}},
  author = {OpenAI},
  url = {https://platform.openai.com/docs/plugins/introduction},
  urldate = {2023-09-25},
  langid = {english}
}

@article{schuett2023,
  title = {Towards Best Practices in {{AGI}} Safety and Governance: {{A}} Survey of Expert Opinion},
  shorttitle = {Towards Best Practices in {{AGI}} Safety and Governance},
  author = {Schuett, Jonas and Dreksler, Noemi and Anderljung, Markus and McCaffary, David and Heim, Lennart and Bluemke, Emma and Garfinkel, Ben},
  date = {2023-05-11},
  eprint = {2305.07153},
  eprinttype = {arxiv},
  eprintclass = {cs},
  doi = {10.48550/arXiv.2305.07153},
  urldate = {2023-09-25},
  pubstate = {preprint},
  journaltitle = {}
}

@article{yu2022,
  title = {Responsible {{Disclosure}} of {{Generative Models Using Scalable Fingerprinting}}},
  author = {Yu, Ning and Skripniuk, Vladislav and Chen, Dingfan and Davis, Larry and Fritz, Mario},
  date = {2022-03-17},
  eprint = {2012.08726},
  eprinttype = {arxiv},
  eprintclass = {cs},
  doi = {10.48550/arXiv.2012.08726},
  urldate = {2023-09-25},
  pubstate = {preprint},
  journaltitle = {}
}

@article{wagner2023,
  title = {Independence by Permission},
  author = {Wagner, Michael W.},
  date = {2023-07-28},
  journaltitle = {Science},
  shortjournal = {Science},
  volume = {381},
  number = {6656},
  pages = {388--391},
  issn = {0036-8075, 1095-9203},
  doi = {10.1126/science.adi2430},
  urldate = {2023-09-25},
  langid = {english}
}

@article{ho2023,
  title = {International {{Institutions}} for {{Advanced AI}}},
  author = {Ho, Lewis and Barnhart, Joslyn and Trager, Robert and Bengio, Yoshua and Brundage, Miles and Carnegie, Allison and Chowdhury, Rumman and Dafoe, Allan and Hadfield, Gillian and Levi, Margaret and Snidal, Duncan},
  date = {2023-07-11},
  eprint = {2307.04699},
  eprinttype = {arxiv},
  eprintclass = {cs},
  doi = {10.48550/arXiv.2307.04699},
  urldate = {2023-09-25},
  pubstate = {preprint},
  journaltitle = {}
}

@article{marcus2023,
  entrysubtype = {magazine},
  title = {The World Needs an International Agency for Artificial Intelligence, Say Two {{AI}} Experts},
  author = {Marcus, Gary and Reuel, Anka},
  date = {2023-04-18},
  journaltitle = {The Economist},
  issn = {0013-0613},
  url = {https://www.economist.com/by-invitation/2023/04/18/the-world-needs-an-international-agency-for-artificial-intelligence-say-two-ai-experts},
  urldate = {2023-09-25}
}

@online{howard2023,
  title = {{{AI Safety}} and the {{Age}} of {{Dislightenment}}: {{Model}} Licensing \& Surveillance Will Likely Be Counterproductive by Concentrating Power in Unsustainable Ways},
  author = {Howard, Jeremy},
  date = {2023-07-10},
  url = {https://www.fast.ai/posts/2023-11-07-dislightenment.html},
  urldate = {2023-09-26},
  langid = {english},
  organization = {{fast.ai}}
}

@online{laion.ai2023a,
  title = {A {{Call}} to {{Protect Open-Source AI}} in {{Europe}}},
  author = {LAION.ai},
  date = {2023-04-28},
  url = {https://laion.ai/notes/letter-to-the-eu-parliament},
  urldate = {2023-09-26},
  langid = {english},
  organization = {{LAION}}
}

@online{scalevirtualevents2022,
  title = {Emad {{Mostaque}} ({{Stability AI}}): {{Democratizing AI}}, {{Stable Diffusion}} \& {{Generative Models}}},
  author = {{Scale Virtual Events}},
  date = {2022-10-23},
  url = {https://exchange.scale.com/public/videos/emad-mostaque-stability-ai-stable-diffusion-open-source},
  urldate = {2023-09-26}
}

@inproceedings{seger2023,
  title = {Democratising {{AI}}: {{Multiple Meanings}}, {{Goals}}, and {{Methods}}},
  shorttitle = {Democratising {{AI}}},
  booktitle = {Proceedings of the 2023 {{AAAI}}/{{ACM Conference}} on {{AI}}, {{Ethics}}, and {{Society}}},
  author = {Seger, Elizabeth and Ovadya, Aviv and Siddarth, Divya and Garfinkel, Ben and Dafoe, Allan},
  date = {2023-08-08},
  pages = {715--722},
  publisher = {{ACM}},
  location = {{Montr\'eal QC Canada}},
  doi = {10.1145/3600211.3604693},
  urldate = {2023-09-26},
  eventtitle = {{{AIES}} '23: {{AAAI}}/{{ACM Conference}} on {{AI}}, {{Ethics}}, and {{Society}}},
  langid = {english}
}

@online{patel2023,
  title = {Google "{{We Have No Moat}}, {{And Neither Does OpenAI}}": {{Leaked Internal Google Document Claims Open Source AI Will Outcompete Google}} and {{OpenAI}}},
  author = {Patel, Dylan and Ahmad, Afzal},
  date = {2023-05-04},
  url = {https://www.semianalysis.com/p/google-we-have-no-moat-and-neither},
  urldate = {2023-09-26},
  langid = {english},
  organization = {{SemiAnalysis}}
}

@incollection{maslej2023,
  title = {Chapter 7: {{Diversity}}},
  booktitle = {The {{AI Index}} 2023 {{Annual Report}}},
  author = {Maslej, Nestor and Fattorini, Loredana and Brynjolfsson, Erik and Etchemendy, John and Ligett, Katrina and Lyons, Terah and Manyika, James and Ngo, Helen and Niebles, Juan Carlos and Parli, Vanessa and Shoham, Yoav and Wald, Russell and Clark, Jack and Perrault, Raymond},
  date = {2023-04},
  publisher = {{Institute for Human-Centered AI, Stanford University}},
  location = {{Stanford, CA}},
  url = {https://aiindex.stanford.edu/wp-content/uploads/2023/04/HAI_AI-Index-Report-2023_CHAPTER_7.pdf}
}

@online{eleutherai2023,
  title = {{{EleutherAI}} Is a Non-Profit {{AI}} Research Lab That Focuses on Interpretability and Alignment of Large Models},
  author = {EleutherAI},
  date = {2023},
  url = {https://www.eleuther.ai/about},
  urldate = {2023-09-26},
  langid = {british}
}

@online{bigscience,
  title = {A One-Year Long Research Workshop on Large Multilingual Models and Datasets},
  author = {BigScience},
  url = {https://bigscience.huggingface.co/},
  urldate = {2023-09-26}
}

@online{kayid2022,
  title = {Bonjour. {\<مرحبا>}. {{Guten}} Tag. {{Hola}}. {{Cohere}}'s {{Multilingual Text Understanding Model}} Is {{Now Available}}},
  author = {Kayid, Amr and Reimers, Nils},
  date = {2022-12-12T15:00:00},
  url = {https://txt.cohere.com/multilingual/},
  urldate = {2023-09-26},
  langid = {english},
  organization = {{Cohere}}
}

@online{beaumont2022,
  title = {Large {{Scale Openclip}}: {{L}}/14, {{H}}/14 and {{G}}/14 Trained on {{LAION-2B}}},
  shorttitle = {Large Scale {{openCLIP}}},
  author = {Beaumont, Romain},
  date = {2022-09-15},
  url = {https://laion.ai/blog/large-openclip},
  urldate = {2023-09-26},
  langid = {english},
  organization = {{LAION}}
}

@software{ilharco2021,
  title = {{{OpenCLIP}}},
  author = {Ilharco, Gabriel and Wortsman, Mitchell and Carlini, Nicholas and Taori, Rohan and Dave, Achal and Shankar, Vaishaal and Namkoong, Hongseok and Miller, John and Hajishirzi, Hannaneh and Farhadi, Ali and Schmidt, Ludwig},
  date = {2021-07-28},
  doi = {10.5281/ZENODO.5143773},
  urldate = {2023-09-26},
  organization = {{Zenodo}},
  version = {0.1}
}

@online{altman2021,
  title = {Moore's {{Law}} for {{Everything}}},
  author = {Altman, Sam},
  date = {2021-03-16},
  url = {https://moores.samaltman.com/},
  urldate = {2023-09-26}
}

@online{miller2021,
  title = {Radical {{Proposal}}: {{Universal Basic Income}} to {{Offset Job Losses Due}} to {{Automation}}},
  shorttitle = {Radical {{Proposal}}},
  author = {Miller, Katharine},
  date = {2021-10-20},
  url = {https://hai.stanford.edu/news/radical-proposal-universal-basic-income-offset-job-losses-due-automation},
  urldate = {2023-09-26},
  langid = {english},
  organization = {{Stanford HAI}}
}

@report{okeefe2020,
  title = {The Windfall Clause: {{Distributing}} the Benefits of {{AI}}},
  author = {O’Keefe, Cullen and Cihon, Peter and Garfinkel, Ben and Flynn, Carrick and Leung, Jade and Dafoe, Allan},
  date = {2020},
  institution = {{Centre for the Governance of AI Research Report. Future of Humanity Institute, University of Oxford}},
  url = {https://www.fhi.ox.ac.uk/wp-content/uploads/Windfall-Clause-Report.pdf},
  langid = {british}
}

@online{bigcode2020,
  title = {Datasets},
  author = {BigCode},
  date = {2020-11-16T13:59:39+01:00},
  url = {https://www.bigcode-project.org/docs/about/the-stack/},
  urldate = {2023-09-26},
  langid = {american},
  organization = {{BigCode}}
}

@online{vincent2022,
  title = {The Scary Truth about {{AI}} Copyright Is Nobody Knows What Will Happen Next},
  author = {Vincent, James},
  date = {2022-11-15T15:00:00},
  url = {https://www.theverge.com/23444685/generative-ai-copyright-infringement-legal-fair-use-training-data},
  urldate = {2023-09-26},
  langid = {american},
  organization = {{The Verge}}
}

@online{polis2023,
  title = {Input {{Crowd}}, {{Output Meaning}}},
  author = {Polis},
  date = {2023},
  url = {https://pol.is/home},
  urldate = {2023-09-26}
}

@online{coy2023,
  title = {Can {{A}}.{{I}}. and {{Democracy Fix Each Other}}?},
  author = {Coy, Peter},
  date = {2023-04-05},
  url = {https://www.nytimes.com/2023/04/05/opinion/artificial-intelligence-democracy-chatgpt.html},
  urldate = {2023-09-26},
  organization = {{The New York Times}}
}

@online{thecollectiveintelligenceproject2023a,
  title = {Alignment {{Assemblies}}},
  author = {{The Collective Intelligence Project}},
  date = {2023},
  url = {https://cip.org/alignmentassemblies},
  urldate = {2023-09-26},
  langid = {canadian},
  organization = {{The Collective Intelligence Project}}
}

@online{costa2022,
  title = {Deliberative Democracy in Action: {{A}} Closer Look at Our Recent Pilot with {{Meta}}},
  author = {Costa, Elisabeth},
  date = {2022-11-07},
  url = {https://www.bi.team/blogs/deliberative-democracy-in-action/},
  urldate = {2023-09-26},
  langid = {british},
  organization = {{The Behavioural Insights Team}}
}

@online{ovadya2023,
  title = {Meta {{Ran}} a {{Giant Experiment}} in {{Governance}}. {{Now It}}'s {{Turning}} to {{AI}}},
  author = {Ovadya, Aviv},
  date = {2023-07-10},
  url = {https://www.wired.com/story/meta-ran-a-giant-experiment-in-governance-now-its-turning-to-ai/},
  urldate = {2023-09-26},
  organization = {{WIRED}}
}

@online{harris2022,
  title = {Improving {{People}}'s {{Experiences Through Community Forums}}},
  author = {Harris, Brent},
  date = {2022-11-16T17:00:16+00:00},
  url = {https://about.fb.com/news/2022/11/improving-peoples-experiences-through-community-forums/},
  urldate = {2023-09-26},
  langid = {american},
  organization = {{Meta}}
}

@online{ovadya2023a,
  title = {‘{{Platform Democracy}}’—a Very Different Way to Govern Big Tech: {{Facebook}} Is Trying \textasciitilde{} It. {{Twitter}}, {{Google}}, {{OpenAI}}, and Other Companies Should Too.},
  author = {Ovadya, Aviv},
  date = {2023-07-10},
  url = {https://reimagine.aviv.me/p/platform-democracy-a-different-way-to-govern},
  urldate = {2023-09-26},
  langid = {english},
  organization = {{Reimagining Technology}}
}

@online{zaremba2023,
  title = {Democratic Inputs to {{AI}}},
  author = {Zaremba, Wojciech and Dhar, Arka and Ahmad, Lama and Eloundou, Tyna and Shibani Santurkar, Shibani and Agarwal, Sandhini and Leung, Jade},
  date = {2023-05-25},
  url = {https://openai.com/blog/democratic-inputs-to-ai},
  urldate = {2023-09-26},
  langid = {american}
}

@online{thewhitehouse2023,
  title = {{{FACT SHEET}}: {{Biden-Harris Administration Secures Voluntary Commitments}} from {{Leading Artificial Intelligence Companies}} to {{Manage}} the {{Risks Posed}} by {{AI}}},
  shorttitle = {{{FACT SHEET}}},
  author = {The White House},
  date = {2023-07-21T09:00:00+00:00},
  url = {https://www.whitehouse.gov/briefing-room/statements-releases/2023/07/21/fact-sheet-biden-harris-administration-secures-voluntary-commitments-from-leading-artificial-intelligence-companies-to-manage-the-risks-posed-by-ai/},
  urldate = {2023-09-26},
  langid = {american},
  organization = {{The White House}}
}

@article{schuett2023a,
  title = {Risk {{Management}} in the {{Artificial Intelligence Act}}},
  author = {Schuett, Jonas},
  date = {2023-02-08},
  journaltitle = {European Journal of Risk Regulation},
  shortjournal = {Eur. j. risk regul.},
  pages = {1--19},
  issn = {1867-299X, 2190-8249},
  doi = {10.1017/err.2023.1},
  urldate = {2023-09-26},
  langid = {english}
}

@report{tabassi2023,
  title = {{{AI Risk Management Framework}}: {{AI RMF}} (1.0)},
  shorttitle = {{{AI Risk Management Framework}}},
  author = {Tabassi, Elham},
  date = {2023},
  number = {error:  NIST AI 100-1},
  pages = {error:  NIST AI 100-1},
  institution = {{National Institute of Standards and Technology}},
  location = {{Gaithersburg, MD}},
  doi = {10.6028/NIST.AI.100-1},
  urldate = {2023-09-26}
}

@online{centerforlong-termcybersecurity2023,
  title = {{{UC Berkeley AI Risk-Management Standards Profile}} for {{General-Purpose AI Systems}} ({{GPAIS}}) and {{Foundation Models}}},
  author = {{Center for Long-Term Cybersecurity}},
  date = {2023-08-29},
  url = {https://cltc.berkeley.edu/seeking-input-and-feedback-ai-risk-management-standards-profile-for-increasingly-multi-purpose-or-general-purpose-ai/},
  urldate = {2023-09-26},
  langid = {american},
  organization = {{CLTC}}
}

@article{barrett2023,
  title = {Actionable {{Guidance}} for {{High-Consequence AI Risk Management}}: {{Towards Standards Addressing AI Catastrophic Risks}}},
  shorttitle = {Actionable {{Guidance}} for {{High-Consequence AI Risk Management}}},
  author = {Barrett, Anthony M. and Hendrycks, Dan and Newman, Jessica and Nonnecke, Brandie},
  date = {2023-02-23},
  eprint = {2206.08966},
  eprinttype = {arxiv},
  eprintclass = {cs},
  doi = {10.48550/arXiv.2206.08966},
  urldate = {2023-09-26},
  pubstate = {preprint},
  journaltitle = {}
}

@report{international2001iaea,
  type = {TECDOC},
  title = {Applications of Probabilistic Safety Assessment ({{PSA}}) for Nuclear Power Plants},
  author = {International Atomic Energy Agency},
  date = {2001},
  number = {1200},
  institution = {{International Atomic Energy Agency}},
  location = {{Vienna}},
  url = {https://www-pub.iaea.org/mtcd/publications/pdf/te_1200_prn.pdf},
  issue = {1200}
}

@report{anthropic2023c,
  title = {Model {{Card}} and {{Evaluations}} for {{Claude Models}}},
  author = {Anthropic},
  date = {2023},
  url = {https://www-files.anthropic.com/production/images/Model-Card-Claude-2.pdf}
}

@inproceedings{raji2019,
  title = {Actionable {{Auditing}}: {{Investigating}} the {{Impact}} of {{Publicly Naming Biased Performance Results}} of {{Commercial AI Products}}},
  shorttitle = {Actionable {{Auditing}}},
  booktitle = {Proceedings of the 2019 {{AAAI}}/{{ACM Conference}} on {{AI}}, {{Ethics}}, and {{Society}}},
  author = {Raji, Inioluwa Deborah and Buolamwini, Joy},
  date = {2019-01-27},
  pages = {429--435},
  publisher = {{ACM}},
  location = {{Honolulu HI USA}},
  doi = {10.1145/3306618.3314244},
  urldate = {2023-09-26},
  eventtitle = {{{AIES}} '19: {{AAAI}}/{{ACM Conference}} on {{AI}}, {{Ethics}}, and {{Society}}},
  isbn = {978-1-4503-6324-2},
  langid = {english}
}

@inproceedings{raji2022,
  title = {Outsider {{Oversight}}: {{Designing}} a {{Third Party Audit Ecosystem}} for {{AI Governance}}},
  shorttitle = {Outsider {{Oversight}}},
  booktitle = {Proceedings of the 2022 {{AAAI}}/{{ACM Conference}} on {{AI}}, {{Ethics}}, and {{Society}}},
  author = {Raji, Inioluwa Deborah and Xu, Peggy and Honigsberg, Colleen and Ho, Daniel},
  date = {2022-07-26},
  pages = {557--571},
  publisher = {{ACM}},
  location = {{Oxford United Kingdom}},
  doi = {10.1145/3514094.3534181},
  urldate = {2023-09-26},
  eventtitle = {{{AIES}} '22: {{AAAI}}/{{ACM Conference}} on {{AI}}, {{Ethics}}, and {{Society}}},
  isbn = {978-1-4503-9247-1},
  langid = {english}
}

@online{stabilityai2022,
  title = {Stable {{Diffusion}} 2.0 {{Release}}},
  author = {{Stability AI}},
  date = {2022-11-24},
  url = {https://stability.ai/blog/stable-diffusion-v2-release},
  urldate = {2023-09-26},
  langid = {british}
}

@online{iso2023,
  title = {{{ISO}}/{{IEC}} 23894:2023},
  shorttitle = {{{ISO}}/{{IEC}} 23894},
  author = {{ISO}},
  date = {2023-02},
  url = {https://www.iso.org/standard/77304.html},
  urldate = {2023-09-26},
  langid = {english}
}

@online{partnershiponaistaff2023,
  title = {{{PAI Is Collaboratively Developing Shared Protocols}} for {{Large-Scale AI Model Safety}}},
  author = {{Partnership on AI Staff}},
  date = {2023-04-06T15:03:52+00:00},
  url = {https://partnershiponai.org/pai-is-collaboratively-developing-shared-protocols-for-large-scale-ai-model-safety/},
  urldate = {2023-09-26},
  langid = {american},
  organization = {{Partnership on AI}}
}

@report{partnershiponaistaff2021,
  title = {Managing the {{Risks}} of {{AI Research}}: {{Six Recommendations}} for {{Responsible Publication}}},
  shorttitle = {Managing the {{Risks}} of {{AI Research}}},
  author = {Partnership on AI Staff},
  date = {2021-05-06},
  url = {https://partnershiponai.org/paper/responsible-publication-recommendations/},
  urldate = {2023-09-26},
  langid = {american}
}

@online{microsoft2023,
  title = {Microsoft, {{Anthropic}}, {{Google}}, and {{OpenAI}} Launch {{Frontier Model Forum}}},
  author = {Microsoft},
  date = {2023-07-26T09:59:56+00:00},
  url = {https://blogs.microsoft.com/on-the-issues/2023/07/26/anthropic-google-microsoft-openai-launch-frontier-model-forum/},
  urldate = {2023-09-26},
  langid = {american},
  organization = {{Microsoft On the Issues}}
}

@book{americanlawinstitute1965,
  title = {Restatement of the {{Law}} ({{Second}}) {{Torts}}},
  author = {{American Law Institute}},
  date = {1965},
  publisher = {{The American Law Institute}},
  location = {{Philadelphia, PA}},
  url = {https://www.ali.org/publications/show/torts/}
}

@book{americanlawinstitute1998,
  title = {Restatement of the {{Law}} ({{Third}}) {{Torts}}: {{Products Liability}}},
  author = {{American Law Institute}},
  date = {1998},
  publisher = {{The American Law Institute}},
  location = {{Philadelphia, PA}},
  url = {https://www.ali.org/publications/show/torts-third/}
}

@article{goldberg2001,
  title = {The {{Restatement}} ({{Third}}) and the {{Place}} of {{Duty}} in {{Negligence Law}}},
  author = {Goldberg, John C. P. and Zipursky, Benjamin C.},
  date = {2001-04-01},
  journaltitle = {Vanderbilt Law Review},
  volume = {54},
  number = {3},
  pages = {657},
  url = {https://scholarship.law.vanderbilt.edu/vlr/vol54/iss3/2}
}

@book{landes1987,
  title = {The {{Economic Structure}} of {{Tort Law}}:},
  shorttitle = {The {{Economic Structure}} of {{Tort Law}}},
  author = {Landes, William M. and Posner, Richard A.},
  date = {1987-05-20},
  publisher = {{Harvard University Press}},
  location = {{Cambridge, MA}},
  isbn = {978-0-674-86403-0},
  pagetotal = {329}
}

@article{hacker2023,
  title = {The {{European AI}} Liability Directives – {{Critique}} of a Half-Hearted Approach and Lessons for the Future},
  author = {Hacker, Philipp},
  date = {2023-11},
  journaltitle = {Computer Law \& Security Review},
  shortjournal = {Computer Law \& Security Review},
  volume = {51},
  pages = {105871},
  issn = {02673649},
  doi = {10.1016/j.clsr.2023.105871},
  urldate = {2023-09-26},
  langid = {english}
}

@online{mulani2023,
  title = {Proposing a {{Foundation Model Information-Sharing Regime}} for the {{UK}} | {{GovAI Blog}}},
  author = {Mulani, Nikhil and Whittlestone, Jess},
  date = {2023-06-16},
  url = {https://www.governance.ai/post/proposing-a-foundation-model-information-sharing-regime-for-the-uk},
  urldate = {2023-09-26},
  langid = {english}
}

@article{anderljung2023b,
  entrysubtype = {magazine},
  title = {How to {{Prevent}} an {{AI Catastrophe}}},
  author = {Anderljung, Markus and Scharre, Paul},
  date = {2023-08-14},
  journaltitle = {Foreign Affairs},
  url = {https://www.foreignaffairs.com/world/how-prevent-ai-catastrophe-artificial-intelligence},
  urldate = {2023-09-26},
  langid = {american}
}

@online{henshall2023,
  title = {The {{Heated Debate Over Who Should Control Access}} to {{AI}}},
  author = {Henshall, Will},
  date = {2023-08-25T19:12:58},
  url = {https://time.com/6308604/meta-ai-access-open-source/},
  urldate = {2023-09-26},
  langid = {english},
  organization = {{Time}}
}

\clearpage

\appendix

\section{AI Model Component Guide}\label{appendixA}
\setlength{\tabcolsep}{4pt}
\begingroup
\setlength\LTleft{-0.9cm}
\renewcommand{\arraystretch}{1.2}
\small
\rowcolors{1}{white}{light-gray}
\begin{longtable}{@{}|>{\arraybackslash\centering}m{2.5cm}|>{\arraybackslash\centering}m{2.2cm}|m{5.3cm}|m{5.25cm}|}
\hiderowcolors
\hline
\multicolumn{4}{|c|}{\small\bfseries Table 6: AI Model Component Guide\rule[-2.5mm]{0pt}{7mm}}\\\hline
\multicolumn{1}{|c|}{\textbf{Component}} & \multicolumn{1}{c|}{\textbf{Subcomponent}} & \multicolumn{1}{c|}{\textbf{Definition}} & \multicolumn{1}{c|}{\begin{tabular}{l}\bfseries What does access to this component \\[-2pt]\bfseries  allow actors to do?\end{tabular}} \\ \hline 
\endfirsthead

\multicolumn{4}{c}%
{{\bfseries \tablename\ \thetable{} -- continued from previous page}} \\
\hline \multicolumn{1}{|c|}{\textbf{Component}} & \multicolumn{1}{c|}{\textbf{Subcomponent}} & \multicolumn{1}{c|}{\textbf{Definition}} & \multicolumn{1}{c|}{\begin{tabular}{l}\bfseries What does access to this component \\[-2pt]\bfseries  allow actors to do?\end{tabular}} \\ \hline 
\endhead

\hline \multicolumn{4}{|r|}{{Continued on next page}} \\ \hline
\endfoot

\hline \hline
\endlastfoot
\showrowcolors

\textbf{Model weights} &  & The variables or numerical values used to specify how the input (e.g., text describing an image) is transformed into the output (e.g., the image itself) & {[}See trained weights{]} \\
 & \textit{Trained weights} & The final values of model weights after they have been updated during training & Alone, nothing; but when combined with the model architecture, any actor can run or fine-tune the optimized model with very low computing costs \\
 & \textit{Model weight snapshots} & The record of the different weight values as they were updated during training & Combined with model architecture, actors could run or fine-tune partially-optimized systems \\
\textbf{Hyperparameters}  &  & The variables used to define other parts of the model, such as model architecture (e.g., the number of layers in the model) and training process (e.g., the strength of regularization in the loss function) & {[}See optimized hyperparameters{]} \\
 & \textit{Optimized hyperparameters} & The hyperparameter values chosen through the hyperparameter optimization process that optimize the efficiency of the training process and increase the model's performance on the training task(s) & Immediately train model more efficiently by skipping the computationally-expensive hyperparameter search; this enables actors to train higher-performance models for a fixed computing cost \\
 & \textit{Methods for hyperparameter optimization} & The techniques used to optimize the hyperparameter for model performance (e.g., grid search, random search, Bayesian optimization); also known as hyperparameter tuning & Leverage known techniques to efficiently find the best model configurations \\
\textbf{Data processing code} &  & The code used to obtain raw training data and convert it into the form necessary for model training & Reproduce the full data pipeline that supplies training data to the model \\
 & \textit{Data cleaning} & The code used to transform the training data into a form more amenable for model training (e.g., normalization, removing invalid data, etc.) & Transform new data into the structure expected by the model and ensure data compatibility \\
 & \textit{Synthetic data creation} & The code used to generate additional, artificial data that is similar to the original training data; synthetic data is useful because training on more data sometimes improves model performance & Generate additional training data with similar statistical properties as the original \\
 & \textit{Data loading} & The code used to transform the cleaned training data into the correct structure / format to be input directly into the model (e.g. transforming data into tensors for training on high-performance chips) & Feed new data into the model seamlessly to enable training \\
\textbf{Training code} &  & The code that defines the model architecture and implements the algorithms used to optimize the model weights during training & Rebuild the model architecture from scratch and train it end-to-end with the same code \\
 & \textit{Model architecture} & The code specifying the structure and design of an AI model, including the types of layers, the connections between them, and any additional components or features that need to be incorporated; it also specifies the types of inputs and outputs to the model, how input data are processed, and how learning happens in the model & Alone, understand better how to train similar models; with trained weights, any actor can run or fine-tune the model \\
 & \textit{Loss function / reward function} & The code that defines the loss function: a mathematical formula that measures model's performance on the training task (e.g. MSE loss); the loss function is critical because minimizing it during training guides the optimization of the model weights & Better understand how to train similar models \\
 & \textit{Saving and loading models} & The code that handles saving the trained model parameters or weights to disk or other storage mediums, allowing the parameters to be loaded and reused for inference or further fine-tuning & Understand better how to distribute trained models \\
 & \textit{Training loop} & The training loop code iterates over the training data; within each iteration, it feeds some input data to the model, computes the loss, and updates the model's weights using the chosen optimization algorithm & Run full end-to-end training from raw data to final model (given training data and model architecture) \\
 & \textit{Hyperparameter optimization code} & The code used to optimize the hyperparameters to improve performance, implementing the methods for hyperparameter optimization (see above) & Discover optimal hyperparameters efficiently and create more capable models faster \\
\textbf{Related models} &  & Some AI systems rely on multiple models, either during the training/fine-tuning process or during inference; for instance, after initial training, many foundation models are fine-tuned via a related Reinforcement Learning from Human Feedback (RLHF) model and, more directly, Meta's CICERO combines a language processing model with a strategic reasoning model & Related models cannot be easily used on their own, but would help actors understand how to integrate different types of AI model into a single system \\
 & \textit{Guidelines for human evaluators in RLHF} & The instructions specifying what kind of feedback human evaluators should provide on the outputs from the foundation model; this feedback is then used in the RLHF training process & Understand how to efficiently obtain high-quality training data from human labelers \\
\textbf{Inference code (prediction or deployment code)} &  & The code that, given the model weights and architecture, implements the trained model; in other words, it runs the AI model and allows it to perform tasks (like writing, classifying images and playing games) & Generate model outputs and use the model directly, understand how to efficiently run the model and how to integrate it into production systems \\
 & \textit{Safety code} & Additional code is often included within the inference code to prevent malicious or harmful use of the model (e.g., preventing users from generating pornographic images) & Understand how developers tried to prevent misuse of the model \\
\textbf{Training strategies} &  & Specific techniques used to train the model (e.g., how long to train the model for); these are specified in the training code but also communicated at a high-level in associated papers and model cards & Understand which techniques boost training efficiency and thus model performance for a fixed computing cost \\
\textbf{Training data} &  & The data used to train and test the model (for instance, pictures for an image recognition model or internet webpages for a large language model) & Understand features of the data used to train the model and, given model architecture and training code, train the model \\
 & \textit{Data labels} & Sometimes, training data are labeled (e.g., a label for a picture could be a caption or description of the image); labels enable evaluation during training about how well the machine learning model is predicting the label, but they are not always necessary depending on the model being trained & Understand how labeling takes place (and whether it is outsourced to a third-party, for example), train or retrain models (depending on the model) \\
 & \textit{Testing data} & To fairly evaluate how well a model performs, its predictions are often evaluated on a new set of testing data that was never used during training; this can be a portion of the original training data that is "held-out" and excluded from training, or a new dataset & Same as training data (but to a lesser extent since there tends to be more training than testing data), evaluate performance when training or retraining models \\
 & \textit{Evaluation Metrics} & Measures against which to assess the performance of the model during training; these metrics may vary depending on the specific task; commonly-used metrics include accuracy, precision, recall, or perplexity & Understand how the model capabilities were assessed, evaluate performance when training or retraining models \\
\textbf{Tacit knowledge} &  & Additional information known only to certain researchers and engineers within AI labs that is often very helpful (and sometimes necessary) to train advanced AI models; for example, Phuong \& Hutter (2022) summarizes some tacit knowledge relating to the Transformer architecture & Train more advanced models more efficiently \\
\textbf{Software stack} &  & A set of software or code libraries that enables the training of an AI model; this includes machine learning frameworks such as PyTorch, TensorFlow and Jax, as well as compilers and optimized libraries like CUDA, cuDNN and Triton that enable training on advanced GPUs & Knowing the version of certain software tools would save time when building training pipelines \\ \hline
\end{longtable}
\endgroup

\end{document}